\documentclass[12pt]{article}

\usepackage{fullpage}
\usepackage{microtype} 
\usepackage[utf8]{inputenc} 
\usepackage[T1]{fontenc}
\usepackage{helvet} 
\usepackage{sectsty} 
\allsectionsfont{\sffamily} 
\makeatletter\def\@seccntformat#1{\protect\makebox[0pt][r]{\csname
the#1\endcsname\hspace{11pt}}}\makeatother 

\makeatletter\renewcommand{\@dotsep}{1000} 

\usepackage{dsfont}
\usepackage{mathtools,amssymb,amsmath,bbold}
\usepackage{euscript}
\usepackage[bbgreekl]{mathbbol}
\DeclareSymbolFontAlphabet{\mathbbm}{bbold}
\DeclareSymbolFontAlphabet{\mathbb}{AMSb}
\usepackage{bm} 
\usepackage[bb=libus, scr=boondoxo]{mathalpha}
\usepackage{braket}
\usepackage{slashed}
\usepackage{multicol}

\usepackage{xcolor}
\usepackage{tcolorbox}
\usepackage{enumitem}
\usepackage{cite}
\usepackage[colorlinks=true,linkcolor=blue,citecolor=purple,urlcolor = blue,linktocpage=true]{hyperref}
\usepackage{accents}
\usepackage{harmony}

\newcommand{\be}{\begin{equation}}
\newcommand{\ee}{\end{equation}}
\newcommand{\bea}{\begin{eqnarray}}
\newcommand{\eea}{\end{eqnarray}}
\newcommand{\bee}{\begin{equation} \begin{aligned}}
\newcommand{\eee}{ \end{aligned} \end{equation}}
\newcommand{\lb}{\label}
\newcommand{\qq}{\qquad}
\newcommand{\sss}{\scriptscriptstyle}

\newcommand{\cL}{\mathcal{L}}

\newcommand{\sR}{\mathscr{R}}

\newcommand{\sD}{\mathscr{D}}

\newcommand{\mK}{\mathsf{K}}

\newcommand{\M}{\mathscr{M}} 
\newcommand{\N}{\mathscr{N}} 
\renewcommand{\H}{\mathscr{H}} 
\newcommand{\sC}{\mathscr{sC}} 

\renewcommand{\S}{\mathscr{S}} 

\newcommand{\Neq}{\stackrel{\sss \N}{=}} 
\newcommand{\EOM}{\hat{=}} 

\newcommand{\bvol}{\bm{\epsilon}} 
\newcommand{\volH}{\bm{\epsilon}_{\sss \bm{\H}}} 
\newcommand{\volN}{\bm{\epsilon}_{\sss \bm{\N}}}
\newcommand{\volM}{\bm{\epsilon}}
\newcommand{\volS}{\bm{\epsilon}_{\sss \bm{\S}}}

\newcommand{\rd}{\mathrm{d}} 
\newcommand{\Lie}{\mathcal{L}} 
\newcommand{\exd}{\bm{\rd}} 
\newcommand{\dH}{\mathbb{d}} 
 \newcommand{\LH}{\mathbb{L}} 
\newcommand{\var}{\bbdelta\hspace{-0.49em}\bbdelta} 

\renewcommand{\bar}{\overline}
\renewcommand{\a}{{\alpha}} 
\newcommand{\ba}{\bar{\alpha}} 
\renewcommand{\v}{v} 
\newcommand{\ac}{\varphi} 
\newcommand{\vor}{w} 
\newcommand{\p}{\wp} 
\newcommand{\E}{\EuScript{E}} 
\newcommand{\NN}{\EuScript{N}} 
\newcommand{\A}{\EuScript{A}} 
\newcommand{\J}{\EuScript{J}} 
\renewcommand{\P}{\EuScript{P}} 
\newcommand{\T}{\EuScript{T}} 
\newcommand{\Wein}{\mathsf{N}} 
\newcommand{\EM}{\mathsf{T}} 
\newcommand{\btheta}{\bar{\theta}{}} 
\newcommand{\s}{\sigma} 
\newcommand{\bs}{\bar{\sigma}{}} 
\newcommand{\RC}{\sR} 
\newcommand{\W}{W} 
\newcommand{\xit}{\hat{\xi}}
\newcommand{\sv}{\mathrm{v}}

\newcommand{\e}{\mathrm{e}} 
\newcommand{\la}{\langle}
\newcommand{\ra}{\rangle}
\newcommand{\pa}{\partial}

\newcommand{\mr}{\mathring}

\begin{document}
\title{\Large{\textsf{\bfseries 
Geometry of Carrollian
\\ 
 Stretched Horizons \\
}}}
\author{Laurent Freidel$^1$ \ \& \ Puttarak Jai-akson$^2$}
\date{\small{\textit{
$^1$Perimeter Institute for Theoretical Physics,\\ 31 Caroline Street North, Waterloo, Ontario, Canada N2L 2Y5\\
$^2$RIKEN iTHEMS, Wako, Saitama 351-0198, Japan\\}}}

\maketitle

\begin{abstract}
In this paper, we present a comprehensive toolbox for studying Carrollian stretched horizons, encompassing their geometry, dynamics, symplectic geometry, symmetries, and corresponding Noether charges. We introduce a precise definition of ruled stretched Carrollian structures (sCarrollian structures) on any surface, generalizing the conventional Carrollian structures of null surfaces, along with the notions of sCarrollian connection and sCarrollian stress tensor. Our approach unifies the sCarrollian (intrinsic) and stretched horizon (embedding) perspectives, providing a universal framework for any causal surface, whether timelike or null. We express the Einstein equations in sCarrollian variables and discuss the phase space symplectic structure of the sCarrollian geometry. Through Noether's theorem, we derive the Einstein equation and canonical charge and compute the evolution of the canonical charge along the transverse (radial) direction. The latter can be interpreted as a spin-2 symmetry charge.

Our framework establishes a novel link between gravity on stretched horizons and Carrollian fluid dynamics and unifies various causal surfaces studied in the literature, including non-expanding and isolated horizons. We expect this work to provide insights into the hydrodynamical description of black holes and the quantization of null surfaces.

\end{abstract}

\thispagestyle{empty}
\newpage
\setcounter{page}{1}


\hrule
\setcounter{tocdepth}{2}
\tableofcontents
\vspace{0.7cm}
\hrule


\section{Introduction}

Stretched horizons are among the most intriguing and essential surfaces in gravitational physics, with a wide range of applications, from the study of physics near black hole horizons to the study of the geometry of asymptotic infinity. In this work, we introduce and study the concept of the \emph{stretched} Carrollian structure, a novel framework that extends Carrollian geometry --- previously thought to exist only on null surfaces --- to include timelike stretched horizons. This unified treatment offers a comprehensive approach, bridging the gap between different types of surfaces and providing a robust toolbox for their analysis. Our framework encompasses detailed examinations of the geometry, dynamics, symplectic structure, symmetries, and charges of stretched horizons, paving the way for new insights and implications. This work builds upon and extends our previous work \cite{Freidel:2022vjq}.


Historically, the concept of stretched horizons first emerged in the context of the membrane paradigm \cite{Damour:1978cg,thorne1986black,Price:1986yy}, where they serve as an effective tool for understanding black hole dynamics by offering a quasi-local description of horizon properties just outside the actual event (null) horizon. A prominent feature of this paradigm is its treatment of the stretched horizon as a fluid-like membrane with physical properties, where the projection of the Einstein equation on the horizon is analogous to hydrodynamical equations. This viewpoint significantly enhances our ability to model and predict black hole behavior. This approach is closely related to the fluid-gravity correspondence \cite{Bhattacharyya:2007vjd, Son:2007vk,Rangamani:2009xk, Hubeny:2010wp,Hubeny:2011hd}, which originally developed within the AdS/CFT framework, connecting gravitational systems to hydrodynamics. The connection between these two approaches has also been investigated in \cite{Bredberg:2010ky,Bredberg:2011jq,Brattan:2011my,Emparan:2013ila}.

One fundamental puzzle of the membrane paradigm was that the limit of the stretched horizon towards the horizons involves divergences due to the infinite redshift effects. Recently, this puzzle was resolved in the work of Donnay and Marteau \cite{Donnay:2019jiz} (see also the earlier argument by Penna \cite{Penna:2018gfx})
who demonstrated that the near-horizon geometry and the Einstein equation imprinted on the horizon can be understood through Carrollian geometry (introduced by Lévy-Leblond \cite{Leblond1965} and Sen Gupta \cite{SenGupta} in their study of the $c \to 0$ limit of relativity) \cite{Duval:2014uoa, 
Duval:2014lpa, Duval:2014uva, Morand:2018tke, Ciambelli:2019lap}. This perspective connected the study of horizon with the recently established Carrollian hydrodynamics \cite{Ciambelli:2018xat,Ciambelli:2018ojf,Petkou:2022bmz,Freidel:2022bai,Armas:2023dcz}. This has sparked a research trend aiming to understand black hole mechanics from the Carrollian perspective, aptly named the Carrollian membrane paradigm \cite{Redondo-Yuste:2022czg, Ciambelli:2023mvj}. This novel perspective has brought the study of null boundaries and stretched horizons closer together.
It also has enabled a profound re-foundation of our understanding of charges, symmetries and symplectic structures on null surfaces and boosted our understanding of the near-horizon geometry of black holes and the dynamics of null surfaces \cite{Donnay:2015abr, Donnay:2016ejv, Chandrasekaran:2018aop, Hopfmuller:2018fni, Jafari:2019bpw, Chandrasekaran:2020wwn, Adami:2021nnf, Adami:2021kvx, Chandrasekaran:2021hxc, Chandrasekaran:2023vzb, Odak:2023pga, Rignon-Bret:2024wlu}. 
This viewpoint has then offered insights into the interplay between gravity and field theories in the ultra-relativistic (or ultra-local) limit.

Recently, it was revealed in \cite{Freidel:2024tpl} that null infinity can be understood, using Penrose compactification \cite{Penrose:1962ij, Geroch:1973am, PenroseRindler1}, as a stretched horizon in the compactified spacetime. The asymptotic stress tensor and the Bondi asymptotic equations of motion were derived from the conservation of the Carrollian stress tensor. This result indicates that the stretched horizon is the proper geometrical structure to understand and unify asymptotic infinity with the geometry of the finite null surface. With hindsight, it was already noticed, first by Wall \cite{Wall:2011hj}, then by Donnay and Giribet \cite{Donnay:2015abr, Donnay:2016ejv} that finite null surfaces carry the same group of symmetry as asymptotic infinity: the BMS group \cite{Bondi:1962px,Sachs:1962wk}. This correspondence between symmetries of asymptotic infinity and null surface was later generalized to extensions of the BMS symmetries \cite{Barnich:2010eb,Campiglia:2014yka,Barnich:2016lyg,Compere:2018ylh, Chandrasekaran:2018aop ,Freidel:2021fxf}. These extensions have been instrumental in developing celestial holography \cite{Kapec:2016jld, Cheung:2016iub, Pasterski:2016qvg ,Pasterski:2017kqt} (see also \cite{Strominger:2017zoo, Raclariu:2021zjz,Pasterski:2021rjz, Pasterski:2021raf, McLoughlin:2022ljp,Pasterski:2023ikd} for reviews and references therein). Furthermore, Carrollian geometry has proven to be a natural language and a valuable tool  relating bulk gravitational theories to Carrollian field theories on the codimension-1 null boundary \cite{Bagchi:2022emh,Campoleoni:2022wmf,Bagchi:2022owq, Donnay:2022aba, Donnay:2022sdg,Bagchi:2019xfx,Gupta:2020dtl,Bagchi:2022eav}. It is also proving to give a new perspective on the quantization of gravity along null surfaces  \cite{Freidel:2023bnj,Ciambelli:2023mir, Wieland:2024dop, Wieland:2024kzw}. See also \cite{Ashtekar:2024bpi} that seek the unification of weakly isolated horizons and null infinity.

Seamlessly connecting stretched horizons (or timelike surfaces in general) with null surfaces requires treating both types of surfaces on the same common ground. This task was usually hindered by the fact that, for the timelike case, one typically works with the induced metric and Levi-Civita connection, while similar notions were lacking for the null case. The surface stress tensor and the connection used to define its conservation laws are radically different in both the timelike case \cite{Brown:1992br, Brown:2000dz} and the null case \cite{Donnay:2015abr, Donnay:2016ejv, Chandrasekaran:2018aop, Hopfmuller:2018fni, Jafari:2019bpw, Chandrasekaran:2020wwn, Adami:2021nnf, Adami:2021kvx, Chandrasekaran:2021hxc}. In the conventional membrane paradigm, ad-hoc renormalization schemes are often required to achieve a non-singular null limit. We have addressed and resolved these issue in \cite{Freidel:2022vjq} by utilizing the embedding perspective and Mars-Senovilla rigging technique \cite{Mars:1993mj, Mars:2013qaa}. 

The present work generalizes and completes the geometric construction of the Carrollian stretched structure laid down in \cite{Freidel:2022vjq}. We precisely define the notion of a stretched Carrollian structure for causal surfaces (null or timelike) from an intrinsic perspective, with emphasis on the Carrollian connection. We also establish the connection between the intrinsic and extrinsic (embedding) pictures.
 Similar results have been obtained by Mars, Sánchez-Pérez, and Manzano \cite{Mars:2022gsa, Manzano:2023oub, Mars:2023hty}, who developed the intrinsic geometry of hypersurfaces of arbitrary causal type, referred to as the geometrical hypersurface data formalism. 
 Additionally, we include the expressions for all components of the Einstein equation, the general symmetries of a stretched horizon beyond the tangential ones, and the corresponding Noether charges. We also discuss how these charges evolve along the direction transverse to the horizon, revealing their interpretation as spin-2 evolution equations and charges. To make this paper self-contained, we provide most of the derivations in the appendices. We expect this work will enrich the geometric toolbox available for the study of stretched horizons and become a go-to resource for anyone looking to study this or related topics. 

Let us also mention some other scenarios beyond those already discussed in which the concept of stretched horizons finds utility. First, in the context of the black hole information paradox, Susskind's works on complementarity utilize stretched horizons to address the paradox by proposing that all information passing through the event horizon is reflected and thus never lost, a perspective that heavily relies on the properties of stretched horizons \cite{Susskind:1993if}. Studies on the mathematical structure of horizons by Chruściel et al. also benefit from the detailed geometric description provided by the stretched horizon (near-horizon) perspective  \cite{Chrusciel:2017vie,Chrusciel:2010gq,Chrusciel:2023onh,Chrusciel:2008js}. Furthermore, the brick wall model proposed by 't Hooft, which attempts to regularize the divergences encountered near the event horizon by introducing a stretched horizon as a cutoff, is another critical application \cite{tHooft:1984kcu}. In quantum gravity, the concept of stretched horizons aids in regularizing infinities and provides a manageable boundary condition for the fields, facilitating more tractable calculations \cite{tHooft:1996rd, Ghosh:2011fc}. This approach is instrumental in the study of horizon entropy and Hawking radiation, where stretched horizons offer a physical and intuitive means of imposing boundary conditions. By encompassing these diverse applications, the concept of stretched horizons is an indispensable tool in advancing our understanding of black hole mechanics and quantum gravity. 

\paragraph{Notations:} In this article, we denote a 4-dimensional spacetime by $\M$, a 3-dimensional timelike stretched horizon by $\H$, a null boundary by $\N$, and a 2-dimensional sphere by $\S$. The notations we will use here are listed below:
\begin{itemize} [topsep=0pt,itemsep=0pt,partopsep=0pt, parsep=0pt]
\item Small Latin letters $a,b,c,...$ are spacetime indices. As usual, they are raised and lowered by a spacetime metric $g_{ab}$ and its inverse $g^{ab}$.
\item Small Latin letters $i,j,k,...$ are indices on 3-dimensional surfaces (both $\H$ and $\N$).
\item The capital letters $A,B,C,...$ are indices on a 2-dimensional horizontal space (or on $\S$). They are raised and lowered by a 2-sphere metric $q_{AB}$ and its inverse $q^{AB}$.
\item Differential forms are presented with boldface letters such as $\bm{k}, \bm{n}, \bm{\omega}, \bm{\epsilon},...$
\item The wedge product between differential forms is denoted by $\wedge$ as usual, while $\odot$ is used to denote a symmetric tensor product.
\item The Cartan exterior derivative on $\M$ is denoted by $\exd$. It is denoted by $\dH$ on $\H$ and $\N$. 
\item The directional derivative along a vector field $V$ is written as $V[ \cdot] = V^a \pa_a (\cdot)$. 
\item The 4-dimensional Levi-Civita connection, 3-dimensional sCarrollian connection and 2-dimensional horizontal connection are respectively denoted by $(\nabla_a, D_i, \sD_A)$.
\end{itemize}

\subsection{Summary of the results}
Let us outline this paper and highlight some key results:

\paragraph{sCarrollian structure:} 
Section \ref{sec:sCarrollian} is devoted to laying down the relevant geometric structures of Carrollian stretched horizons\footnote{We usually shorten it to just stretched horizon}. We begin with the introduction of a \emph{ruled stretched Carrollian structure} (sCarrollian\footnote{We thank Luca Ciambelli for suggesting this name.} for abbreviation) defined by the data $\sC = (\v^i, k_i, h_{ij}, \rho)$, where $\v^i$ is a vertical vector, $k_i$ is a ruling that serves as an Ehresmann connection providing the notion of horizontality, $h_{ij}$ is the metric on the stretched horizon, and $\rho = -\tfrac12 h_{ij}\v^i\v^j$ is called the \emph{stretching}. We then define a non-metric sCarrollian connection $D_i$ and, from there, various derivatives, such as $D_i \v^j$ and $D_i k_j$, as well as the generalized news tensor and the sCarrollian fluid stress tensor $\EM_i{}^j$ derived from them \cite{Chandrasekaran:2021hxc}. The conservation laws of the stress tensor are obtained by projecting the Einstein equation onto the stretched horizon. Furthermore, these objects encode information about conjugate momenta in the phase space of the sCarrollian fluid \cite{Freidel:2022bai}, which can also be derived from the gravitational (covariant) phase space \cite{Freidel:2022vjq,Freidel:2024tpl}. These properties solidify the correspondence between gravity and Carrollian fluid dynamics: the Carrollian membrane paradigm.

There are two key points regarding this definition of the sCarrollian structure. First, a stretched horizon is causal and all our geometric objects presented here apply equally to both timelike and null surfaces. The conventional (null) ruled Carrollian structure \cite{Duval:2014uoa,Duval:2014uva,Ciambelli:2019lap,Donnay:2019jiz,Ciambelli:2023mir}, comprising $(\v^i, k_i, q_{ij})$, can be viewed as a special limit of the timelike horizon, and all equations remain non-singular in this limit. The null case corresponds to the vanishing of the stretching $\rho \stackrel{\sss \mathrm{null}}{=} 0$, and the metric $h_{ij} = q_{ij} - 2\rho k_i k_j \stackrel{\sss \mathrm{null}}{=} q_{ij}$ is degenerate in the vertical direction. Our construction thus provides a unified treatment for (s)Carrollian geometries for general causal surfaces.

The second point, which was not clearly emphasized in \cite{Freidel:2022vjq}, is that the sCarrollian structure, as well as the connection and the fluid stress tensor, are \emph{intrinsic} data on the horizon \cite{Ciambelli:2023mir}. This is fundamentally different from the standard Brown-York construction, where the metric, connection, and stress tensor all descend from the geometry of the ambient spacetime in which the stretched horizon is embedded, thereby depending on extrinsic data. Here, the intrinsic picture is more akin to the construction of null surfaces \cite{Hopfmuller:2016scf,Hopfmuller:2018fni,Chandrasekaran:2021hxc,Chandrasekaran:2020wwn}. One important implication is that we can now discuss gravitational radiation, encoded in the connection's shear tensor $\bs_{ij} = q_{\la i}{}^k q_{j \ra}{}^l D_k k_l$, defined from  the intrinsic data (see the discussion in \cite{Freidel:2024tpl}). This perspective, where a choice of connection defines the notion of radiation, aligns with the work of Ashtekar and Streubel \cite{Ashtekar:1981bq} and later Herfray \cite{Herfray:2020rvq,Herfray:2021xyp} in the context of asymptotic null infinity.

\paragraph{Rigging structure:}

There are two perspectives one can take in describing the sCarrollian geometry of a stretched horizon: the intrinsic picture and the embedding picture. The former treats the stretched horizon as a stand-alone object intrinsically endowed with a sCarrollian structure. The latter relies on embedding a stretched horizon in an ambient spacetime of higher dimension, where the sCarrollian geometry is inherited from the spacetime geometry through the embedding. Remarkably, the two perspectives turn out to be equivalent! 

In Section \ref{sec:Rigging}, we consider a family of sCarrollian structures $\sC_r=\left(v^i(r), k_j(r), h_{ij}(r), \rho(r) \right)$ labeled by the function $r$. The stretched horizon with different $r$ can now be considered the leaves of the $r = \mathrm{constant}$ foliation of the surrounding spacetime. By utilizing the Mars-Senovilla rigging technique \cite{Mars:1993mj,Mars:2013qaa} (see also \cite{axioms10040284,Chandrasekaran:2021hxc} for situations involving null surfaces), we explicitly establish the correspondence between the two viewpoints and derive the relations between the sCarrollian structure and the rigging structure, including the sCarrollian connection and the sCarrollian stress tensor (see also \cite{Mars:2022gsa, Manzano:2023oub, Mars:2023hty}). We also discuss the arbitrariness in the embedding of the sCarrollian structure into the spacetime, leading to more general gauge fixings than those adopted in \cite{Freidel:2022vjq}. Additionally, we discuss the adapted coordinates, although we will not rely heavily on the choice of coordinates, thus presenting the results in a covariant manner.

\paragraph{Einstein equation:} Section \ref{sec: Einstein} discusses the Einstein equation $G_{ab} = 0$. In this section, we present all components of the Einstein equation expressed in terms of sCarrollian variables and highlight the simplifications that occur when considering the case of a null surface. We also explain that some equations exhibit dual interpretations, meaning they can be viewed as either time evolution (along $v^a$) equations or radial evolution (along $k^a$) equations.

Our constructions and equations are applicable to any general causal surface, encompassing all situations that are considered subclasses of causal surfaces. These include the notions of non-expanding horizons, weakly isolated horizons, and isolated horizons developed by Ashtekar and collaborators \cite{Ashtekar:2000hw, Ashtekar:2001jb, Lewandowski:2006mx, Ashtekar:2021wld, Ashtekar:2024bpi, Ashtekar:2024mme}. We demonstrate in Section \ref{sec: Einstein} that the equations relevant to these situations can be easily derived from our general equations, thereby connecting our work to theirs. The radial evolution equation is also useful in the (in)stability analysis of black holes under linear dynamical perturbations \cite{Hollands:2012sf}, and we provide some comments on this aspect as well. Finally, we also propose, through a choice of ruling, the definition of a \emph{hydrodynamical horizon}, which identifies more closely gravitational and fluid-like degrees of freedom.

\paragraph{Symmetries and charges:} In Section \ref{sec: Einstein-symmetries}, we discuss the symmetries of the Carrollian stretched horizon and the corresponding Noether charges. We start from the canonical pre-symplectic potential $\Theta^{\mathrm{can}}_{\H}$ of the horizon, which can be expressed in terms of the sCarrollian structure and the components of the fluid stress tensor, $\EM_i{}^j=-  k_i \left( \E \v^j + \J^j \right)+ \p_i \v^j + \left( \T_i{}^j + \P q_i{}^j \right)$, as
\begin{align}
\Theta_{\H}^{\mathrm{can}}   = - \int_{\H} \bigg( 
\left( \E \v^i + \J^i \right)  \delta k_i
+ \p_i \delta \v^i  -
\tfrac{1}{2}(\T^{ij} + \P q^{ij}) \delta q_{ij} 
   +\btheta \delta \rho \bigg) \volH.  \lb{Theta-can}
\end{align}
While this potential can be defined intrinsically, we show that it is also derivable using the covariant phase space formalism by treating the rigging structure as a background structure, from the pre-symplectic potential of the Einstein-Hilbert theory. These two potentials differ by a total variation term and a corner term \cite{Parattu:2015gga, Hopfmuller:2016scf, Hopfmuller:2018fni,Chandrasekaran:2021hxc,Freidel:2022vjq, Freidel:2024tpl}. 

Symmetry transformations preserving the background rigging structure comprise vertical (time) translations $T$, horizontal diffeomorphism $X^A$, transverse (radial) translations $R$, rescaling symmetry $W$, and shift symmetry $Z_A$. The former two transformations constitute the  diffeomorphism tangent to the surface, $\xit = T \v + X^Ae_A$,  which transform sCarrollian tensors covariantly. Their canonical charges contain the tangential components of the Einstein equation $G_i{}^an_a = 0$ as the constraints, which can be dually interpreted as the conservation laws for the stress tensor, $D_j \EM_i{}^j = 0$. The charge aspects are the Carrollian energy and the Carrollian momentum \cite{Donnay:2019jiz,Chandrasekaran:2021hxc,Freidel:2022vjq,Ciambelli:2023mir},
\begin{align}
Q^{\mathrm{can}}_{\xit} = - \int_{\H} \xit^b G_b{}^a n_a \volH + \int_{\pa \H} \xit^b  \EM_b{}^a  \bvol_a.
\end{align}
For the transverse translation $\xi_{\perp} = R k$, the constraint is $R_{kn} = -\frac{1}{2} q^{AB} G_{AB} + \rho G_{kk} = 0$, governing the radial development of the news scalar, $\Wein$.
Lastly, the rescaling and Lorentz shift symmetry, labelled respectively by $(W, Z_A)$ preserve the sCarrollian metric $h_{ij}$. The latter is gauge while the former is considered an edge mode symmetry whose charge aspect is given by the area.
These transformations $(R, W, Z_A)$ are also responsible for the non-covariance (anomaly) of the sCarrollian structure.

It is important to appreciate the results: the derivation of the Einstein equations $G_{in} = 0$ and $R_{kn} = 0$ from symmetries. While one might argue that these results are expected from the Einstein-Hilbert theory, we emphasize that the canonical potential $\Theta^{\mathrm{can}}_\H$ used in the derivation of the Noether charges are purely sCarrollian and intrinsic, making no reference to the bulk Einstein-Hilbert theory. In fact, as shown in \cite{Freidel:2022bai}, the expression \eqref{Theta-can} of $\Theta^{\mathrm{can}}_\H$ can be obtained from the Carrollian limit, $c \to 0$, of any relativistic theory with the inclusion of the stretching term $\btheta \delta \rho$. Therefore, $\Theta^{\mathrm{can}}_\H$ must take the same form for any sCarrollian theory. From this perspective, we should interpret our result in this way: Starting from the sCarrollian pre-symplectic potential on the stretched horizon, the Einstein equation emerges holographically!

Knowledge about the evolutions, symmetries, and charges is instrumental for the quantization of gravitational sub-regions bounded by null surfaces. This perspective has already been pursued in a series of works by Reisenberger et al. \cite{Reisenberger:2007pq, Reisenberger:2007ku,Reisenberger:2012zq,Fuchs:2017jyk,Reisenberger:2018xkn}, and more recently by Ciambelli, Leigh, and one of the authors in \cite{Ciambelli:2023mir}, as well as by Wieland in \cite{Wieland:2024dop,Wieland:2024kzw}.

\paragraph{Radial evolution and Spin-2:} 

Our consideration in Section \ref{sec:spin-2} addresses the spin-2 sector of the Einstein equation and the associated Noether charge. The existence and conservation laws of spin-2 and higher charges have been studied in the context of null infinity, with applications to soft theorems and celestial holography \cite{Freidel:2021qpz,Freidel:2021dfs,Freidel:2021ytz}. However, their analog in the finite, general null boundary case has rarely been explored. We aim to elucidate the method for obtaining the spin-2 charge and its evolution equation. Interestingly, these charges are revealed when considering the radial evolution of the canonical charge, specifically the one associated with tangential diffeomorphism, $Q^{\mathrm{can}}_{\xit}$. 

More precisely, we show that on the null surface $\N$, the radial evolution of the charge can be expressed as the contraction
\begin{align}
\Lie_k Q^{\mathrm{can}}_{\xit} &= I_{\xit} \widehat{\Theta}_\N + \left( I_{\xit} \widehat{\Theta}_{\pa \N} - \int_{\pa\N} \iota_{\xit}\bm{\mathscr{L}} \right) \ \EOM \  I_{\xit} \widehat{\Theta}_{\pa \N} - \int_{\pa\N} \iota_{\xit}\bm{\mathscr{L}},
\end{align}
where we defined $\bm{\mathscr{L}} := \mathscr{L} \volN$ and $\mathscr{L}$ is, up to a boundary term, the Einstein-Hilbert Lagrangian (see \eqref{LEH}). We also defined the new potentials 
\begin{alignat}{2}
&\widehat{\Theta}_\N &&:=  -\int_\N \left(  G_{\v k} \delta k_v -   G_\v{}^ A\delta k_A - G_{k A}\delta \v^A + \tfrac12 G^{AB}\delta q_{AB}\right) \volN \\
&\widehat{\Theta}_{\pa\N} &&:= - \int_{\S_\sv} \left( \tfrac12\btheta^{AB} \delta q_{AB} + \p^A \delta k_A \right)\volS.
\end{alignat}
We clearly see now that the variations of elements of the Carrollian structure on $\N$ become the field-dependent and dynamical transformation parameters. Furthermore, from the bulk perspective, this feature can be regarded as stemming from the Bianchi identity $\nabla_b G_a{}^b = 0$. This result is intriguing, but we leave its interpretation and further exploration for future works.

\section{Carrollian Stretched Horizon} \lb{sec:sCarrollian}

In this section, we study the geometry of stretched horizons, denoted by $\H$. We assume that $\H$ is equipped with a kinematical Carrollian structure, i.e., $\H$ is a line bundle $\pi: \H \to \S$ over $\S$, which is restricted to be a 2-dimensional sphere. We denote general coordinates on $\H$ with $\{y^i\}$ and coordinates on $\S$ by $\sigma^A$.


This Carrollian fiber bundle \cite{Duval:2014uoa,Duval:2014uva,Duval:2014lpa,Ciambelli:2019lap, Freidel:2022bai, Freidel:2022vjq}, 
defines a one-dimensional vertical subspace of $T\H$ spanned 
by a Carrollian vector $\v = \v^i \pa_i$.
A \emph{ruling} of $\H$ consists of a  decomposition of $T\H$ into a direct sum of vertical and horizontal components. Such a ruling  is characterized by a choice of  Ehresmann connection $\bm{k} = k_i \exd y^i$. 
A stretched Horizon is in addition equipped with a causal metric $h_{ij}$. The \emph{stretching} of $\H$ is encoded into a positive scalar function $\rho \geq 0$ which proportional to the norm-square of the Carrollian vector.
The metric at each point of $\H$ is either null or timelike, depending on whether the stretching vanishes or not at the given point. 
The purpose of this geometrical structure is to describe on the same footing timelike, null or mixed type causal hypersurfaces. This allows us to understand, in a smooth manner, the limit from stretched to null horizons.   
By definition, a stretched horizon is a $3$-dimensional manifold equipped with a \emph{stretched Carrollian structure}.


\subsection{Stretched Carrollian structure}
 
More formally, a stretched Carrollian structure (sCarrollian hereafter) on the manifold $\H$ consists of the data $\sC = (\v^i, k_j, h_{ij}, \rho)$; where 
 $\v = \v^i \pa_i \in T\H$ is a vector, called the \emph{Carrollian vector}, belonging to the kernel of the differential map $\exd \pi$, while $\bm{k} = k_i \exd y^j \in T^*\H$ is an Ehresmann connection 1-form dual to $\v$ that provides a \emph{ruling} of $\H$. The tensor $h_{ij}$ is a metric on $\H$, and the positive function $\rho \geq 0$, which defines the \emph{stretching} of $\H$, is related to the norm-square of the Carrollian vector, $2\rho := - h_{ij}  \v^i \v^j$. The relationship between the stretched Carrollian data $(\v^i, k_j, h_{ij}, \rho)$ is 
\be\label{sCarrollian}
\v^i k_i  = 1, \qq \text{and} \qq h_{ij} \v^j = -2\rho k_i.
\ee 
The Carrollian vector allows us to define the notion of \emph{horizontal} forms, $\bm \alpha \in \Omega(\H)$, which are such that $\alpha_i v^i=0$.
Similarly, the Ehresmann connection defines a notion of \emph{horizontality}, where a vector $X\in T\H$ is horizontal if and only if $X^ik_i=0$. 
It also allows the definition of a dual metric, a symmetric 2-tensor $q^{ij}$ of corank one. By definition, $q^{ij}$ is such that $h_{ij}q^{jk}$ is a codimension-1 projector onto horizontal forms and vectors:
\be 
q^{ij} k_j=0, \qq \text{and} \qq q_i{}^k:= h_{ij}q^{jk}= \delta_i^k - k_i v^k.
\ee 
We see that the pair $(q^{ij},k_j)$ complements the pair $(h_{ij},v^j)$. The definition shows that $k$ is null since $q^{ij}k_ik_j=0$. 
In the following, the dual metric $q^{ij}$ is used to raise indices, while $h_{ij}$ is used to lower them.

The horizontal projection of the metric is denoted by $ q_{ij}:= q_i{}^kq_j{}^l h_{kl}$. It can be used to decompose the metric $h_{ij}$ as\footnote{Its inverse exists only when $\rho\neq0$, in which case it is given by $h^{ij}= q^{ij}-\frac{v^iv^j}{2\rho}$. The whole purpose of the construction is to avoid the need to use the inverse of $h$ in the definition of the connection. This eliminates any potential singularity linked with the limit  $\rho\to 0$.} 
\be 
h_{ij} = q_{ij}-2\rho k_i k_j.
\ee
When the stretching vanishes, we have that 
$h_{ij}=q_{ij}$ is degenerate in the direction of the Carrollian vector $\v$ since $h_{ij}v^j=0$ and the pair $(q_{ij}, v^j)$ then defines a weak Carrollian structure \cite{Duval:2014uoa,Ciambelli:2019lap}, while the triplet $(q_{ij}, v^j, k_i)$ denotes a ruled Carrollian structure \cite{Ciambelli:2023mir}.
As we will see, the sCarrollian structure parameterizes the configuration variables of the gravitational phase space of the stretched horizon. The last geometrical element needed for our construction is the volume form on $\H$,
\be 
\volH =  \bm{k} \wedge \volS, \qq \text{where} \qq \volS = \pi^* (\sqrt{q} \rd^2 \s)
\ee 
is horizontal and denotes the area form associated with $q_{ij}$. It satisfies $\dH \volS= \theta \volH$, where $\theta= q^{ij}\theta_{ij}$ is the expansion, where we denoted by $\dH $ the Cartan differential operator on $\H$.

Two types of transformations preserve the metric $h_{ij}$. The first is a \emph{rescaling} symmetry that preserves the Carrollian fiber bundle $\pi: \H \to \S $. A scalar $\phi$ parameterizes it and the infinitesimal transformation is given by 
\be \label{rescalingT}
   \delta_\phi \rho = 2\phi \rho, \qquad
   \delta_\phi k_i = -\phi k_i, \qquad 
\delta_\phi \v^i = \phi \v^i, \qquad \text{and} \qq \delta_\phi q^{ij}= 0.
\ee 
In the following, we say that a tensor $O_s$ is weight $s$ under rescaling if $\delta_\phi O_s=s\phi O_s$.

The second type of transformation is a \emph{shift} of the Ehresmann connection $\bm{k}$, representing a Lorentz boost transformation. It fixes the stretching and changes the Carrollian bundle unless $\rho=0$. The transformation is labelled by a horizontal vector $\zeta^i$ with  corresponding form $\zeta_i=h_{ij}\zeta^j$ and is given by 
\be \label{shiftT}
\delta_\zeta \rho = 0, \qquad
   \delta_\zeta k_i = \zeta_i, \qquad  
\delta_\zeta \v^i = -2\rho \zeta^i,
\qquad \text{and} \qq
\delta_\zeta q^{ij}= -(v^i \zeta^j + v^j \zeta^i).
\ee 
This transformation obviously preserves the relations $ h_{ij} v^j + 2\rho k_i=0$ and $q^{ij} k_j=0$.


\subsubsection{Frames, volume form and expansion}\label{sec:Framesand}
The stretched Carrollian geometry elements define several essential geometrical tensors: the Carrollian expansion tensor $\theta_{ij}$, acceleration $\varphi_i$, and vorticity $\vor_{ij}$. Their definitions are given respectively by
\be \label{Liev}
\theta_{(ij)}:= \frac12 \cL_v q_{ij},\qquad 
\varphi_i := -\cL_v k_i, \qquad \text{and} \qq
\vor_{ij} := - q_i{}^k q_{j}{}^l (\pa_k k_l - \pa_l k_k).
\ee 
These tensors are horizontal, and the last two enter the decomposition of the differential of $\bm{k}$. Denoting by $\bm{\ac}= \varphi_i \rd y^i$ and $\bm{\vor} = \frac12 \vor_{ij} \rd y^i \wedge \rd y^j $ the acceleration and vorticity forms, respectively, we have that the Ehresmann curvature is 
\be \label{gauge1}
\dH \bm{k} = -\left( \bm{k} \wedge \bm{\ac} +\bm{\vor}\right),
\ee 
where $\dH=\rd y^i \pa_i$ is the Cartan differential on $\H$.
Given the Carrollian bundle, we can pull back the coordinate forms on $\S$ and define a horizontal coframe basis $\bm{e}^A$, given by the pull-back $\bm{e}^A = \pi^* (\dH \sigma^A)$, where $\sigma^A$ are coordinates on $\S$.
Such horizontal coframes are, by construction, such that 
\be 
\dH \bm{e}^A =0.
\ee
The dual horizontal frame vectors are denoted by $e_A$. They are such that $\iota_{e_A} \bm{k}=0$ and $\iota_{e_A} \bm{e}^B=\delta_A{}^B$. Their contraction $e_i{}^A e_A{}^j =q_i{}^j$ gives the horizontal projector. In what follows, we use that a horizontal vector $X$ and a horizontal form $\alpha$ can be written as $X^i = X^A e_A{}^i $ and $ \alpha_i = e_i{}^A\alpha_A $, where $(X^A,\alpha_A) $ are the components in these bases. In particular, we define the horizontal metric components to be $q_{AB}:=h(e_A,e_B)$.

The Lie bracket of the frame basis $(v, e_A)$ is given by 
\be 
[v,e_A]=\varphi_A v, \qquad \text{and} \qq
[e_A,e_B] = \vor_{AB} v,
\ee
where $\varphi_A = e_A{}^i \varphi_i$ and 
$\vor_{AB}= e_A{}^i e_B{}^j\vor_{ij}$.
The area form can be written in terms of the coframe as 
\be 
\volS= \frac{1}{2}\sqrt{q} \ \varepsilon_{AB} \bm{e}^A \wedge \bm{e}^B,
\ee 
where $\varepsilon_{AB}$ denoted the 2-dimensional Levi-Civita symbol.
The Stokes theorem expressed in terms of Carrollian objects can be found in Appendix \ref{App:Stokes}

\subsubsection{sCarrollian connection} 

By definition, a sCarrollian connection $D_i$ is a torsionless connection compatible with the sCarrollian structure in the sense that it satisfies the following metricity condition:
\be\label{Dh}
D_i h_{jk}= -(\mK_{ij}k_k +\mK_{ik}k_j).
\ee 
where $\mK_{ij}$ is a weight-$1$ symmetric tensor on $\H$ which is \emph{invariant} under shift $\delta_\zeta$. Contracting the definition \eqref{Dh} with 
$v^j v^k$ and using \eqref{sCarrollian}, we get that 
\be \label{Kva}
\mK_{iv}= (D_i  - 2  \omega_i) \rho, \quad\mathrm{where}\quad 
\omega_i := k_j D_i v^j.
\ee 
The 1-form $\omega_i$ transforms as an abelian connection, $\delta_\phi \omega_i = D_i \phi$, under rescaling and is therefore called the \emph{rescaling connection}. It ensures that
 $(D_i-s\omega_i)O_s $ is weight $s$ under rescaling when $O_s$ is weight $s$. It also controls the derivative of the volume form $(D_i +\omega_i) \volH = 0$.
 From the definition \eqref{Dh} of the sCarrollian connection, we can evaluate the relationship between $\mK_{ij}$ and the Lie derivative of $h_{ij}$ along $v$:
\begin{align}
\mK_{ij} &= \frac12 \cL_vh_{ij} + D_i(\rho k_j ) + D_j(\rho k_i).\label{KLieh}
\end{align}
This equation simplifies when $\rho=0$, in which $\mK_{ij}$ is entirely determined by the Lie derivative of $h_{ij}$ along $v$. 

We can also express the Lie derivative of the metric along any horizontal vector $X^i$, i.e., satisfying $k_iX^i=0$, in terms of the sCarrollian connection. It is simply given by the usual expression,
\be 
\cL_{X} h_{ij}= D_iX_j+D_jX_i,
\ee 
where $X_i = h_{ij}X^j$. 


Finally, denoting $\Gamma_{ij}^k$ the connection symbol: $D_i T_j{}^k
:= \pa_i T_j{}^k - \Gamma_{ij}^l T_l{}^k +\Gamma_{il}^k T_j{}^l$, we can derive from its definition the transformation of the sCarrollian connection under rescaling and shifts,\footnote{We acknowledge important discussions with L. Ciambelli and R. G. Leigh about the shift symmetry. The results presented here about its role are also part of an upcoming work \cite{Freidel:coming}. }
\be 
\delta_\phi \Gamma_{ij}^k
= 0, \qquad \text{and} \qq
\delta_\zeta \Gamma_{ij}^k = \mK_{ij}\zeta^k.
\ee 
Furthermore, these transformations imply the conditions $\delta_\zeta \mK_{ij}=0$ and $\delta_\phi \mK_{ij}=\phi \mK_{ij}$.

\subsubsection{Horizontal connection}

Besides the sCarrollian connection, $D_i$, it will be useful also to introduce a horizontal covariant derivative $\sD_A$. 
This connection, called the Levi-Civita-Carroll covariant derivative \cite{Ciambelli:2018xat}, preserves the horizontal metric $q_{AB}$. It acts on a horizontal tensor $T_i{}^j = e_i{}^A T_A{}^B e_B{}^j$ as the projection of the sCarrollian connection:
\be 
\sD_C T_A{}^B = 
e_C{}^k e_A{}^i D_k T_i{}^j e_j{}^B. 
\ee 
From the definition of the sCarrollian connection \eqref{Dh}, it is direct to see that the horizontal connection preserves the horizontal metric, i.e., $\sD_A q_{BC}=0$. This implies that one can express the horizontal derivative as 
\begin{align}
\sD_C T_A{}^B = e_C[T_A{}^B] + \Gamma^B_{CD} T_A{}^D - \Gamma^D_{CA} T_D{}^B, \lb{sD-def}
\end{align}
and extend it in a usual way to any arbitrary horizontal tensor. The torsion-free Christoffel-Carroll symbols, $\Gamma^A_{BC} = \Gamma^A_{CB}$, is defined as\footnote{The  connection between the 
the relation between the horizontal Christoffel symbol and the sCarrollian connection coefficients is 
\begin{align}
\Gamma^A_{BC} 
= e_B{}^j e_C^k\Gamma^i_{jk}e_i^A +  e_B[e_C{}^i]e_i{}^A.
\end{align}} 
\begin{align}
\Gamma^A_{BC} = \frac{1}{2}q^{AD} \left( e_B[q_{DC}] + e_C[q_{BD}] - e_D[q_{BC}] \right). \lb{Gamma}
\end{align}
 
For a scalar field $F$, we have $\sD_A F = e_A[F]$ and 
\begin{align}
D_i F = \v[F] k_i + (\sD_A F) e_i{}^A,
\end{align}
This article will occasionally use the spacetime indices for horizontal derivatives. For example, $\sD_k T_i{}^k := e_k{}^C e_i{}^A (\sD_C T_A{}^B) e_j{}^b$. 
We will also denote a spacetime Lie derivative of a spacetime tensor $T_a{}^b$ along a spacetime vector $\ell$ and projected onto the horizontal subspace of $\H$ as
\begin{align}
\Lie_\ell T_A{}^B := e_A{}^a  (\Lie_\ell T_a{}^b ) e_b{}^B. \lb{hor-Lie-def}
\end{align}
The reader can find more details about derivatives and integrations in Appendix \ref{App:derivative}. 

\subsection{Connection coefficients and sCarrollian stress tensor}\label{sec:Conection}

As we have already described, the geometry of the Carrollian stretched horizon $\H$ is captured by the sCarrollian structure $\left(v^i, k_j, h_{ij}, \rho \right)$. They serve as canonical fields in the (covariant) phase space of gravity on $\H$. Next, we discuss their complement: the phase space conjugate momenta. 
These conjugate canonical momenta are given, 
in the context of membrane hydrodynamics, by the component of   sCarrollian fluid stress tensor \cite{Freidel:2022vjq}. In the sCarrollian construction, the fluid stress tensor which generalizes to stretched horizon the Brown-York stress tensor is defined as 
\begin{align}
\EM_i{}^j =  -  k_i \left( \E \v^j + \J^j \right)+ \p_i \v^j + \left( \T_i{}^j + \P q_i{}^j \right), \lb{sCT}
\end{align}
It comprises fluid energy density $\E$, fluid pressure $\P$, fluid momentum $\p_i =  e_i{}^A\p_A$,  heat current $\J^i = \J^A e_A{}^i$, and viscous stress tensor $\T_{ij} = e_i{}^A e_j{}^B\T_{AB}$ that is symmetric and traceless. All the vectors and tensors $(\p_i,\J^i, \T_i{}^j)$ are horizontal.
The structure of the sCarrollian stress tensor can be understood from the usual relativistic stress tensor which is given in $(-,+,+,+)$ signature by the following form,
\begin{align}
\EM^{ij} = \frac{1}{c^2} \E \v^i \v^j + \frac{1}{c^2}\J^i\v^j + \frac{1}{c^2} \v^i \J^j + (\T^{ij} + \P q^{ij}).
\end{align}
Upon identifying $c^2=2\rho$ and defining $k_i = -\frac{1}{c^2} h_{ij} v^j$ (so that $\v_i \v^i= -c^2$) and $\J_i := c^2\p_i$ we get that $T_i{}^j = h_{ik} T^{kj} $ possesses the structure \eqref{sCT}. With this structure, the Carrollian limit $c\to 0$ is no longer singular \cite{Donnay:2019jiz,Chandrasekaran:2021hxc,Freidel:2022bai}.

In agreement with the membrane paradigm idea \cite{Damour:1978cg,thorne1986black,Price:1986yy}, the structure of a stretched horizon requires that we identify the fluid stress tensor in terms of components of the sCarrollian connection.
To state the correspondence between fluid dynamics and geometry, it is essential to introduce the $(1,1)$ tensor $\Wein_i{}^j$  given by 
\be \label{WeinT}
\Wein_i{}^j := D_i \v^j  + 2\rho  (D_i k_k) q^{kj}.
\ee
This tensor is referred to as the \emph{generalized news tensor}.\footnote{see \cite{Freidel:2024tpl} for the connection with the usual notion of news in asymptotically flat spacetimes.}
It reduces to the Weingarten map in the null case. The sCarrollian fluid stress tensor is then defined to be given by 
\be \label{CEM2}
\EM_i{}^j := \Wein_i{}^j - \Wein \delta_i{}^j, \qquad \Wein := \Wein_{i}{}^i.
\ee
One of the main property of this stress tensor is that its covariant conservation equation is, when  the stretching $\rho$ is constant, the vacuum Einstein equation \cite{Freidel:2022vjq}.  
In appendix \ref{App:SCarrolcon} we establish the correspondence between the generalized news tensor $\Wein_i{}^j$ and the extrinsic curvature tensor $\mK_{ij}$. One finds that 
\be \label{KN}
\mK_{ij} = \Wein_i{}^k h_{kj} +  (D_i \rho) k_j. 
\ee 
Since $\mK_{ij}$ is symmetric, this establishes a relationship between different components of $\Wein_i{}^j$.
Under rescaling and shifts, the news tensor $\Wein_i{}^j$ transforms as 
\be \label{boost}
\delta_\phi\Wein_i{}^j = \phi \Wein_i{}^j + (D_i\phi) v^j, \qquad \text{and} \qq
\delta_\zeta \Wein_i{}^j =- D_i\rho \zeta^j.
\ee 

We now complete our description of the sCarrollian connection by providing the horizontal decomposition of $(D_iv^j,D_ik_j, \Wein_i{}^j)$ in terms of the components of the sCarrollian fluid stress tensor.

Firstly, the horizontal decomposition of the rescaling connection is given by 
\be 
\omega_i =\kappa k_i + \p_i,
\ee 
where $\p_i$ is the fluid momenta and $\kappa := k_i D_v v^i$ is the normal acceleration.

We can unravel the expressions of the actions of the connection on the basis fields $(v^i,k_i)$ 
(expressions for the covariant derivatives are given in Appendix \ref{App:derivative}),
\begin{align}
D_i \v^j= \ & \theta_i{}^j + (\p_i + \kappa k_i) \v^j + k_i\A^j, \lb{D-l} \\
D_i k_j = \ & \btheta_{ij} - (\p_i + \kappa k_i) k_j  - k_i (\p_j + \ac_j), \lb{D-k}\\
\Wein_i{}^j =\ & \theta_i{}^j + 2\rho \btheta_{i}{}^j + (\p_i + \kappa k_i) \v^j - k_i \J^j,
 \\
\mK_{ij}=\ & \theta_{ij}+2\rho \bar\theta_{ij} + (D_i \rho-2 \rho \omega_i) k_j - k_i \J_j .
\end{align}
The vector $\A^i :=  (D_v v^j) q_j{}^i$ represents horizontal acceleration. It evaluates to 
\be \A^i = q^{ij}(D_j  + 2\varphi_j) \rho.
\ee
In addition, the symmetry of $\mK_{ij}$ implies 
that the current $\J_i$ is given by the derivative
\bea
 \J^i = q^{ij}( - D_j  + 2\p_j) \rho.
\eea 
We see that both $\J^i$ and $\A^i$ vanish in the null case. The symmetry of $\mK_{ij}$ also dictates that 
the horizontal tensor 
$\NN_{ij}:= \theta_{ij}+2\rho \bar\theta_{ij}$ is symmetric. Its trace 
$\NN = q^{ij}\NN_{ij}$ defines the fluid energy and its traceless component is the fluid viscous stress\footnote{In arbitrary spacetime dimension $d$ (where $\H$ is of $(d-1)$ dimensions), the second expression reads $\T_i{}^j = \NN_i{}^j - \frac{1}{d-2} \E q_i{}^j$.}
\be 
\NN =\E, \qquad \text{and} \qq
\T_i{}^j = \NN_i{}^j - \tfrac{1}{2}\E q_i{}^j.
\ee 
Finally, the fluid pressure is related to the normal acceleration\footnote{In $d$-dimensional spacetime, this reads 
$\P = - \left( \kappa +\frac{(d-3)}{(d-2)}\NN \right)$.} through 
\be 
\P = - \left(\kappa +\frac{\NN}{2} \right). 
\ee 
One can also provide the value of the derivative of the co-metric $q^{ij}$. It is given by 
\be 
D_i q^{jk}
= - \left( \mathsf{\bar{K}}_i{}^j v^k + \mathsf{\bar{K}}_i{}^k v^j\right),
\ee 
where we introduced the dual curvature tensor,
\be \label{KT}
\mathsf{\bar{K}}_i{}^j := 
(D_i k_k) q^{kj} = \bar\theta_i{}^j - k_i (\p^j+\varphi^j).
\ee

Let us note finally that the sCarrollian connection is entirely determined by the values of $(\omega_i, \bar\theta_{(ij)})$, or equivalently by the symmetric components $D_{(i}k_{j)}$.  
For each ruled sCarrollian structure $\sC$, there exists a \emph{unique} canonical connection $\mathring{D}_i$ associated to it, given by the conditions $\mathring\btheta_{(ij)}=0$, $\mathring\p_i+\tfrac12\varphi_i=0$ and $\mathring \kappa =0$.
This connection is such that\footnote{ In general we have that $D_{(i} k_{j)} = \ \btheta_{(ij)} - (\p+\tfrac12  \ac)_{(i}k_{j)}   - 
\kappa k_i k_j$.} 
$\mathring{D}_{(i} k_{j)}
=0$.

The formalism that we have developed so far allows for the freedom to redefine the Ehresmann connection through rescaling \eqref{rescalingT} and shifting \eqref{shiftT}. In the gravity literature, it is customary to restrict this freedom by imposing that $\bm{k}= \exd \sv$ is a closed form, i.e., $k_i$ is chosen to be hypersurface orthogonal ($\vor_{ij}=0$) and acceleration-free ($\ac_i=0$)\footnote{Under a  shift $\zeta$, the acceleration, and vorticity transform as
\begin{alignat}{3}
&\delta_\zeta \ac_i &&= - (D_\v \zeta_i +\theta_{(ij)} \zeta^j) + (\rho \vor_{ij} - k_i \A_j)\zeta^j &&\Neq - \Lie_\v \zeta_i \nonumber  \\
&\delta_\zeta \vor_{ij}&& = -2(\sD -\ac)_{[i}\zeta_{j]} - 4\rho k_{[i} \vor_{j] \zeta} &&\Neq -2(\sD -\ac)_{[i}\zeta_{j]} \nonumber 
\end{alignat}
}. One can further restrict the time dependence of the foliation by imposing that the null congruence is affinely parameterized, i.e., $\kappa=0$. In total, we have the conditions
\be 
\ac_i=0,\qquad \vor_{ij}=0, \qquad \text{and} \qquad \kappa=0,
\ee 
corresponding to the choice of Gaussian null coordinates \cite{Hollands:2006rj, Rahman:2019bmk}.
With this choice, we still have the freedom to perform an exact shift plus rescaling  $\delta_T k_i = -D_i T = -\v[T]k_i  -\sD_i T$, corresponding to a change of foliation.  We will see in Section \ref{sec:symmetry} that such a transformation can be reabsorbed by a change of foliation associated with the infinitesimal diffeomorphism along $\xi=T \v$. 
As shown in \cite{Ashtekar:2001jb}, we can fix this foliation freedom by demanding the condition $\sD_A \p^A=0$.

We want to emphasize now that there exists another way to fix the choice of ruling, which is more hydrodynamical. This choice allows $k_a$ to possess vorticity. To see this, we first note that
under the rescaling symmetry, we have $\delta_\phi \btheta_{ij} = -\phi \btheta_{ij}$. Therefore, we see that we can fix this symmetry by imposing that the value of $\btheta$ is a constant.
The sign of this constant determines the nature of the horizons, since $\mathrm{sign}(\theta\btheta)$ is positive if the surface is trapped.

Then, under the shift \eqref{shiftT}, we have $\delta_\zeta D_i k_j = D_i \zeta_j$: hence, the transverse expansion tensor transforms as
\bee 
\delta_\zeta \btheta_{ij} &=
\sD_i \zeta_j + \zeta_i (\p_j + \ac_j)+\p_i  \zeta_j +2\rho ( k_i \btheta_{\zeta j} +
  k_j  \btheta_{i\zeta } ),\cr
  \delta_\zeta \bar\sigma_{ij} &=
\sD_{\langle i} \zeta_{j\rangle} + 2\zeta_{\langle i} (\p_{j\rangle} +\tfrac12 \ac_{j\rangle}) + 4\rho  k_{\langle i} \bar\sigma_{j\rangle \zeta}.
\eee 
On a null surface, these transformations simplify. Less obvious is the fact that the shift symmetry can be used to fix the value of $\bs_{ij}$ to zero. First, $\bs_{ij}$ being traceless contains two degrees of freedom on the sphere, the same as $\zeta_i$. It is a classic theorem of 2-dimensional geometry that we can always uniformize a 2-dimensional metric and locally write it as a conformally flat metric. The proof of uniformization utilizes that we can always find a vector $\zeta^i$ such that $T_{ij} = \sD_{\langle i} \zeta_{j\rangle}$ for any traceless tensor $T_{ij}$. From this theorem, one can conclude that the condition $\bar\sigma_{ij}=0$ can be reached when the equilibrium condition $\p_i+\tfrac12 \ac_i=0$ is satisfied. We expect that it can also be achieved at least in the neighborhood of this equilibrium condition.

This shows that, on a null surface, the shift and rescaling symmetries can be fixed in two different manners: either in terms of a foliation or in terms of a ``hydrodynamical'' condition on the transverse expansion tensor.


\section{Rigging Structure} \lb{sec:Rigging}

The purpose of our work is to study not only the geometry of a single sCarrollian structure, but rather the geometry of a family $\sC_r=\left(v^i(r), k_j(r), h_{ij}(r), \rho(r) \right)$ of such structures on $\H$, labelled by a parameter $r$. In the following, we assume that $r\geq 0$ and that we choose $\rho=0$ when $r=0$ and $\rho\neq 0$ when $r>\epsilon$. In other words, $\sC_0$ is a Carrollian structure that describes the geometry of a null surface and $\sC_r$ for $r>\epsilon$ describes the sCarrollian geometry of timelike surfaces, that approaches the null structure as $r\to 0$.

There is some arbitrariness in the way we label the family of sCarrollian structures $\sC_r$. Utilizing the rescaling and shift symmetries, we can impose that the Ehresmann connection is independent of $r$:
\be \label{kr}
\pa_r \bm{k}=0.
\ee 
This choice insures that the variation in $r$ of the basis vectors $(v, e_A)$ are horizontal. We parameterize this variation by a vector $\Upsilon^A$ and a matrix $\Omega_A{}^B$ as follows
\be \label{varC}
\pa_r v = \Upsilon^A e_A, \qquad \text{and} \qquad \pa_r e_A  = - \Omega_A{}^B e_B.
\ee 

\begin{figure}[t]
\centering
\includegraphics[scale=0.3]{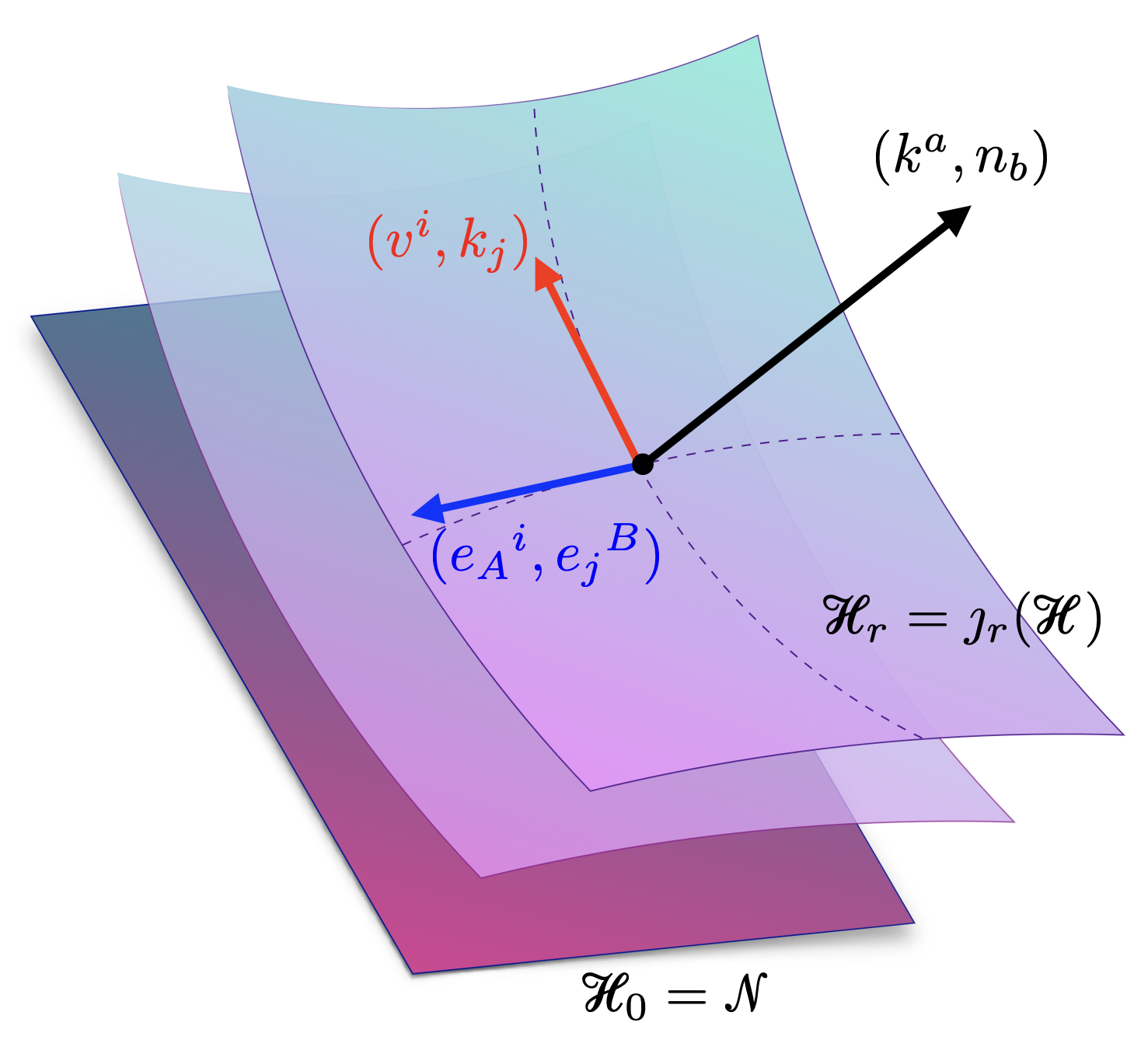} 
\caption{In the embedding picture, the Carrollian stretched horizons are hypersurfaces in the spacetime $\H_r = \jmath_r(\H)$, such that the null boundary $\N = \H_0$ can be viewed as the limit $r \to 0$ of the sequence of stretched horizons. These surfaces are endowed with the rigging structure $(k^a,n_b)$ and the intrinsic sCarrollian structure $(\v^i,k_j,h_{ij},\rho)$. Additionally, the horizontal frame $(e_A{}^i, e_j{}^B)$ is defined from the projector $q_i{}^j = e_i{}^Ae_A{}^j$ and $e_A{}^i e_i{}^B = \delta_A^B$.}\lb{stretched}
\end{figure}

\paragraph{Stretched horizon embedding:}
The perspective we advocate in our work \cite{Freidel:2022vjq, Freidel:2023bnj, Ciambelli:2023mir} (also see the works by Mars et al. \cite{Mars:2022gsa,Mars:2023hty,Manzano:2023oub}) is to understand the family of sCarrollian structures as a foliation of spacetime near a given null hypersurface in terms of causal hypersurfaces. This establishes an \emph{equivalence} between the stretched horizon and the sCarrollian perspectives.

In what follows, we consider an ambient spacetime to be a 4-dimensional Lorentzian manifold $(\M, g, \nabla)$, where $g = g_{ab} \exd x^a \odot \exd x^b$ is a Lorentzian metric and $\nabla$ is a Levi-Civita connection on $\M$.

Given $\sC_r$, we introduce a foliation of $\M$ as an embedding $\jmath:[0,1[ \times \H \to \M$,  which corresponds  to a family of embeddings $\jmath: r\to \jmath_r$ where $\jmath_r: \H \to \M$. The image hypersurfaces $\H_r:=j_r(\H)$ are hypersurfaces associated with the constant value $r(x)=r$ for a scalar field $r(x)$ on $\M$ (see Figure \ref{stretched}). We assume that this foliation is such that 
the hypersurface $r(x)=0 $ is a null surface (or boundary), denoted by $\N$, and the hypersurfaces $\H_r$ with $r>0$ are causal\footnote{By causal, we mean that it is locally timelike or null.} hypersurfaces.

The foliation normal form, denoted by $\bm{n}= n_a \rd x^a \in \Omega(\M)$, and the transverse vector, denoted by $k = k^a\pa_a \in \mathfrak{X}(\M)$. are given respectively by\footnote{This means that $k=\pa_r$.} 
\be 
\bm{n}= \rd r, \qquad \text{and} \qq  k^a = \pa_r \jmath^a.
\ee 

The connection between the 
spacetime metric $g_{ab}$ and the sCarrollian structure $\sC=(v^i, k_j, h_{ij}, \rho)$ is obtained by demanding the following conditions:
\begin{enumerate}[label = \textbf{\textit{\roman*}})]
\item The pair $(h_{ij}, k_i)$ is given by the pullback  of $(g_{ab}, g_{ab}k^b)$ by $\jmath$.
\item The vector $k$ is null, that is $g(k,k)=0$.
\item The stretching parameter $\rho$ determines the causal nature of $j_r(\H)$ through
$ g(n,n) =2\rho$.
\item The pushforward of $v$ by $\jmath$ is equal to the vector $v^a\pa_a|_{\jmath(H)} $ where $v^a=g^{ab} n_b - 2\rho k^a $.
\end{enumerate}
Using that $\jmath$ is represented, in coordinates, as a map $(r,y^i)\to x^a=\jmath(r,y^i)=\jmath_r^a(y)$, this explicitly means that
\be
h_{ij}=\pa_i \jmath^a \pa_j \jmath^b g_{ab}\circ \jmath, 
\qquad
k_i = \pa_i \jmath^a \pa_r \jmath^b g_{ab}\circ \jmath, \qquad 
0 = \pa_r \jmath^a \pa_r \jmath^b g_{ab}\circ \jmath .
\ee 
while we also have $g^{ab}n_b= v^a +2\rho k^b $ where $v^a = v^i \pa_i \jmath^a$.




\subsection{A (background) null rigging structure:}

The previous discussion shows that the stretched horizon $(\H,h, v, \bm{k}, \rho) $ can be viewed as a \emph{null rigging structure} $( g, \bm{n}, k)$, on the manifold $\M= \jmath([0,1[\times \H)$, where  $\bm{n} =  \exd r$, is a closed\footnote{One can indeed choose a more general foliation-defining normal form such that $\bm{n} = \e^{\ba} \exd r$ for a generic spacetime function $\ba$. 
The scale $\ba$ does not enter the bulk canonical phase space of gravity (see the discussions in \cite{Hopfmuller:2016scf,Hopfmuller:2018fni,Freidel:2022vjq}) and its value is completely gauge, thus allowing us to work in a gauge where $\ba = 0$ without forsaking any physical degrees of freedom.} normal form, and  $k = k^a \pa_a$  is a non-vanishing null \emph{rigging} vector field dual to the normal form and transverse to $\H$. This, by construction, imposes the conditions
\be\label{extn}
\exd \bm{n} = 0, \qq k^a n_a =1, \qq \text{and} \qq g_{ab} k^a k^b =0.
\ee 
In this work, the rigging structure is considered a background structure, inferring that both $\bm{n}$ and $k$ as well as the conditions \eqref{extn} are kept fixed under diffeomorphism symmetries. It is also worth mentioning that a notable advantage of this particular choice of rigging is that geometrical objects such as the induced connection, the surface stress tensor, and the conservation laws built upon it all admit non-singular null-limits, hence treating timelike and null surfaces on the same footing. 

We define the 3-dimensional \emph{rigging projector}, 
\be \Pi: T\M \to T\H,
\ee 
whose components are $\Pi_a{}^b := \delta_a^b - n_a k^b$, satisfying the transversality conditions, $\Pi_a{}^b n_b = 0$ and $k^a\Pi_a{}^b =0$. 
The null rigging structure induces on $\jmath(H)$ a sCarrollian structures. The sCarrollian vector and Ehresmann connection are respectively given by the following definitions:
\begin{align}
\v^i = n_b g^{bc } \Pi_c{}^i, \qq k_i = \Pi_i{}^a g_{ab} k^b,
\end{align}
while the sCarrollian metric and its dual are given by 
\be 
h_{ij} = \Pi_i{}^a \Pi_j{}^b g_{ab}, \qquad q^{ij} = g^{ab} \Pi_a{}^i \Pi_b{}^j. 
\ee  
The tensors $(\v^a, k_a,h_{ab})$ are the lift of $(\v^i,k_i,h_{ij})$ on $\jmath(\H)\subset M $ and satisfy the sCarrollian conditions \eqref{sCarrollian}. 
Notice that, when working with the null rigging, the 1-form $k_a=g_{ab}k^b$ is already tangent to the hypersurface as one can check that $\Pi_a{}^b k_b = k_a$. By construction, the sCarrollian vector $\v^a$ obeys $\v^a n_a = 0$. And one can easily verify the relation
\begin{align} 
\v^a = n^a - 2\rho k^a, \qquad \text{where} \qquad 2\rho = n_a n^a = -\v_a \v^a. \lb{l-def}
\end{align}
The horizontal (codimension-$2$) projector is related to the rigging (codimension-$1$) projector by \be \Pi_a{}^b = q_a{}^b + k_a \v^b.
\ee
The spacetime metric decomposes as $g_{ab}  = q_{ab}  + k_a \v_b + n_a k_b$. This means that it and its inverse can be decomposed in terms of the sCarrollian metric and its dual as 
\begin{align}
g_{ab} &
= h_{ab} +  2n_{(a} k_{b)}   \cr
g^{ab} &= q^{ab}+ 2v^{(a} k^{b)} + 2\rho k^a k^b   . \lb{4Dmetric}
\end{align}

\subsection{Rigging and sCarrollian connections}

We have observed that the connection between the Carrollian and embedding perspective is given through the rigging projector. A sCarrollian tensor $T_i{}^j$ corresponds to a spacetime tensor $T_a{}^b$ such that $k^aT_a{}^b=0=T_a{}^bn_b$. For such tensors, the correspondence between the spacetime Levi-Civita connection $\nabla$ and the sCarrollian connection is given by 
\be 
D_i T_j{}^k= \Pi_i{}^a \Pi_j{}^b (\nabla_a T_b{}^c) \Pi_c{}^k.
\ee 
With this definition, it is direct to see that $D_i\Pi_j{}^k=0$, which is consistent with the fact that $\Pi_i{}^j$ represents the sCarrollian identity map. 
We can also give an embedding perspective to the connection coefficients introduced earlier. The rescaling connection can be expressed as 
\be 
\omega_i =  \Pi_i{}^a ( k^b\nabla_a n_b),
\ee 
while the tensors $(\mathsf{K},\Wein)$ are given by 
\be 
\mathsf{K}_{ij}= \Pi_i{}^a \Pi_j{}^b (\nabla_a n_b), \qquad \text{and} \qquad
\Wein_i{}^j = \Pi_i{}^a  (\nabla_a n^b) \Pi_b{}^j.
\ee  
From this expression, it is obvious that $\mK_{ij}$ is symmetric.  It  represents the extrinsic curvature of the hypersurface $\jmath(\H)$, since  $\mathsf{K}_{ab}  = \frac{1}{2}  \Pi_a{}^c \Pi_b^d \Lie_n g_{cd}$.
The relationship \eqref{KN} between the extrinsic and news tensor can be easily derived from these definitions.

\subsection{Frame and coordinates}

In the setup just described, it is natural to choose coordinates $x^a = (r, y^i)$, on the ambient spacetime $\M$, adapted to the sCarrollian ones, where $y^i = (\sv,z^A)$ are general coordinates on the surface $\H$ radially extended throughout the surrounding spacetime by keeping their value fixed along the flow generated by the rigging vector $k = \pa_r$, in a way that resembles the construction of the Gaussian null coordinates that has been extensively utilized in the description of near-horizon geometry of black holes \cite{Hollands:2006rj, Rahman:2019bmk, Donnay:2016ejv, Donnay:2019jiz} and also geometry of general finite-location null surfaces
\cite{Hopfmuller:2016scf,Hopfmuller:2018fni,Adami:2021nnf,Adami:2021kvx}.  In these coordinates, we have that  $\Pi_i{}^j=\pa_i \jmath^j = \delta_i{}^j$.
The basis vector fields and their corresponding dual basis 1-form fields expressed in these coordinates are enumerated below:
\begin{equation}
\lb{vector-form}
\begin{alignedat}{6}
&\text{\sffamily \it Carrollian vector}  \ \ &&\v &&= \e^{-\a} \mathrm{D}_\sv  \ \ &&\text{\sffamily \it Ehresmann connection} \ \ &&\bm{k} &&= \e^\a (\bm{\rd} \sv  - \beta_A \bm{e}^A) \\
&\text{\sffamily \it Null rigging vector} \ \ && k &&= \pa_r  \ \ &&\text{\sffamily \it Normal form}  \ \ && \bm{n} &&= \exd r  \\
&\text{\sffamily \it Horizontal frame} \ \  && e_A &&= (J^{-1})_A{}^B \tfrac{\pa}{\pa z^B} + \beta_A \mathrm{D}_\sv \ \  &&\text{\sffamily \it Horizontal coframe} \ \ && \bm{e}^A &&= J_B{}^A (\bm{\rd }z^B - V^B \bm{\rd} \sv) 
\end{alignedat}
\end{equation}
where we defined $\mathrm{D}_\sv := \pa_\sv + V^A \tfrac{\pa}{\pa z^A}$. In these coordinates, the metric reads\footnote{Here, the coordinate $\sv$ is the advanced time, that is $\sv = t + r$ in flat space.}
\be 
\rd s^2 = 2 \e^\a \left(\bm{\rd} \sv  - \beta_A \bm{e}^A\right) \odot\left( \rd r - \rho \e^\a  \left(\bm{\rd} \sv  - \beta_B \bm{e}^B\right) \right)
+ q_{AB} \bm{e}^A \odot \bm{e}^B. \lb{metric-coord}
\ee
The spacetime metric is parameterized by a \emph{scale factor} $\alpha(x)$, a \emph{Carrollian connection} $\beta_A(x)$, a \emph{velocity field} $V^A(x)$, a \emph{Jacobian matrix} $J_A{}^B(x)$, and the sphere metric $q_{AB} (x)$.  It is important for the canonical analysis to assume that the coordinates are field-independent $\delta x^a =0$ and therefore keep the fields $(\alpha,\beta_A, V^A)$ free to vary. In the Carrollian literature, it is customary to choose a field-dependent coordinate system which imposes $V^A=0$ and $J_A{}^B=\delta_A^B$ and even $\alpha=0$. Such choices restrict the variations of the Carrollian vectors to $\delta v^i=0$.

\subsubsection{Gauge fixings}

We can now provide further details on the coordinate description of the sCarrollian geometry elements and the conditions they impose on their spacetime embedding. These conditions are as follows:

\paragraph{$\bm{i)}$ \emph{Geometric Carrollian structure}:} As we have seen in section \ref{sec:Framesand}, the Carrollian fiber bundle structures provides a horizontal coframe $\bm{e}^A$ whose differential vanishes  when pulled back to $\H$, i.e., $\dH \bm{e}^A := \jmath^* (\exd \bm{e}^A)= 0$, or in components, $(\dH \bm{e}^A)_{ab} = \Pi_a{}^c \Pi_b{}^d (\exd \bm{e}^A)_{cd} =0$. Here $\rd$ and $\dH$ denotes respectively the differentials on $\M$ and $\H$. This grants $\exd \bm{e}^A$ a general expression,
\begin{align}
\exd \bm{e}^A = - \Upsilon^A  \bm{n} \wedge \bm{k} +\Omega_B{}^A \bm{n} \wedge \bm{e}^B, \lb{gauge2}
\end{align}
where $\Upsilon^A (x)$ and $\Omega_B{}^A (x)$ are generic spacetime functions which enter the definition of the radial derivative of the sCarrollian structure \eqref{varC}. The special case where $\exd \bm{e}^A =0$ throughout the entire spacetime $\M$, as assumed in \cite{Freidel:2022vjq}, fixes $\Upsilon^A = 0$ and $\Omega_B{}^A =0$. These two variables do not appear explicitly in the gravitational phase space, and their values can be deemed gauge choices. Nonetheless, they become essential when discussing the radial evolution of some geometric quantities on $\H$ and when deriving a general gauge-preserving diffeomorphism. Thus, we will allow $\Upsilon^A$ and $\Omega_B{}^A$ to have non-zero values.

By computing the exterior derivative of the coframe $\bm{e}^A$, whose form is given in \eqref{vector-form}, one can derive the following equations,
\begin{subequations}
\begin{align}
\Upsilon^A & = \e^{-\alpha} J_B{}^A k[V^B], \lb{Upsilon}\\
\Omega_B{}^A & = (J^{-1})_B{}^C k[J_C{}^A] - \e^{\alpha} \beta_B \Upsilon^A, \lb{Omega}\\
\pa_A J_B{}^C & = \pa_B J_A{}^C, \lb{fiber1}\\
\mathrm{D}_\sv J_B{}^A &= - \left(\pa_B V^C\right) J_C{}^A. \lb{fiber2}
\end{align}
\end{subequations}
The first two equations relate $\Upsilon^A$ and $\Omega_B{}^A$ with the radial derivative of $V^A$ and $J_B{}^A$ while the last two equations, essentially corresponding to the condition that $\dH \bm{e}^A = 0$, can be seen as following from the fiber bundle property\footnote{Basically, we have that $\bm{e}^A=\dH \pi^A$ where $\pi:\H\to \S$ is the Carrollian projection map. We can solve the conditions \eqref{fiber1} and \eqref{fiber2} for the Jacobian matrix and the velocity field. The solutions are
\begin{align*}
J_A{}^B = \frac{\pa \pi^B}{\pa z^A}, \qquad \text{and} \qquad V^B J_B{}^A = - \frac{\pa\pi^A}{\pa \sv},
\end{align*}
where $\pi^A (x)$ can be seen as a radial-dependent transition map from the coordinates $y^i = (\sv,z^A)$ on $\H$ to the coordinates $\s^A$ on $\S$. This property is precisely where their name, the Jacobian matrix, and velocity field came from.} of the stretched horizon $\H$ (see \cite{Freidel:2022bai}).

 \paragraph{$\bm{ii)}$ \emph{Affinely parameterized null rays}:} The condition \eqref{kr} imposed on the Ehresmann connection means we are choosing the null rigging vector $k$ to be a generator of null geodesic congruences, and demand that  $\nabla_k k = 0$. This condition is equivalent to demanding $\Lie_k \bm{k} = \iota_k \exd \bm{k}=0$, inferring that the curvature of the Ehresmann connection is completely tangent to the surface, meaning that $\dH \bm{k} := \jmath^* (\exd \bm{k}) = \exd \bm{k}$. In general, one has $\nabla_k k = \bar{\kappa} k + \bar{a}^A e_A$ where $\bar{\kappa}$ and $\bar{a}^A$ are components of the acceleration of $k$. Choosing the affine null generator means 
 \begin{align}
\bar{\kappa} = k[\a] + \e^\a \beta_A \Upsilon^A = 0, \qquad \text{and} \qquad - \e^{-\a}\bar{a}_A  = k[\beta_A] + \Omega_A{}^B\beta_B = 0, \lb{constraint}
\end{align}
where we also used \eqref{gauge2}. One reason we are allowed to do so is that the inaffinity $\bar{\kappa}$ is not a part of the gravitational phase space \cite{Hopfmuller:2016scf,Freidel:2022vjq}. The radial coordinate $r$ also serves as the affine parameter of the null geodesics generated by a vector field $k$. These equations combined with \eqref{Upsilon} and \eqref{Omega} relate the radial derivative of different components of the spacetime metric. 

The Ehresmann curvature can thus be parameterized as
\begin{align}
\exd \bm{k} = -\bm{k} \wedge \bm{\ac} -\bm{\vor} =  -\ac_A \bm{k} \wedge \bm{e}^A - \tfrac{1}{2}\vor_{AB} \bm{e}^A \wedge \bm{e}^B, 
\end{align}
where, in the context of Carrollian fluid dynamics \cite{Duval:2014uoa,Ciambelli:2018ojf,Freidel:2022bai}, the quantities $\bm{\ac} = \ac_A \bm{e}^A$ and $\bm{\vor} = \frac{1}{2}\vor_{AB} \bm{e}^A \wedge \bm{e}^B$ play a role of \emph{Carrollian acceleration} and \emph{Carrollian vorticity}, respectively. In components, they are given by 
\begin{align} 
\ac_A (x) = \mathrm{D}_\sv \beta_A + \sD_A\alpha, \qquad \text{and} \qquad \vor_{AB}(x)  = \e^\alpha \left(\sD_A \beta_B - \sD_B \beta_A  \right). \lb{accel-vorti}
\end{align}
It is worth noting that the presence of the vorticity $\vor_{AB}$ dictates the non-integrability of the horizontal subspace according to the Frobenius theorem. \\

The gauge conditions \eqref{gauge1} and \eqref{gauge2} can be expressed in terms of the basis vector field  Lie brackets  as  
\begin{align}
[\v, e_A] = \ac_A \v, \qquad \qquad [e_A, e_B] = \vor_{AB} \v, \cr\qquad [k, \v] = \Upsilon^A e_A, \qquad  \qquad [e_A, k] = \Omega_A{}^B e_B. \lb{com-4}
\end{align} 
The first two Lie brackets are the Carrollian commutation relations \cite{Ciambelli:2018xat, Ciambelli:2019lap}. 
Furthermore, the Jacobi identities of these Lie brackets (also simply corresponds to the nilpotent properties, $\exd^2 \bm{k} =0$ and $\exd^2 \bm{e}^A =0$) impose some conditions on the components $(\ac_A, \vor_{AB}, \Upsilon^A, \Omega_A{}^B)$. These conditions are listed in Appendix \ref{App:Jacobi}.
From all these identities, we can evaluate how the horizontal frames gets transported under the radial evolution. We have that 
\be 
\Lie_k e_A = -\Omega_A{}^B e_B, \qq \text{and} \qq \quad\Lie_k\bm{e}^A= -\Upsilon^A \bm{k} +\bm{e}^B \Omega_B{}^A. 
\ee 
We can then show that the radial evolution of a horizontal tensor $T_a{}^b = e_a{}^AT_A{}^B e_B{}^b$ is given by (using the notation \eqref{hor-Lie-def})
\begin{equation}
\begin{aligned}
\Lie_k T_A{}^B := e_A{}^a \left( \Lie_k T_a{}^b \right) e_b{}^B  &= e_A{}^a \Lie_k \left( e_a{}^C T_C{}^D e_D{}^b \right) e_b{}^B \\
& = k[T_A{}^B] + \Omega_A{}^C T_C{}^B - T_A{}^C\Omega_C{}^B.
\end{aligned}
\end{equation}
This formula, which will be used many times, can be straightforwardly generalized to a horizontal tensor of arbitrary degree. \\

Let us provide some remarks: First, when both $\Upsilon^A =0$ and $\Omega_A{}^B =0$, the remaining components $(\alpha,\beta_A, V^A, J_A{}^B)$ are independent of the radial coordinate $r$ (see the equations \eqref{Upsilon}, \eqref{Omega} and \eqref{constraint}), and the tangent vector fields $(\v, e_A)$ to the surface $\H$ are Lie-transported along the radial direction, so that $\Lie_k \v =0$ and $\Lie_k e_A =0$ (and $\Lie_k \bm{k} =0$ in the gauge we chose). One could say that this is the special case when the sCarrollian structure induced on $\H$ is partially independent of the radial coordinate $r$, even though the remaining data $(\rho,q_{ab})$ can still depend on $r$. 

The second remark concerns the Jacobian matrix $J_A{}^B$ that does not generally appear in most literature. As far as the spacetime metric \eqref{4Dmetric} is concerned, one can always reabsorb $J_A{}^B$ (or set $J_A{}^B = \delta_A^B$) into a new set of variables, $\tilde{\beta}{}_A := J_A{}^B \beta_B$ and $\tilde{q}_{AB} := J_A{}^C J_B{}^D q_{CD}$, and a new horizontal frame $\tilde{e}_A := J_A{}^B e_B = \pa_A+\tilde{\beta}_A \mathrm{D}_\sv$ and a coframe $\tilde{\bm{e}}{}^A := \bm{e}^B(J^{-1})_B{}^A = \exd z^A - V^A \exd \sv$. The new frame and coframe fields still obey the same orthogonality conditions as the original ones. By doing so, one will be rid of the Jacobian matrix in their consideration. However, it no longer makes manifest a geometric Carrollian structure in the geometry of the surface $\H$, and the condition \eqref{gauge2} cannot be chosen for the new field $\tilde{\bm{e}}{}^A$, without imposing further restriction on the fields. While this method might work better in practicality, we refrain from doing so and choose to retain the Carrollian geometry perspective. Furthermore, the main problem with such a frame is that it is no longer a coordinate frame even when we choose $\vor_{AB} = 0$. More precisely, the commutator $[\tilde{e}_A ,\tilde{e}_B]$ does not vanish if $V^A$ does not vanish. It only makes sense to put $J_A{}^B=\delta_A^B$ in the adapted coordinates $z^A=\sigma^A$ where $V^A=0$.




\subsection{Connection components}

We can now express the different tensors entering the decomposition of the sCarrollian connection in terms of the coordinate components.
The key ingredient is the Koszul formula, which expresses the Levi-Civita connection coefficients along vector fields $(X,Y,Z)$ as 
\begin{align}
2 g( \nabla_X Y, Z) = &X\left[ g(Y,Z) \right] + Y\left[ g(X,Z) \right]  - Z\left[ g(X,Y) \right] \cr & + g([X,Y],Z) +g([Z,X], Y) - g([Y,Z],X).    
\end{align}
We use the expression \eqref{com-4} for the basis vectors $(v,k,e_A) $ Lie brackets. 

\paragraph{$\bm{i)}$ Viscous stress tensor and energy:} We first consider the spin-2 components contained in the decomposition \eqref{D-l} and \eqref{D-k}, consisting of the \emph{expansion tensor}, $\theta_{ab} = e_a{}^A e_b{}^B\theta_{AB}$ and the \emph{extrinsic curvature tensor}, $\btheta_{ab} = e_a{}^A e_b{}^B\btheta_{AB}$. The former computes the change of the sphere metric along the tangent vector $\v$, 
\begin{align}
\theta_{AB} := g( \nabla_{e_A} \v,e_B) = \tfrac{1}{2}\v[q_{AB}] + \rho \vor_{AB}. \lb{theta}
\end{align}
Notice that, on the general Carrollian stretched horizon $\H$, the expansion tensor is not symmetric due to the presence of the Carrollian vorticity $\vor_{AB}$. Since $\rho \Neq 0$, it is always symmetric on the null surface. The symmetric components, $\theta_{(AB)} = \frac{1}{2}\Lie_\v q_{AB}$, can be further separated into the trace-free part $\s_{AB}$, which we refer to as the \textit{tangential shear}, and the trace $\theta$, called the \textit{tangential expansion}, as follows
\begin{align}
\s_{AB} := \theta_{(AB)} - \tfrac{1}{2} \theta q_{AB}, \qquad \text{and} \qquad \theta := q^{AB} \theta_{AB} = \v[\ln \sqrt{q}].  
\end{align}

The extrinsic curvature tensor $\btheta_{AB}$ is defined similarly, and it is related to the change of the sphere metric in the transverse direction, 
\begin{equation}
\lb{theta-bar}
\begin{aligned}
\btheta_{AB} := g( \nabla_{e_A} k,e_B) &= \tfrac{1}{2}k[q_{AB}] +\Omega_{(AB)} -\tfrac{1}{2} \vor_{AB} \\
& = \tfrac{1}{2} \left( \Lie_k q_{AB}  - \vor_{AB} \right)
\end{aligned}
\end{equation}
where we defined $\Omega_{AB} := \Omega_A{}^C q_{BC}$. Interestingly, this tensor is not symmetric even on the null surface $\N$. However,
the tensor $\NN_{AB}$ 
\be
\NN_{AB}:= \theta_{AB} + 2 \rho \bar\theta_{AB} = \T_{AB} +\frac12 q_{AB} \E, \lb{E-AB-def}
\ee
is always symmetric. Given $\btheta_{(AB)} = \frac{1}{2}\Lie_k q_{AB}$, its symmetric trace-free part $\bs_{AB}$ and its trace $\btheta$ are respectively called the \textit{transverse shear} and the \textit{transverse expansion}, and are given by
\begin{align}
\bs_{AB} := \btheta_{(AB)} - \tfrac{1}{2} \btheta q_{AB}, \qquad \text{and} \qquad \btheta := q^{AB} \btheta_{AB} = k[\ln \sqrt{q}] + \Omega, \lb{sigma-bar}
\end{align}
where $\Omega = \Omega_A{}^A$ is simply the trace. Invoking the definitions (\ref{sCT},\ref{CEM2}) of the Carrollian stress tensor, one can easily verify that the viscous stress tensor $\T_{AB}$ is given by  
\begin{align}
\T_{AB} = \NN_{\la AB \ra}  = \s_{AB} + 2\rho \bs_{AB}. \lb{shear-energy-gen}
\end{align}
Note that their values on $\N$ are determined by the tangential objects $\s_{AB}$ and $\theta$. 

\paragraph{$\bm{ii)}$ Fluid momentum and heat current:} The spin-1 part comprises the Carrollian fluid momentum $\p_A$ and the Carrollian heat current $\J^A$. The fluid momentum, corresponding geometrically to the Hájí$\mathrm{\check{c}}$ek 1-form field, is given by 
\begin{align}
\p_A := g(\nabla_{e_A} \v,k) = \tfrac{1}{2}\left( \Upsilon_A -\ac_A\right), \lb{momentum-gen}
\end{align}
where $\Upsilon_A = q_{AB} \Upsilon^B$. This relation follows from the commutators \eqref{com-4} and the Koszul identity
\be
 2g(k,\nabla_{e_A} \v) = g([k,\v], e_A) + g([k,e_A], \v) - g([\v,e_A],k).
\ee The relation between $\p_A, \ac_A$ and $\Upsilon_A$ will be called-for numerous times in subsequent computations. Let us also comment that the special case when $\Upsilon^A = 0$, which always holds in the co-moving coordinates where the velocity field vanishes, implying the relation $\p_A = - \frac{1}{2}\ac_A$ \cite{Ciambelli:2019lap}. The Carrollian heat current is defined as
\begin{align}
\J_A := - g( \nabla_\v n,e_A)  =  (- \sD_A + 2\p_A)\rho.  \lb{heat-gen}
\end{align}
This heat current vanishes for the null case. 

\paragraph{$\bm{iii)}$ Energy and pressure:} The Energy evaluates the trace of $\E=\N_A{}^A$. fluid pressure $\P$ is another key scalar entering the definition of the Carrollian stress tensor. To evaluate it, we define the \textit{surface gravity} (or normal acceleration of $\v$) as 
\begin{align}
\kappa := g(k, \nabla_\v \v) = -\tfrac{1}{2}k\left[g(v,v)\right] = k[\rho], \lb{kappa}
\end{align} 
which, again, followed from the Koszul formula\footnote{The non-affinity $\bar{\kappa}$ of $k$ (defined from $\nabla_k k = \bar{\kappa} k$) modifies the Lie bracket; $[k,\v] = -\bar{\kappa} \v + \Upsilon^A e_A$. In such case, the surface gravity is given by $\kappa = k[\rho] + 2\rho \bar{\kappa}$.}.
It measures the inaffinity of geodesics that lie on the null surface $\N$ and are generated by the vertical vector $\v$. On the general stretched horizon $\H$, the acceleration $D_\v \v^i $ also contains the horizontal components, $\A_A = \sD_A \rho + 2\rho \ac_A$ (see \eqref{D-l} and  \cite{Freidel:2022vjq}).
To summarize, three scalars $(\P,\E,\Wein)$ play a role in the expression of Einstein equations and the canonical charges. They are 
\begin{align}
\E &=\theta + 2\rho\bar\theta, \\
\P &= -\left( \kappa + \tfrac{1}{2}\E \right),  \lb{pressure-gen} \\
\Wein &=  \tfrac{1}{2}\E - \P = \E +\kappa . \lb{Ndef}
\end{align}


\section{Einstein Equations on Carrollian Stretched Horizons} \lb{sec: Einstein}
The geometry and the canonical momenta of the Carrollian stretched horizon $\H$ evolve according to the vacuum Einstein equations\footnote{In our work, we restrict ourselves to the vacuum case, without matter field and cosmological constant. We also work in the unit $8\pi G_{\sss \mathrm{Newton}} =1$.},
${G}_{ab} =0$, where, as defined, ${G}_{ab}$ are components of the Einstein tensor pulling back onto $\H$. 
To express these equations in the basis $(n_a, \v^a, k^a, e_A{}^a)$, we use (see \eqref{hor-Lie-def}) that $\cL_\ell T_A{}^B$ denotes the Lie derivative of a spacetime tensor $T_a{}^b$ along a spacetime vector $\ell=(v,k,n)$ projected onto the horizontal subspace of $\H$ as
\begin{align}
\Lie_\ell T_A{}^B := e_A{}^a  (\Lie_\ell T_a{}^b ) e_b{}^B. \lb{hor-Lie-def2}
\end{align}

It is convenient to decompose the Einstein equation into sCarrollian components that describe evolution along $\H$ and radial evolution equations that describe how the sCarrollian data depends on $r$.

These sCarrollian components, expressed in the basis $(n_a, \v^a, k^a, e_A{}^a)$, are\footnote{The notation is $\Lie_\v \bs_{\la AB \ra} = e_{\la A}{}^a e_{B \ra}{}^b \Lie_\v \bs_{ab} = \Lie_\v \bs_{AB} - \frac{1}{2}(q^{CD}\Lie_\v \bs_{CD}) q_{AB}$} \\
\noindent\fbox{%
\parbox{\textwidth}{%
\centering \textbf{Einstein Evolution on $\H$}
\begin{subequations}
\lb{EinsteinEv}
\begin{align}
-G_{\v n} &= \Lie_\v\E+(\E+\P)\theta + (\sD_A+2\ac_A)\J^A+ \T^{AB}\s_{AB} - \btheta \Lie_\v\rho \\
G_{A n} &= \Lie_\v \p_A + \theta\p_A + \E \ac_A + \J^B\vor_{BA} + \left( \sD_B  +\ac_B \right) (\T_A{}^B +\P \delta_A^B) + \btheta \sD_A \rho \\
G_{k n} & = \Lie_\v \btheta+\Wein \btheta +(\sD_A +\p_A + \ac_A ) (\p^A + \ac^A) - \tfrac{1}{2} \RC  - \rho \left( \btheta_{AB} \btheta^{AB} + \btheta^2 \right) \lb{G-kn} \\ 
-G_{\langle A B \rangle} 
& = 2 \Lie_{\v +\rho k}\bs_{\la AB \ra} - 2(\E + \P) \bs_{AB} + \btheta \T_{AB}+2 (\sD+\ac +\p)_{\la A} (\p+\ac)_{B \ra} - \RC_{\la AB \ra} 
\end{align}
\end{subequations}
}}\\
\noindent where $\RC_{AB}$ and $\RC = q^{AB}\RC_{AB}$ are the \emph{Ricci-Carroll tensor} and its trace \cite{Ciambelli:2018xat}, respectively. Their definitions and their properties are given in Appendix \ref{app:Riemann}. In four spacetime dimension, $\RC_{\langle AB\rangle }=0$ and the last equation simplifies. 
The first three equations describe the evolution of $(\E,\p_A,\bar\theta)$ along $v$. The first two equations are the sCarrollian analog of the Raychaudhuri and Damour equations \cite{Adami:2021kvx}, which describe the conservation of energy and momenta of the sCarrollian fluid, while $\btheta$ can be seen as the charge aspect for radial super-translation. 
The last equation is an evolution equation only when $\rho=0$. Otherwise, it relates $\rho \Lie_k \bar{\sigma}_{AB}$ to the sCarrollian components. We will show later that $\bar\sigma_{AB}$ can be understood as a spin-2 charge aspect.
Next, we display the transverse evolution equations, which govern how the fields evolve off $\H$.
\noindent\fbox{%
\parbox{\textwidth}{%
\centering \textbf{Transverse evolution off $\H$}
\begin{subequations}
\lb{Einstein}
\begin{align}
-G_{k k}  &=  \Lie_k\btheta  + \btheta_{AB} \btheta^{BA}\\
-G_{A k}
& = \Lie_k\p_A +\btheta \p_A
 + \vor_{AB}(\p^B+ \ac^B)  -\sD_B \left( \btheta_A{}^B - \btheta \delta_A^B \right) \\
 G_{ \v k} 
 & = \Lie_k \E + \btheta \E -\kappa\btheta - (\sD -\p)_A \p^A -\tfrac{1}{2}\RC + \rho \left(\btheta_{AB}(\btheta^{BA} +\vor^{AB}) - \btheta^2\right).\\
\tfrac{1}{2}q^{AB} G_{AB} 
& = \Lie_k \Wein +(\E+\P)\btheta   - (\sD_A +2\p_A+2\ac_A) \p^A +\T^{AB}\bs_{BA} + \rho G_{k k}  \\
-G_{\langle A B \rangle} 
& = 2(\Lie_k \T_{\la AB \ra} -  \rho \Lie_k \bs_{\la AB\ra}) + 2(\E+\P)\bs_{AB} - \btheta \T_{AB} - 2(\sD -\p)_{\la A} \p_{B \ra} - \RC_{\la AB \ra}
\end{align}
\end{subequations}
\vspace{-0.5cm}}}\\
\\
We recall that $\btheta_{[AB]}= -\tfrac12 \vor_{AB}$.
 Other components of the Einstein tensor, such as $G_{\v k}$ and $G_{A \v}$, are linear combinations of these components. We provide the derivation of all equations in Appendix \ref{app-Einstein}.
 
Remarks are in order here:

\paragraph{Vertical vs Radial:} The components of the Einstein equation \eqref{Einstein} control either the vertical evolution (containing $\Lie_\v$) or the radial evolution (containing $\Lie_k$) or the mixture of both of the sCarrollian fluid variables. It is, however, crucial to appreciate that the two notions of evolution are interchangeable for some components of the Einstein equation. This interesting aspect lies in the relation (or equivalent) between the radial derivative $\Lie_k \theta_{(AB)}$ and vertical derivative $\Lie_\v \btheta_{(AB)}$. One can show that (see the formulae in Appendix \ref{App:Lie})\footnote{We used the Lie derivatives of the Carrollian metric: $
\Lie_\v q_{ab} = 2\theta_{(ab)} + 2 \Upsilon_{(a} n_{b)}$ and $\Lie_k q_{ab} = 2\btheta_{(ab)} -2 \Upsilon_{(a} k_{b)}$.}
\begin{equation}
\lb{k-thetaAB}
\begin{aligned}
\Lie_k \theta_{(AB)} = \tfrac{1}{2}e_A{}^a e_B{}^b\Lie_k \left( \Lie_\v q_{ab} - \Upsilon_{(a} n_{b)} \right)
 &= \tfrac{1}{2}e_A{}^a e_B{}^b \left( \Lie_\v \Lie_k q_{ab} + \Lie_\Upsilon q_{ab} \right)  \\
  &= e_A{}^a e_B{}^b \left( \Lie_\v \left(\btheta_{(ab)} - \Upsilon_{(a} k_{b)}\right) + \tfrac{1}{2}\Lie_\Upsilon q_{ab} \right)  \\
 & =  \Lie_\v \btheta_{(AB)} + (\sD + \ac )_{(A}\Upsilon_{B)},
\end{aligned}
\end{equation}
where we used $[\Lie_k, \Lie_\v] = \Lie_{[k,\v]} = \Lie_\Upsilon$ as following from the Lie bracket \eqref{com-4} and $\cL_v k_a=-\varphi_a$, $\cL_v n_a=0$.  We also recall that $\Upsilon_A =2\p_A+\ac_A$ (see  \eqref{momentum-gen}). From the above equation, one can easily derive the following descendent relations (derivation of these identities is given in Appendix \ref{App:RadT}): 
\begin{subequations}
\label{eq:RadT}
\begin{align} 
\Lie_k \theta & = \Lie_\v \btheta + (\sD_A + \ac_A) \Upsilon^A \label{eq:RadT-1}\\
\Lie_k \E & = \Lie_n \btheta - (\E+2\P) \btheta + (\sD_A + \ac_A) \Upsilon^A \label{eq:RadT-2}\\
\Lie_k \s_{AB} & = \Lie_\v\bs_{AB} + \btheta \s_{AB} - \theta \bs_{AB} + (\sD + \ac)_{\langle A} \Upsilon_{B \rangle} \label{eq:RadT-3}\\
\Lie_k \NN_{AB} & = \Lie_n \btheta_{(AB)} - (\E+2\P) \btheta_{(AB)} + (\sD + \ac)_{(A} \Upsilon_{B)} \label{eq:RadT-4}\\
\Lie_k \T_{AB} & = \Lie_n \bs_{AB} - 2(\E+\P) \bs_{AB} + \btheta \T_{AB} + (\sD + \ac)_{ \la A} \Upsilon_{B \ra}. \label{eq:RadT-5}
\end{align}
\end{subequations}
Nice consistency checks for these identities can be performed:
If we take the difference of the first two equations, we get 
\be
\cL_k (2\rho \btheta)= 2\rho \Lie_{k} \btheta -(\E+2\P) 2\kappa \btheta = 2\rho \Lie_k\btheta + 2 \kappa \btheta.
\ee  which is satisfied as $\kappa= k[\rho]$.
We used that $\E=\theta+2\rho\btheta$ and  $\E+2\P=-2\kappa$.
Taking the difference between the last and next-to-last equations, we obtain
\begin{align}
\Lie_k(2\rho \bs_{AB}) &= 2\rho \Lie_k \bs_{AB} - 2(\E+\P)\bs_{AB}
+ 2\rho \btheta \bs_{AB} + \theta \bs_{AB} \\
& = 2\rho \Lie_k \bs_{AB} - 2(\tfrac{1}{2}\E+\P)\bs_{AB} \\
& = 2\rho \Lie_k \bs_{AB} + 2\kappa\bs_{AB}.
\end{align}

Using these relations, one can express the Einstein tensor either as the vertical (time) or radial or mixed evolution equations. To illustrate this, the $G_{\la AB \ra}$ components can be expressed in various ways as follows: 
\begin{equation}
\begin{aligned}
-G_{\langle A B \rangle} 
& = \Lie_\v \bs_{\la AB \ra} + \Lie_k \T_{\la AB\ra} +(\sD +2\p + \ac )_{\la A}\ac_{B \ra} + 2\p_{\la A} \p_{B \ra} - \RC_{\la AB \ra} \\
& = 2 (\Lie_\v +\rho \Lie_k)\bs_{\la AB \ra} - 2(\E + \P) \bs_{AB} + \btheta \T_{AB}+2 (\sD+\ac +\p)_{\la A} (\p+\ac)_{B \ra} - \RC_{\la AB \ra} \\
& = 2(\Lie_k \T_{\la AB \ra} -  \rho \Lie_k \bs_{\la AB\ra}) + 2(\E+\P)\bs_{AB} - \btheta \T_{AB} - 2(\sD -\p)_{\la A} \p_{B \ra} - \RC_{\la AB \ra}.
\end{aligned}
\end{equation}
Similarly, the trace part can be written as
\begin{equation}
\begin{aligned}
& \tfrac{1}{2}q^{AB} G_{AB} \\
& \ = \Lie_k \Wein +(\E+\P)\btheta   - (\sD_A +2\p_A+2\ac_A) \p^A +\T^{AB}\bs_{BA} + \rho G_{k k} \\
& \ = (\Lie_\v+\rho \Lie_k-\P)\btheta +  \Lie_k\kappa  + (\sD_A +\ac_A) (\p^A+\ac^A)-\p_A (2\p^A+\ac^A)+\T^{AB}\bs_{BA} - \rho \btheta^{AB}\btheta_{BA}.
\end{aligned}
\end{equation}
Finally, for $G_{kn}$, one gets
\begin{equation}
\begin{aligned} \label{G-kn2}
G_{k n} & = \Lie_\v \btheta+\Wein \btheta +(\sD_A +\p_A + \ac_A ) (\p^A + \ac^A) - \tfrac{1}{2} \RC  - \rho \left( \btheta_{AB} \btheta^{AB} + \btheta^2 \right) \cr
& = \Lie_k \theta +\Wein \btheta -(\sD_A - \p_A  ) \p^A  - \tfrac{1}{2} \RC  - \rho \left( \btheta_{AB} \btheta^{AB} + \btheta^2 \right).
\end{aligned}
\end{equation}

\subsection{Ricci scalar}

The Ricci tensor, specifically the component $R_{k n}$, will be important in our subsequent derivation of the Einstein equation from symmetries. In four dimensions, it is related to the trace of the horizontal components, $q^{AB} {G}_{AB}$, and the component $G_{kk}$ of the Einstein tensor\footnote{One can show the relation $q^{AB}G_{AB} = - (R_{\v k} + R_{k n}) = -2R_{k n}+ 2\rho G_{kk}$ using that $\v = n - 2\rho k$.} and is given by
\begin{equation}
\begin{aligned}
-R_{k n} = \tfrac{1}{2} q^{AB} G_{AB}  -\rho G_{kk} = \Lie_k \Wein +(\E+\P)\btheta   - (\sD_A +2\p_A+2\ac_A) \p^A +\T^{AB}\bs_{BA}. \lb{R-kn}
\end{aligned}
\end{equation}

Following from this expression and recalling the expression \eqref{G-kn} for $G_{kn}$, we derive the Ricci scalar in terms of sCarrollian quantities, 
\begin{equation}
\lb{RicciScalar}
\begin{aligned}
\tfrac{1}{2}R   = \ & R_{kn} -G_{kn}
\\
 = \ & \tfrac{1}{2}\RC  - \P\btheta  +\p_A\p^A - \T^{AB}\bs_{AB} + \rho(\btheta_{AB}\btheta^{AB} +\btheta^2) \\
& - (\Lie_k +\btheta) \Wein - (\Lie_\v + \E)\btheta - (\sD_A +\ac_A)\ac^A.
\end{aligned}
\end{equation} 
Let us recall that $(V[f] + \nabla_a V^a) \volM = \exd (f V^a \iota_a \volM)$ for a spacetime function $f$ and a vector $V^a$. Let us also remember the spacetime divergence of the basis vectors, $\nabla_a k^a =\btheta$, $\nabla_a \v^a = \theta $, and $\nabla_a \ac^a = (\sD_A+\ac_A) \ac^A$, and the relation $\E = \theta + 2\rho \btheta$. Using these, we find that the Einstein-Hilbert Lagrangian, $\bm{L}^{\sss \mathrm{EH}} = \frac{1}{2}R \volM$, can now be written as 
\begin{equation}
\begin{aligned}\label{LEH}
\bm{L}^{\sss \mathrm{EH}} = \mathscr{L} \volM + \exd\bm{\ell}. 
\end{aligned}
\end{equation} 
The bulk Lagrangian and the boundary Lagrangian are given by 
\begin{alignat}{2}
&\mathscr{L} &&= \tfrac{1}{2}\RC - \P\btheta +\p_A \p^A  - \T^{AB}\bs_{AB} + \rho(\btheta_{AB}\btheta^{AB} -\btheta^2) \lb{L-bulk} \\
&\bm{\ell} &&= -(\Wein k^a + \btheta \v^a  +\ac^a)  \iota_a \volM. \lb{L-bdy}
\end{alignat}
The boundary term reduces to $\Wein \volH$ when pulled back on $\H$,
since $\volH = - \iota_k \volM$. The bulk term can also be written in terms of the sCarrollian tensor \eqref{WeinT} as 
\be 
\mathscr{L} = \frac12 \RC - \left(\Wein_i{}^j -\rho {\bar\mK}_i{}^j\right) \left(\bar\mK{}^i{}_j -\delta^i{}_j \bar\mK\right),
\ee 
where we denoted
\begin{align}
\bar{\mK}_i{}^j := (D_i k_k)q^{kj}
, \qquad \text{and} \qquad
{\bar\mK}{}^i{}_j := q^{ik}(D_k k_j), 
\end{align}
and $\bar\mK= \bar\mK_i{}^i=\btheta$ is their trace. This splitting of the bulk Einstein-Hilbert Lagrangian into bulk and boundary terms which are expressible in terms of sCarrollian objects,  is in the same spirit as the splitting of the Einstein-Hilbert Lagrangian into the ADM Lagrangian and the Gibbons-Hawking-York boundary term \cite{Wald:1984rg,Hawking:1996ww} (see also the discussion in \cite{Freidel:2020xyx} and references therein).

\subsection{Einstein equations on null surfaces and horizons} 

The Einstein equations on any null surface, a special case of \eqref{Einstein} where $\rho =0$ and $\v^a = n^a$, are particularly interesting. The expression for these equations simplifies due to the fact that, on the null surface $\N$, we have $\E \Neq \theta$, $\T_{AB} \Neq \s_{AB}$, and $\J^A \Neq 0$.
However, we need to remember that $\Lie_k \E\neq  \Lie_k \theta$ at $ \N$. Indeed, since $k[\rho]=\kappa$ we have $\Lie_k \E \Neq \Lie_k \theta+ 2 \kappa \btheta$ 
and 
$\Lie_k \T_{AB} \Neq \Lie_k \sigma_{AB}+ 2 \kappa \bs_{AB}$. After making those simplifications, these equations are
\begin{subequations}
\lb{Einstein-null}
\begin{align}
-G_{\v \v} &\Neq \Lie_\v\theta+(\theta+\P)\theta + \T^{AB}\s_{AB}  \\
G_{A \v} &\Neq \Lie_\v \p_A + \theta\p_A + \theta \ac_A  + \left( \sD_B  +\ac_B \right) (\T_A{}^B +\P \delta_A^B) \\
G_{k \v} & \Neq \Lie_\v \btheta+\btheta\Wein  +(\sD_A +\p_A + \ac_A ) (\p^A + \ac^A) - \tfrac{1}{2} \RC   \nonumber \\
& \Neq \Lie_k \theta + \btheta \Wein  - (\sD_A -\p_A) \p^A - \tfrac{1}{2} \RC \label{Gkv}\\
-G_{\langle A B \rangle} 
& \Neq 2 \Lie_\v \bs_{\la AB \ra} - 2(\theta + \P) \bs_{AB} + \btheta \T_{AB}+2 (\sD+\ac +\p)_{\la A} (\p+\ac)_{B \ra} - \RC_{\la AB \ra} \nonumber \\
& \Neq  2\Lie_k \T_{\la AB \ra}  + 2(\theta+\P)\bs_{AB} - \btheta \T_{AB} - 2(\sD -\p)_{\la A} \p_{B \ra} - \RC_{\la AB \ra} \\
\tfrac{1}{2}q^{AB} G_{AB} 
& \Neq \Lie_\v \btheta + \Lie_k\kappa -\P\btheta + (\sD_A +\ac_A) (\p^A+\ac^A)-\p_A (2\p^A+\ac^A)+\T^{AB}\bs_{BA} \\
& \Neq \Lie_k \Wein +(\theta+\P)\btheta   - (\sD_A +2\p_A+2\ac_A) \p^A +\T^{AB}\bs_{BA} \nonumber \\
-G_{k k}  & \Neq  \Lie_k\btheta  + \btheta_{AB} \btheta^{BA}\\
-G_{A k}
& \Neq \Lie_k\p_A +\btheta \p_A
 + \vor_{AB}(\p^B+ \ac^B)  -\sD_B \left( \btheta_A{}^B - \btheta \delta_A^B \right).
\end{align}
\end{subequations}
Note that the pressure and the news scalar are respectively given by $\P=-(\frac12 \theta+\kappa)$ and $\Wein = \theta+\kappa$.
We can also extract from these equations an interesting balance equation
\bee 
\tfrac12 q^{AB} G_{AB} - G_{kn} 
&\Neq \Lie_k\kappa +\tfrac{1}{2}\RC
-\tfrac12 \theta\btheta   - (3\p_A+2\ac_A) \p^A 
 +\sigma^{AB}\bs_{BA},
\eee
where we simply used $(\P +\kappa)=-\tfrac12\E \Neq \tfrac12\theta$. 
Note that the equations written here are central in the analysis of \cite{Sachs:1962wk, Gourgoulhon:2005ng, rendall1990reduction,hayward1993dual, Chandrasekaran:2023vzb} using null surfaces as initial data. 

\subsubsection{Non-expanding, Weakly isolated, and Isolated horizons}
Some interesting classes of null surfaces deserve further discussion, including non-expanding horizons, weakly isolated horizons, and isolated horizons \cite{Ashtekar:2000hw,Ashtekar:2001jb,Lewandowski:2006mx} (see also  \cite{Ashtekar:2021wld,Ashtekar:2024bpi,Ashtekar:2024mme} for recent developments about these horizons). Non-expanding horizons (NEH) are given by the condition $\theta_{ij}=0$, weakly isolated horizons (WIH) are given by the extra conditions $\cL_\v \omega_i=0$, and isolated horizons (IH) are obtained by demanding the additional condition $[\cL_\v, D_i]=0$.

For an NEH, the vanishing of the expansion tensor implies that the sCarrollian connection becomes metric and satisfies 
\begin{align}
D_i q_{jk}  \stackrel{\sss \mathrm{NEH}}{=} 0, \qq \text{and} \qq D_i \v^j \stackrel{\sss \mathrm{NEH}}{=} \omega_i \v^j.
\end{align}
The component $G_{\v\v}$ vanishes identically. Other evolution equations simplifies to (we have that $\P\stackrel{\sss \mathrm{NEH}}{=} -\kappa$ and $\Wein\stackrel{\sss \mathrm{NEH}}{=}\kappa$ and $\T_{AB}\stackrel{\sss \mathrm{NEH}}{=}0$)
\begin{subequations}
\lb{NEH}
\begin{align}
G_{A \v} & \stackrel{\sss \mathrm{NEH}}{=} \Lie_\v \p_A   - (\sD_A+\ac_A)\kappa \\
G_{k \v} & \stackrel{\sss \mathrm{NEH}}{=}  \Lie_\v \btheta+\btheta\kappa  +(\sD_A +\p_A + \ac_A ) (\p^A + \ac^A) - \tfrac{1}{2} \RC   \\
-G_{\langle A B \rangle} 
& \stackrel{\sss \mathrm{NEH}}{=} 2 \Lie_\v \bs_{\la AB \ra} + 2\kappa \bs_{AB} +2 (\sD+\ac +\p)_{\la A} (\p+\ac)_{B \ra} - \RC_{\la AB \ra}  \\
\tfrac{1}{2}q^{AB} G_{AB} 
& \stackrel{\sss \mathrm{NEH}}{=}  \Lie_\v \btheta + \Lie_k\kappa +\kappa \btheta + (\sD_A +\ac_A) (\p^A+\ac^A)-\p_A (2\p^A+\ac^A).
\end{align}
\end{subequations}
It then follows from $G_{k\v}$ and $q^{AB} G_{AB}$ that 
\bee 
\tfrac12 q^{AB} G_{AB} - G_{k\v} 
& \stackrel{\sss \mathrm{NEH}}{=} \Lie_k\kappa +\tfrac{1}{2}\RC - (3\p_A+2\ac_A) \p^A.
\eee
 
Using the Einstein equations $G_{A\v} =0$, we can show that the evolution of the rescaling connection, $\omega_i = \p_i + \kappa k_i$, can be written as 
\be
\Lie_\v \omega_i = q_i{}^j \left( \Lie_\v \p_j - \kappa \ac_j \right) + (\Lie_\v \kappa) k_i = D_i \kappa.  \lb{Lie-omega}
\ee 
This relation, which can be written as $\Lie_\v \bm{\omega} \stackrel{\sss \mathrm{NEH}}{=} \dH \kappa$, infers that the curl $\dH \bm{\omega}$ is conserved\footnote{This follows from $\Lie_\v \dH \bm{\omega} = \dH \Lie_\v \bm{\omega} \stackrel{\sss \mathrm{NEH}}{=} \dH^2 \kappa =0$.}. In particular, on NEH, both $\RC$ and $\dH \bm{\omega}$, which are respectively the real and the imaginary part of the Newman-Penrose Weyl scalar $\Psi_2$, are conserved (see \cite{Ashtekar:2021wld,Ashtekar:2024bpi} for their relations with multipole moments),
\begin{align}
\Lie_\v \RC = \Lie_\v \dH \bm{\omega} =0.
\end{align}
Under rescaling, we have that $\delta_\phi \omega_i = D_i \phi$, $\delta \kappa = \phi \kappa + \cL_v \phi$, and 
\bee
\delta_\phi (\cL_v \omega_i- D_i \kappa) &= \phi (\cL_v \omega_i- D_i \kappa).
\eee
This equality shows, as expected, that the conservation law \eqref{Lie-omega} is invariant under rescaling. The transformation of $\kappa$ means that we can always choose a gauge where the WIH condition $\Lie_\v \omega_i \stackrel{\sss \mathrm{WIH}}{=} 0$ is achieved. In this frame, the surface gravity is constant $D_i\kappa \stackrel{\sss \mathrm{WIH}}{=} 0$.  When considering the case $\v^i \omega_i = \kappa \stackrel{\sss \mathrm{WIH}}{=} 0$, the WIH is called \emph{extremal}\footnote{In the language of black hole thermodynamics, such a surface has zero temperature}. 

For an NEH, we have 
\begin{align}
\bar{R}_{i \v \v}{}^j \stackrel{\sss \mathrm{NEH}}{=}  (D_\v \omega_i + \kappa \omega_i - D_i \kappa) \v^j \stackrel{\sss \mathrm{NEH}}{=}  (\Lie_\v \omega_i - D_i \kappa )\v^j  \stackrel{\sss \mathrm{NEH}}{=} 0.
\end{align}
where $\bar{R}_{i j k }{}^\ell$ is the curvature tensor of the sCarrollian connection and $\bar{R}_{i \v \v}{}^\ell$ is its projection along $v^j $ and $v^k$.  

\paragraph{Isolated Horizon}
For an IH, the additional condition, $[\cL_\v, D_i] \stackrel{\sss \mathrm{IH}}{=} 0$, is needed. 
On the one hand, this commutator applied to $k_i$ can be evaluated in terms of the curvature tensor
\bee \label{vDcom}
[\cL_\v, D_i] k_j 
&\stackrel{\sss \mathrm{NEH}}{=} - R_{ivjk} - (D_i + \omega_i)\omega_j,
\eee 
where $\bar{R}_{i\v jk}$ is the curvature tensor of the sCarrollian connection $D_i$\footnote{It equals the projection of the Riemann tensor onto $\N$: $ \bar{R}_{i\v jk} \Neq \Pi_i{}^a \Pi_j{}^b R_{a\v bk} = R_{i\v j k}$ pulled-back to $\N$.}. Note that the left-hand side of \eqref{vDcom} is symmetric, which implies $R_{ij vk}=- 2D_{[i}\omega_{j]}$.
We can, on the other hand, evaluate directly this commutator using \eqref{D-k} and we get 
\be\label{vDcom2}
[\cL_\v, D_i] k_j = \Lie_\v \btheta_{ij}  + (D_i + \omega_i) \ac_j + \ac_i (\p_j + \ac_j)  - (\Lie_\v \omega_i) k_j - k_i \Lie_\v (\p_j+\ac_j).
\ee 
Projecting this equation along $v^i$ or $v^j$, one recovers the WIH condition: 
$\v^j [\cL_\v, D_i] k_j = \v^j[\cL_\v, D_j] k_i \stackrel{\sss \mathrm{NEH}}{=} - \Lie_v \omega_i \stackrel{\sss \mathrm{WIH}}{=} 0$\footnote{This means $\bar{R}_{i\v\v k} \stackrel{\sss \mathrm{NEH}}{=} \Lie_\v \omega_i - D_i \kappa = 0$.}.
This last equality means
$R_{i\v j k} \stackrel{\sss \mathrm{NEH}}{=} e_i{}^A e_j{}^B R_{A\v B k}$. 
Symmetrizing this identity and assuming that $R_{AB}=0$ implies that
$R_{(i|\v| j) k} \stackrel{\sss \mathrm{NEH}}{=}
 -\tfrac12 q^{CD} R_{CADB} e_i{}^A e_j{}^B.
$
Using the Gauss-Codazzi relation \eqref{GC1} for an NEH, and that $\RC_{(AB)} = \tfrac{1}{2}\RC q_{AB}$ in four dimensions, we obtain $R_{(i|\v |j) k} \stackrel{\sss \mathrm{NEH}}{=} - \tfrac14 \RC q_{ij}$, which can be used to simplify \eqref{vDcom}.
The horizontal components  of \eqref{vDcom2}read 
\begin{equation}
\begin{aligned} 
q_i{}^k q_j{}^l [\Lie_\v, D_k]k_l &= \Lie_\v \btheta_{ij} + \p_j\ac_j + \ac_i(\p_j +\ac_j) + \sD_i \ac_j \\
& = \Lie_\v \btheta_{(ij)}  + (\sD+2\p +\ac)_{(i} \ac_{j)} \stackrel{\sss \mathrm{IH}}{=} 0
\end{aligned}
\end{equation}
where we recalled $\sD_i \ac_j = e_i{}^A e_j{}^B \sD_A \ac_B$. The second equality followed from $\btheta_{[ij]} = -\tfrac12 \vor_{ij}$ and the Jacobi identity \eqref{Jacobi-1}, $\Lie_\v \vor_{ij} = 2\sD_{[i}\ac_{j]}$. See also \cite{Rignon-Bret:2024gcx, Ashtekar:2024mme} where this equation appears as a definition for the news tensor, in the context of asymptotic infinity. 

The Einstein equations $G_{k \v}$ and $G_{\la AB\ra}$ therefore dictate 
\begin{align}
\kappa \btheta_{(AB)} - \tfrac{1}{2}\RC_{(AB)} + (\sD+\p)_{( A} \p_{B )} \stackrel{\sss \mathrm{IH}}{=} 0. 
\end{align}
This is indeed an interesting result. It suggests that the isolated horizon exhibits the behavior of viscous fluids: that is, for $\kappa$ non-zero constant, the transverse shear is given by the gradient of fluid momentum,
\begin{align}
\bs_{AB} \stackrel{\sss \mathrm{IH}}{=} - \frac{1}{\kappa} (\sD+\p)_{\la A} \p_{B \ra}
\end{align}
where $\RC_{\la AB \ra} = 0$ for a 2-sphere $\S$ in 4-dimensional spacetime.
One can also check that the condition $[\Lie_\v, D_i] k_j \stackrel{\sss \mathrm{IH}}{=} 0$ is sufficient and that $[\Lie_\v, D_i] e_j{}^A \stackrel{\sss \mathrm{IH}}{=} 0$ does not provide any additional constraints\footnote{One can check that $[\Lie_\v, D_i] e_j{}^A \stackrel{\sss \mathrm{IH}}{=} 0$ by using the fact that, on NEH, we have $\Lie_\v \bm{e}^A = 0$, $\v^j D_i e_j{}^A = - D_i \v^j e_j{}^A = 0$, and $\v[\Gamma^A_{BC}] =0$.}. 


\subsubsection{Hydrodynamical horizons}
We have already seen in Section \ref{sec:Conection} that it was possible to fix the rescaling and the shift symmetry by choosing a  \emph{hydrodynamical frame} obtained by imposing that  $\bar\sigma_{ab}\Neq 0$ and $\bar\theta\Neq \frac1{\eta}$ where $\eta$ is constant on $\N$. In this frame, the Einstein equations simplify. They become
\begin{subequations}
\begin{align}
-G_{\v \v} &\Neq \Lie_\v\theta+(\theta+\P)\theta + \s^{AB}\s_{AB}  \\
G_{A \v} &\Neq (\Lie_\v +\theta) \p_A  + 
(\theta +\P) \ac_A  + \sD_A \P + \left( \sD_B  +\ac_B \right)\s_A{}^B \\
G_{k \v} & \Neq \tfrac{1}{2\eta} \left( \theta-2\P\right) +(\sD_A +\p_A + \ac_A ) (\p^A + \ac^A) - \tfrac{1}{2} \RC \\
G_{\langle A B \rangle} 
& \Neq   -\tfrac{1}{\eta}  \s_{AB}- 2 (\sD+\ac +\p)_{\la A} (\p+\ac)_{B \ra} 
\end{align}
\end{subequations}
where we have used that $\RC_{\la AB \ra} =0$ in four dimensions.
From these equations, we see that imposing the vacuum Einstein equations means that  
$\sigma_{AB}$ is determined from the  combination  of the fluid momenta and acceleration $\p_A+\ac_A$ and that $\eta$ plays the role of the shear viscosity. While $\P$ is determined from the divergence of $\p_A+\ac_A$. Once we choose a relationship between $\p_A$ and $\ac_A$ then the gravitational dynamics becomes, in this frame, similar to a  hydrodynamical flow equation (see also \cite{Redondo-Yuste:2022czg}). 

\subsubsection{Radial evolution and linear (in)stability of black holes}
One of the prime examples of a non-expanding horizon is a black hole horizon. There has been a longstanding question of the (in)stability of a black hole horizon when subjected to linear classical perturbations. In the case of stationary and axisymmetric spacetimes, Hollands and Wald \cite{Hollands:2012sf} (see also \cite{Keir:2013jga} for the case of non-vacuum solutions) have developed criteria for determining black holes (and the landscape of black objects) instability, which shares a close connection with thermodynamic instability.

One important point in their construction is that, to the first order in perturbations, an event horizon (which is an NEH) must continue to exist and that, using gauge transformations, the expansion $\theta$ remains zero \cite{Sorkin:1995dc}. Proving that this is always possible requires the Einstein equation that controls the radial evolution of the expansion $\Lie_{k} \theta$, namely $G_{\v k} =0$. Here, we will neither discuss nor provide a detailed explanation of their derivation. Instead, we will verify that the key equation in their work (equation (8) there) can be obtained from our result.

To make the comparison, one works in \emph{Gaussian null coordinates} where the spacetime metric near the black hole horizon admits a simple form
\begin{align}
g (\lambda) &= 2 \exd \sv \odot \left( \exd r - r^2 \mathfrak{a}(\lambda) \exd \sv - r \mathfrak{b}_A (\lambda) \exd z^A \right) + q_{AB} (\lambda) \exd z^A \odot \exd z^B \\
& = 2 \exd \sv \odot \left( \exd r - r^2 (\mathfrak{a} + \tfrac12 \mathfrak{b}\cdot \mathfrak{b}) \exd \sv \right) + q_{AB} \left(\exd z^A - r\mathfrak{b}^A \exd \sv \right)\odot\left(\exd z^B - r\mathfrak{b}^B \exd \sv \right)
\lb{GNC}
\end{align}
where the parameter $\lambda$ labels a family of perturbed metrics such that $g(0)$ is the unperturbed (background) metric whose (in)stability one seeks to analyze. 

To verify this, we must first match our metric \eqref{metric-coord} with the one expressed in the Gaussian null form \eqref{GNC}. The matching conditions are 
\begin{align}
\alpha =0, \ \ \  \beta_A = 0, \ \ \ V_A = r\mathfrak{b}_A,  \ \ \ J_A{}^B = \delta_A^B, \ \ \ \text{and} \ \ \ \rho = r^2(\mathfrak{a} + \tfrac12 \mathfrak{b}_A \mathfrak{b}^A).
\end{align}
It then implies that on the horizon, $\Upsilon_A = k[V_A]= \mathfrak{b}_A $ and $\kappa = \pa_r\rho$. Since $\bm{k}=\rd \sv$ in these coordinates, we have on the horizon that 
\be 
\kappa\Neq0, \quad \ac_A \Neq 0,\quad \vor_{AB}=0, \qq \text{and} \qq \p_A = \tfrac12 \Upsilon_A \Neq  \tfrac12 \mathfrak{b}_A.
\ee 
One starts with the $G_{k\v} = 0$ equation \eqref{Gkv} evaluated on the NEH, which reads
\begin{align}
G_{k\v} \stackrel{\sss \mathrm{NEH}}{=} \Lie_k \theta + \kappa\btheta   - (\sD_A -\p_A) \p^A - \tfrac{1}{2} \RC =0, \lb{NEH-radial}
\end{align}
To get the stability equation, one looks for a deformation of the metric under the  vector field
$\xi = f \pa_r$.
This corresponds to a rescaling $k \to f k$ with a function $f(\sv,z^A)$ on the horizon. This rescaling  transforms $\theta \to f^{-1} \theta$ and $\p_A \to \p_A - \sD_A \ln f$. The radial equation \eqref{NEH-radial} then becomes, 
\begin{align}
f\pa_r \theta = - (\sD^A - \mathfrak{b}^A)\sD_A f + \tfrac12 f \left( \RC(q) + \sD_A \mathfrak{b}^A - \tfrac12 \mathfrak{b}^A \mathfrak{b}_A \right),\lb{stability}
\end{align}
which is the key equation\footnote{In \cite{Hollands:2012sf} they have an extra term $\theta_{ab}\theta^{ab}$ which vanishes for horizons.} of \cite{Hollands:2012sf}. 
Following the result of \cite{Andersson:2005gq}, one has that the linear differential operator on the right-hand side is invertible. 
This ensures that the expansion for the perturbed surface $\theta(\lambda)$ can be brought back to zero using the diffeomorphism (gauge) transformation generated by $\xi$. Interested readers can find the details in \cite{Hollands:2012sf}.

\section{Einstein Equations from Symmetries} \lb{sec: Einstein-symmetries}

The primary objective of this work is to demonstrate the correspondence between the Einstein equation, which also relates to the sCarrollian fluid conservation laws, and the symmetries of the Carrollian stretched horizon. 

To accomplish this task, we employ the covariant phase space formalism: a modern-day language of the Noether theorem. At the heart of our consideration lies a Noether current (a spacetime 3-form), denoted by $\bm{J}_\xi$, associated with a symmetry generated by a vector field $\xi$. In theories characterized by diffeomorphism covariance, such as the Einstein-Hilbert gravity under our scrutiny, the expression for the Noether current is given by the following equation:
\begin{align}\label{Noether}
\bm{J}_\xi := I_\xi \bm{\Theta} - \iota_\xi \bm{L} = \bm{c}_\xi + \exd \bm{q}_\xi.
\end{align}
Here, $\bm{\Theta}$ represented the theory's pre-symplectic potential (a 3-form), $\bm{L}$ was the theory's bulk Lagrangian, and $I_\xi$ signified a field space interior product with a Hamiltonian vector field corresponding to a spacetime vector field $\xi$. Notably, the Noether current encompasses the \emph{constraint} term, $\bm{c}_\xi = \xi^a G_a{}^b \epsilon_b$, which vanishes on-shell, while the Noether \emph{charge aspect} (a 2-form) is denoted by $\bm{q}_\xi$. The Noether charge is the integral over the surface $\H$ of the current:
\begin{align}
Q_\xi := \int_\H \bm{J}_\xi = C_\xi + \int_{\pa \H} \bm{q}_\xi, \qq \text{where} \qq C_\xi := \int_\H \bm{c}_\xi.
\end{align}
This framework is general and relies on the bulk theory encoded in the Lagrangian $\bm{L}$. However, in this work, we are adopting a different perspective. We aim to derive the Noether charges and conservation laws solely from the sCarrollian data on the stretched horizon $\H$. This approach aligns with the principle of holography. First, we need to define the pre-symplectic potential for the sCarrollian geometry on $\H$.

\subsection{Gravitational pre-symplectic potential} \lb{sec: potential}

The covariant phase space analysis of Einstein-Hilbert gravity has undergone extensive scrutiny across various types of boundaries \cite{Harlow:2019yfa, Donnelly:2016auv,Adami:2021nnf,Hopfmuller:2016scf,Shi:2020csw}. In the case concerning a timelike Carrollian stretched horizon $\H$, we adopt the expression for the gravitational pre-symplectic potential derived in \cite{Freidel:2022vjq}. The pre-symplectic potential of Einstein gravity,  denoted $\Theta_\H^{\sss \mathrm{EH}}$ when pulled-back on $\H$, can be decomposed into three distinct terms (a short derivation is provided in Appendix \ref{App:potential}):
\be
\Theta_{\H}^{\sss \mathrm{EH}}= \Theta_{\H}^{\mathrm{can}} +\delta L_{\H} + \Theta_{\pa \H}. \lb{potential}
\ee
The total variation term, which depends on a  Lagrangian intrinsic to $\H$, $L_\H := \int_\H \bm{\ell}_\H$, and the corner term, $\Theta_{\pa \H} := \int_{\pa \H} \bm{\vartheta}_{\pa \H}$, are given by
\be 
L_{\H}  = \int_{\H} \Wein \volH , \qquad \text{and} \qquad \Theta_{\pa\H}  = -\frac{1}{2}   \int_{\pa\H } \delta v^a {\bvol}_a,
\ee
where we recalled that the area forms are ${\bvol}_a = \iota_{\pa_a}\volH$ and $\volS = \iota_v \volH$, and the expression \eqref{Ndef} for the news scalar is $\Wein = \tfrac{1}{2}\E-\P$.
Here, $\bm{\ell}_\H$ is precisely the boundary Lagrangian \eqref{L-bdy} pulled back on $\H$. The canonical term is given by the sCarrollian canonical pairs,
\be
\Theta_{\H}^{\mathrm{can}}   = - \int_{\H} \bigg[ 
\left( \E \v^a + \J^a \right)  \delta k_a
+ \p_a \delta \v^a  -
\tfrac{1}{2}(\T^{ab} + \P q^{ab}) \delta q_{ab} 
   +\btheta \delta \rho \bigg] \volH. 
\ee
A similar form of the canonical potential is given in \cite{Ciambelli:2023mir} for the null case, and in \cite{Freidel:2024tpl} for the case concerning asymptotic null infinities. 

The variation of the components constituting the sCarrollian structure $(\v^a, k_a,q_{ab},\rho)$ can be decomposed into several distinct components, denoted by
\be 
\delta k_\v := \v^a \delta k_a,\ \ \ \ \delta k_A  := e_A{}^a \delta k_a, \ \ \ \
 \delta v^A  :=  \delta \v^a e_a{}^A, \ \ \ \ \text{and} \ \ \ \ \delta q_{AB} := e_A{}^a e_B{}^b \delta q_{ab}.
\ee
This means we can decompose the sCarrollian variations as follows:
\begin{subequations}
\begin{alignat}{2}
&\delta  k_a &&= k_a \delta k_\v + e_a{}^A \delta k_A, \\ 
&\delta  \v^a &&= -(\delta k_\v)  v^a + \delta v^A  e_A{}^a, \\ 
&\delta q_{ab}  &&= -2   k_{(a} e_{b) A} \delta v^A + e_a{}^A e_b{}^B \delta q_{AB}. \qquad
\end{alignat}
\end{subequations} 
Furthermore, they can be explicitly expressed in terms of the variations of the sCarrollian coefficients (see equation \eqref{vector-form}) as\footnote{In our previous work \cite{Freidel:2022vjq}, we used a rather non-standard notations, that are $\var \alpha = \delta k_\v$, $\e^\a \var \beta_A = -\delta k_A$, $\e^{-\a} \var V^A = \delta \v^A$, and $\var q_{AB} = \delta q_{AB}$.}
\begin{subequations}\label{var}
\begin{align}
\delta k_\v & = \delta \a + \e^\a\beta_A \delta \v^A , \\
\delta k_A  &  = \e^\a\big(   \e^\a(\beta \cdot \delta \v)\beta_A-(J^{-1})_A{}^C\delta ( J_C{}^B\beta_B ) \big), \\
\delta q_{AB} & = (J^{-1})_A{}^C (J^{-1})_B{}^D\delta (J_C{}^E J_D{}^F q_{EF}) - 2q_{C(A} \beta_{B)} \e^\a\delta \v^C, \\
\delta \v^A & = \e^{-\a} (\delta V^B)J_B{}^A. 
\end{align}
\end{subequations}

The gauge we have chosen is such that the rigging structure $(\bm{n}, k)$ is treated as a background structure, inferring that the variation leaves invariant the rigging structure and also the rigging projector,
\be
\delta n_a=0,\qquad \delta k^a =0,\qquad \text{and} \qquad \delta \Pi_a{}^b=0. 
\ee
It also implies that the variational coefficients \eqref{var} enter the decomposition of the metric variation as follows
\be
\delta g_{ab} = - 2  k_ak_b \delta \rho  + 2 v_{(a} k_{b)} \delta k_\v 
- k_{(a} e_{b) B} \delta v^B + 
2 v_{(a} e_{b)}{}^B \delta k_B
+ e_{(a}{}^A e_{b)}{}^B \delta q_{AB},
\ee
where $ e_{bA} = e_b{}^B q_{B A}$. The canonical pre-symplectic potential can be written in terms of these coefficients as
\be 
\Theta_{\H}^{\mathrm{can}}  = -\int_{\H}\bigg[  \E \delta k_\v +  \J^A \delta k_A + \p_A  \delta \v^A   - \tfrac{1}{2}(\T^{AB} + \P q^{AB}) \delta q_{AB} +\btheta \delta \rho \bigg]\volH.  \lb{potential-can}
\ee
Note that although we derived the canonical pre-symplectic potential from the Einstein-Hilbert theory, the canonical structure should hold for any theory. Specifically, this form of the canonical pre-symplectic potential aligns with variations of any Carrollian action (derived from the $c \to 0$ limit), augmented by the $\btheta \delta \rho$ term originating from the stretching \cite{Freidel:2022bai,Ciambelli:2018ojf}. The appearance of the sCarrollian canonical pre-symplectic potential in the pre-symplectic potential for the Einstein-Hilbert theory strongly supports the correspondence between Carrollian hydrodynamics and gravity at the stretched horizon.
\subsection{Symmetry transformations} \lb{sec:symmetry}

We now discuss the symmetries and how they act on the fields. The symmetries are generated by a diffeomorphism vector field,
\begin{align}
\xi = T \v  + X^A e_A + R k. \lb{diffeo}
\end{align}
The action of the symmetries on the phase space variables is denoted by $\delta_\xi$. This action preserves the rigging structure $(n_a,k^a,\Pi_a{}^b)$, which means that it generally differs form the Lie derivative action. Demanding that the symmetry transformations preserve the gauge conditions: 
\begin{align}
k^ak_a=0,\qq  \Lie_k  \bm{k}=0,\qq \text{and} \qq \dH e^A=0. \lb{gauge-constraint}
\end{align}
The former two conditions impose the following radial evolution equations for the symmetry parameters\footnote{We use that 
$\pa_r X^a = \Lie_k X^a=(k[X^A]-X^B \Omega_B{}^A) e_A{}^a$.}, 
\begin{equation}
\begin{alignedat}{3}
&\pa_r T=0,  \qquad &&\pa_r X^a = Z^a - T\Upsilon^a ,\qquad \pa_rR= \W, \\
&\pa_r \W = 2\p_a  Z^a, \qquad &&\pa_r Z_a = Z^b \vor_{ba}. \lb{radial-constraint}
\end{alignedat}
\end{equation}
where we denote $ X^a= X^A e_A{}^a$ and similarly for $Z^a= q^{ab} Z_b$ and $\Upsilon^a$.\footnote{Preserving the last condition, $\dH \bm{e}^A =0$, does not impose an additional constraint on the transformation parameters but instead determines how the horizontal basis fields $(e_A{}^a, e_a{}^A)$ transform. Their transformations turn out to be non-local (see equations \eqref{delta-coframe} and \eqref{delta-frame} in Appendix \ref{App:symmetries}). } Note that the first three radial derivatives appear in the Lie bracket, 
\be 
[k,\xi]
= k[T] \v  + k[R] k  + Z^A e_A.
\ee This indicates that the diffeomorphism $\xi$ preserving the background rigging structure is characterized by the parameters $(T,R,\W, X^A,Z^A)$ on $\H$ obeying the linear radial constraint equations \eqref{radial-constraint}. The parametrization we have chosen emphasize that both $R$ and its radial derivative $\pa_rR=W$,   serve as free and independent transformation parameters when evaluated on the horizon $\H$.
This is encoded in the relationship
\be \W= k[R],
\ee
which means that $\W$ labels rescaling transformations $r\pa_r$.

Elements of the sCarrollian structure transform under these symmetries as follows (the detailed derivation is provided in Appendix \ref{App:symmetries}): 
\begin{subequations}
\lb{sCarrollian-transform}
\begin{alignat}{3}
&\delta_\xi \rho &&= (T \v  + X^A e_A)[\rho]  - (\v- \kappa )[R]-2\rho \W, \\
&\delta_\xi k_\v &&= \v[T] + X^A\ac_A+ \W,\\
&\delta_\xi k_A &&= (\sD_A-  \ac_A)T +\vor_{AB} X^B  + Z_A,\\
&\delta_\xi \v^A &&=- \v[X^A] -(\sD^A- \Upsilon^A) R - 2\rho Z^A, \\
&\delta_\xi q_{AB} &&= 2(T \theta_{(AB)} + \sD_{(A} X_{B)} + R \btheta_{(AB)} ), \\
&\delta_\xi \volH &&= \left( (\v +\theta)[T] +  (\sD_{A} + \ac_A) X^A + R\btheta + \W \right) \volH. \label{volH}
\end{alignat}
\end{subequations} 
When the hypersurface is null, $\rho \Neq 0$. Demanding that the symmetry transformation preserves the null-ness imposes $(\v- \kappa )[R] \Neq 0$. Quite remarkably, it does not impose that $R$ vanishes. If, in addition, we demand that the Carrollian bundle is preserved, this imposes $\delta_\xi v^A=0$, providing us with the additional relationship, 
\be 
v[X^A]+ (\sD^A -\Upsilon^A) R\Neq 0.
\ee
 This means that there is an additional symmetry of null surface associated with time independent ($(\v- \kappa )[R] \Neq 0$) radial translation! Note that it is only when $(\sD_A-\Upsilon_A)R=0$  and $Z_A=0$ that $\delta_\xi v^A=0$ implies the Carrollian condition $\v[X^A]=0$.

When the Gaussian null coordinates condition is imposed, we have that  $\delta k_v=0$, which relates $v[T]$ with $W$.
Finally, let us note that we can always impose that $\delta k_i=0$ by choosing $(W,Z_A)$ in terms of $(T,R,X^A)$. This corresponds to the prime phase space of \cite{Ciambelli:2023mir}.
In particular, when $\ac_i = 0$ and $\vor_{ij} =0$, the variation of $k_i$ is given by 
\be 
\delta_\xi k_i = D_i T + W k_i +Z_i.
\ee 
Choosing the rescaling and shift to be given by $W = -v[T]$ and $Z_i = -\sD_i T$ imposes $\delta_\xi k_i =0$.

\subsection{Quasi-Killing symmetry}

Up to this point, our focus has been on symmetries generated by general diffeomorphism that preserves the rigging structure. We now turn our attention to a noteworthy special case known as the \emph{quasi-Killing symmetry}, which has been employed in the context of gravitational algebras and dynamical entropy \cite{Faulkner:2024gst,Hollands:2024vbe}. This symmetry preserves the metric only on the null boundary $\N$

Any diffeomorphism vector field \eqref{diffeo} that is, when evaluated on $\N$ (where $r =0$), parallel to the null generator of $\N$ requires $(R,X^A) \Neq 0$. From \eqref{sCarrollian-transform}, this also infers that $\delta_\xi \rho \Neq 0$ and $\delta_\xi \v^A \Neq 0$. The variation of the metric, $\delta_\xi g_{ab} = \Lie_\xi g_{ab}$, evaluated on $\N$ is then given by 
\begin{equation}
\begin{aligned}
\delta_\xi g_{ab} &\Neq e_a{}^A e_b{}^B(\delta_\xi q_{AB}) + 2n_{(a} \delta_\xi k_{b)} \\
& = 2T\theta_{ab} + 2\left(v[T] +W \right)n_{(a}k_{b)} + \left( (\sD_A -\ac_A)T + Z_A  \right) 2n_{(a} e_{b)}{}^A.
\end{aligned}
\end{equation}
Demanding $\delta_\xi g_{ab} \Neq 0$ imposes the following conditions
\begin{align}
\theta_{(ab)} \Neq 0, \qq W \Neq - \v[T] , \qq \text{and} \qq Z_A \Neq -\left(\sD_A - \ac_A\right)T. 
\end{align}
The first condition dictates that the null surface $\N$ is non-expanding and shear-free. The remaining two conditions relate the symmetry parameters $(W,Z)$ to $T$. Hence, the  =quasi-Killing vector field considered here is labelled by the function $T(\sv,z^A)$. The components $(R,X^A)$ of the vector field $\xi_T$ outside $\N$ can be obtained by integrating $(W,Z^A)$ with respect to the radial coordinate $r$ (see \eqref{radial-constraint}). The resulting vector field $\xi$ is expressed as a power series expansion in small-$r$ by
\begin{equation}
\begin{aligned}
\xi_T^a & = T \v^a  - r\v[T] k^a - r\left(\sD^a +2\p^a \right)T  + \mathcal{O}(r^2) \qq \text{or} \\
\xi_{Ta} & = T \v_a - r \left(D_a T  + 2T\p_a\right) + \mathcal{O}(r^2).\label{xiT}
\end{aligned}
\end{equation}
The result concurs with \cite{Faulkner:2024gst}\footnote{In \cite{Faulkner:2024gst}, they work with the vector $\ell = \pa_\sv = v - 2 \p^a \pa_a $, hence \eqref{xiT} reads $\xi = T \ell - rD_a T$.}.
Note that if we impose that the inaffinity of $\v^a$ vanishes on $\N$ and demand that  
the quasi-Killing symmetry preserves the condition  $\delta_\xi \kappa \Neq 0$ \cite{Chandrasekaran:2018aop}. One can show using the variation $\delta_\xi \rho$ given in \eqref{sCarrollian-transform} that the variation $\delta_\xi \kappa$ under the quasi-Killing symmetry is 
\begin{align}
\delta_\xi \kappa \Neq - \Lie_\v W = -\Lie_\v \Lie_\v T =0.
\end{align}
Preserving $\kappa$ additionally imposes $\Lie_\v W = \Lie_\v \Lie_\v T =0$. This means that $\xi_T$  belongs to the BMSW group of symmetries  \cite{Freidel:2021fxf}. 


\subsection{Anomaly}

The diffeomorphism vector field $\xi$ can be split into the tangential component, $\xit = T \v + X^Ae_A$, and transverse component, $\xi_{\perp} = R k$. Additionally, we have identified a \emph{rescaling symmetry} parameter $W=k[R]$ and a \emph{shift symmetry} represented by the parameter $Z^A = \mathcal{L}_k X^A - T \Upsilon^A$. As we will demonstrate in the subsequent section, the shift symmetry, a gauge transformation while the rescaling symmetry is an edge mode symmetry.

As firstly emphasized in \cite{Hopfmuller:2018fni} the phase space action $\delta_\xi$ of the symmetry generators can be decomposed into  the Lie action $\cL_\xi$ and the anomaly action\footnote{In the context of covariant phase space, the anomaly operator is defined as $\Delta_\xi := \delta_\xi - \Lie_\xi - I_{\hat{\delta \xi}}$, where the last term compensates the possible field-dependent of $\xi$.} $\Delta_\xi$. We can evaluate the anomaly operator when acting on the rigging projector
\be 
\Delta_\xi \Pi_a{}^b = (D_a W)k^b-n_aZ^b,
\ee 
which vanishes when both $W =0$ and $Z^A =0$.
Accordingly, we  have that the tangential diffeomorphism  which is independent of $r$, i.e., for which $W=Z^A=R=0$, is not anomalous. For such diffeomorphism, we have $\Delta_{\xit}(h_{ab},v^a,k_b)=0$. We can also check that its action on the news and stress tensor vanishes:
\begin{align}
\Delta_{\xit} \Wein_a{}^b = 0, \qquad \text{and} \qquad \Delta_{\xit} \EM_a{}^b = 0.
\end{align}
The \emph{transverse translation} $\xi_\perp$ and the shift transformation, arising from the embedding of $\H$ in the ambient spacetime $\M$, contribute to the deviation of the variation $\delta_\xi$ from the Lie derivative $\Lie_\xi$, as indicated in equation \eqref{delta-news-tensor}. In other words, they render the symmetry transformation anomalous. 
In fact, transformation of sCarrollian structure is also anomalous due to $(R,\W,Z^A)$, as we can show that 
\begin{subequations}
\begin{alignat}{3}
& \Delta_\xi \rho && =- \left( \v[R] + 2\rho \W\right) \\
& \Delta_\xi \v^a && = - \W \v^a + \v[R]k^a - (\sD^a R + 2\rho Z^a) \\
& \Delta_\xi k_a && = \W k_a + Z_a \\
& \Delta_\xi q_{ab} && = (\sD_a R + 2\rho Z_a) k_b - n_a Z_b + (a \leftrightarrow b).
\end{alignat}
\end{subequations}

\subsubsection{Transformation of the news and the stress tensor}
Variation under these symmetries of the sCarrollian stress tensor is also interesting, although it plays little role in this paper. We can show that (see Appendix \ref{app:var-em}) the news tensor and the news scalar transform under the symmetries $\xi = T \v + X^A e_A + R k$, whose components satisfy the constraints \eqref{radial-constraint}, as
\begin{equation}
\lb{delta-news-tensor}
\begin{split}
\delta_\xi \Wein_a{}^b 
= \ & \LH_\xi \Wein_a{}^b - \left(\v^b D_a +\Wein_a{}^b\right)\W -  \left(\btheta_a{}^b+ k_a(\sD^b - 2\p^b) \right)\v[R]  +\sD_a R \p^b   \cr
&   - \sD_a \sD^b R + \left( \btheta_a{}^c - k_a (\p^c + \ac^c) \right)(\sD_c R)  \v^b - k_a \left(\theta^{bc}\sD_c R  \right) - D_a \rho Z^b, \\
= \ & \LH_\xi \Wein_a{}^b - \left(\v^b D_a +\Wein_a{}^b\right)\W - D_a \rho Z^b 
-\left( \sD_a \sD^b R + \btheta_a{}^b \v[R]  -\sD_a R \p^b\right)
\cr
&    
+ \left(\btheta_a{}^c\sD_c R  - k_a (\p^c + \ac^c)(\sD_c R)\right)  \v^b  -k_a\left[(\sD^b - 2\p^b)\v[R]  +  \left(\theta^{bc}\sD_c R  \right) \right] , 
\\
\delta_\xi \Wein =\ & \Lie_\xi \Wein- \left(\v +\Wein \right)[\W] - \btheta\v[R]  - (\sD_a + \ac_a) \sD^a R  - Z^a \sD_a \rho
\end{split}
\end{equation}
where we denoted $\sD_a \sD^b R = e_a{}^A (\sD_A \sD^B R) e_B{}^b$ and we defined the spacetime Lie derivative projected onto the stretched horizon $\H$ as
\begin{align}
\LH_\xi \Wein_a{}^b := \Pi_a{}^c (\Lie_\xi \Wein_c{}^d) \Pi_d{}^b 
=  \Lie_{\xit} \Wein_a{}^b + R \cL_k\Wein_a{}^b,
\end{align}
and it can be generalized to a tensor of arbitrary degree. Combining this transformation with \eqref{volH}, we can show, with the help of the Stokes theorem \eqref{Stokes0}, that 
\begin{flalign}
\delta_\xi (\Wein \volH) = & \left( R \left((k+\btheta)[\Wein]+(v+\theta)[\bar\theta] \right)  + \W \theta- Z^A \sD_A \rho\right) \volH \cr
& + \dH \left( (T \Wein- R\bar\theta -\W)\volS +( X^A  \Wein - \sD^A R) \bvol_A \right), 
\end{flalign}
where we used that
\be 
-\bar\theta v[R]= 
-v[\bar\theta R] + v[\bar\theta] R
= -(v +\theta)[\bar\theta R] + (v+\theta)[\bar\theta] R. 
\ee 

Knowing how the news $\Wein_a{}^b$ transforms, we can evaluate the transformation of the components of the sCarrollian fluid stress tensor. We present the derivation and the results in Appendix \ref{app:var-em}. 

Lastly, the anomaly of the news scalar $\Wein$ will be important in the subsequent discussion of the canonical charge. It is given by
\begin{align}
\Delta_\xi \Wein   &=  - \left(\v +\Wein \right)[W] - \btheta\v[R]    - (\sD_a + \ac_a) \sD^a R  - Z^a \sD_a \rho \\
\Delta_\xi (\Wein\volH)  &=  \left(W\theta - \btheta\v[R] - Z^A \sD_A \rho \right)\volH - \dH \left( W \volS + \sD^A R \bvol_A \right),\lb{Delta-News-1}
\end{align}
where we used $\Delta_\xi\volH = W \volH$ and the Stokes theorem \eqref{Stokes} to derive the anomaly $\Delta_\xi (\Wein\volH)$.

\subsection{Canonical charge}

After outlining the field variations under the diffeomorphism that preserves the background structure, we now proceed to construct the corresponding Noether charges associated with these symmetries. First, we need to define what it means for transformations to be symmetries at the level of phase space. In what follows, the transformations $\xi$ are considered symmetries of the sCarrollian canonical phase space if its contraction (in field space) with the canonical pre-symplectic potential \eqref{potential-can} satisfy the following relation
\begin{align}
\boxed{
I_\xi \Theta^{\mathrm{can}}_\H = C^\mathrm{can}_\xi - \int_\H \left(\Delta_\xi  \bm{\ell}_\H + \iota_\xi \exd \bm{\ell}_\H \right)  + \int_{\pa \H} \bm{q}^{\mathrm{can}}_\xi. \lb{symmetries-def}
}
\end{align}
In this equation, $C^\mathrm{can}_\xi$ denotes the \emph{bulk canonical constraint} (on the surface $\H$) which vanishes on-shell, $\bm{q}^{\mathrm{can}}_\xi$ is the canonical charge \emph{aspect} associated with the symmetry, and $\bm{\ell}_\H$ is the boundary Lagrangian pulled-back to $\H$ and is given by
\begin{align}
\bm{\ell}_\H = \Wein \volH \qq \text{and that} \qq L_\H := \int_\H \bm{\ell}_\H = \int_\H \Wein \volH
\end{align}
is precisely the boundary term appearing in the gravitational pre-symplectic potential \eqref{potential}.

When $\xi=\xit$ is tangential to $\H$, the anomaly vanishes and the term $\iota_\xi \exd \bm{\ell}_\H$ also vanishes,\footnote{The anomaly is non-zero for the transformations $(W,Z,R)$ while the term $\iota_\xi \exd \bm{\ell}_{\H}$ is non-trivial only for the radial translation $R$.} hence \eqref{symmetries-def} simply reads 
$Q^{\mathrm{can}}_{\hat{\xi}} =I_\xi \Theta^{\mathrm{can}}_\H = C^\mathrm{can}_\xi + \int_{\pa \H} \bm{q}^{\mathrm{can}}_\xi\hat{=} \int_{\pa \H} \bm{q}^{\mathrm{can}}_\xi$. In this case, we see that the contraction of the symmetry generator is a corner term on-shell. This is the hallmark of a local symmetry.

In the general case, the canonical charge is given by 
\begin{align}
Q^{\mathrm{can}}_\xi :=  C^\mathrm{can}_\xi  + \int_{\pa \H} \bm{q}^{\mathrm{can}}_\xi = I_\xi \Theta^{\mathrm{can}}_\H + \int_\H \left(\Delta_\xi  \bm{\ell}_\H + \iota_\xi \exd \bm{\ell}_\H \right). \lb{Q-can}
\end{align}
The relation between the canonical charge $Q^{\mathrm{can}}_\xi$ and the one derived from the Einstein-Hilbert gravity is discussed in Appendix \ref{App:charges}.

Let us remark that we can alternatively write \eqref{symmetries-def} as the contraction with the new canonical pre-symplectic potential enhanced with the total variation term, $\widetilde{\Theta}^{\mathrm{can}}_\H := \Theta^{\mathrm{can}}_\H + \delta L_\H$. 
${\Theta}^{\mathrm{can}}_\H$ and $\widetilde{\Theta}^{\mathrm{can}}_\H$ are related by a change of polarization and therefore lead to the same symplectic form.
By using $(\Delta_\xi + \iota_\xi \exd)\bm{\ell}_\H = (\delta_\xi - \exd \iota_\xi)\bm{\ell}_\H$, we can then write \eqref{symmetries-def} as
\be 
I_\xi \widetilde{\Theta}^{\mathrm{can}}_\H  = C^\mathrm{can}_\xi + \int_{\pa \H} \widetilde{\bm{q}}^{\mathrm{can}}_\xi,
\ee 
where $\widetilde{\bm{q}}^{\mathrm{can}}_\xi = \bm{q}^{\mathrm{can}}_\xi + \iota_\xi \bm{\ell}_\H$ is the new charge aspect for the new potential. 

From the definition \eqref{symmetries-def}, one can already distinguish two types of symmetries: the \emph{geometrical symmetries} which are non-anomalous  and the \emph{generalized symmetries} which are anomalous. The geometrical symmetries are symmetries for which $I_\xi \Theta^{\mathrm{can}}_\H$ are given by the corner integral when evaluated on-shell. The tangential diffeomorphism $\xit = T\v +X^A e_A$ belongs to this type of symmetry. The generalized symmetries, on the other hand, receive additional contributions from the term $\Delta_\xi  \bm{\ell}_\H + \iota_\xi \exd \bm{\ell}_\H$, signifying that $I_\xi \Theta^{\mathrm{can}}_\H$ is not a pure corner term while  $I_\xi \Theta^{\mathrm{can}}_\H + \delta_\xi L_{\H}$ is. The rescaling symmetry $W$, the shift symmetry $Z_A$, and the transverse translation $R$ fall under this category.

For the symmetries in consideration, generated by $(T,X^A,\W, Z^A, R)$, the canonical charge can be generally decomposed as the sum 
\be
Q^{\mathrm{can}}_\xi = Q^{\mathrm{can}}_{T}+Q^{\mathrm{can}}_{X} + Q^{\mathrm{can}}_{Z}+Q^{\mathrm{can}}_{\W} +Q^{\mathrm{can}}_{R}.
\ee
We will evaluate each term individually.

\subsubsection{Time (vertical) translation}

The time translation is generated by the diffeomorphism in the direction of the sCarrollian vector field $\xi = T \v$ obeying the condition $\pa_r T =0$ \eqref{radial-constraint}. It acts on the sCarrollian fields as follows:
\begin{flalign}
&\delta_T  k_a = \v[T]k_a + (\sD_a-\ac_a)T,  
\ \ \ \ \delta_T  \v^a = - \v[T] v^a  ,
\ \ \ \ \delta_T  \rho = T \v[\rho] ,
\ \ \ \ \delta_T q_{ab}  = 2 T \theta_{(ab)} . &&
\end{flalign}

To compute the canonical charge, we first evaluate the field space contraction of the canonical pre-symplectic potential \eqref{potential-can} with this symmetry transformation. The result is
\begin{equation}
\begin{aligned}
I_T \Theta^{\mathrm{can}}_{\H} &=  -\int_{\H}\bigg[  \E \v[T]  + \J^A  (\sD_A-\ac_A)T   
- T\left( {\T}^{ab} + {\P} q^{ab} \right) \theta_{ab} + \btheta \v [\rho] \bigg]\volH. 
\end{aligned}
\end{equation}
Since the boundary Lagrangian is covariant under time translation, meaning $\Delta_T \bm{\ell}_\H = 0$, and after performing integration by parts and applying Stokes' theorem \eqref{Stokes}, the canonical time translation charge is given by:
\begin{equation}
\lb{Q-T-off}
\begin{aligned}
Q_{T}^{\mathrm{can}} = \ &  \int_{\H} T \bigg[   (\v+\theta)[\E]  + \P \theta  + (\sD_A + 2\ac_A )\J^A     
+  \T^{AB} \sigma_{AB}   - \btheta \v [\rho]  \bigg]\volH \\
&  - \int_{\pa \H} \bigg( T \E  \volS + T \J^A  \bvol_{ A}  \bigg) \\
= \ &- \int_{\H} T G_{\v n} \volH + \int_{\pa \H} T \EM_\v{}^a \bvol_a,
\end{aligned}
\end{equation} 
where we recalled the expression \eqref{EinsteinEv} for the Einstein tensor $G_{\v n}$ and the energy momentum tensor \eqref{sCT}, $T_\v{}^a = - (\E \v^a + \J^a)$. We see that the bulk constraint term, $G_{\v n} =0$, is the energy evolution equation of sCarrollian hydrodynamics (i.e., the Raychaudhuri equation). The on-shell charge is the energy current (in agreement with \cite{Freidel:2022vjq,Ciambelli:2023mir}),
\be 
Q_{T}^{\mathrm{can}} \ \EOM
- \int_{\pa \H} T\bigg(  \E  \volS +  \J^A  \bvol_{ A}  \bigg)
=  \int_{\pa \H}  T \EM_{v}{}^a \bvol_a  . \lb{Q-T}
\ee


\subsubsection{Horizontal diffeomorphism}
The horizontal diffeomorphism is generated by the vector field $\xi = X^A e_A$ satisfying $\Lie_k X^A = 0$ (when $T=0$ and $Z^A =0$). The symmetry acts on the sCarrollian fields as follows:
\begin{equation}
\begin{aligned}
\delta_X  k_a &= k_a (X^A\ac_A)  + \vor_{ab} X^b    ,\\
\delta_X  \v^a &= - (X^A\ac_A) \v^a - \v[X^A] e_A{}^a  ,\\
\delta_X \rho &= X[\rho] ,\\
\delta_X q_{ab} & = 2 \sD_{(a} X_{b)} .
\end{aligned}
\end{equation}  
Similar to the time translation charge, as the anomaly vanishes $\Delta_X \bm{\ell}_\H =0$, the corresponding canonical charge is obtained from the contraction $I_X \Theta^{\mathrm{can}}_\H$. With the help of Stokes theorem \eqref{Stokes}, one can show that 
\begin{equation}
\lb{Q-X}
\begin{aligned}
Q^{\mathrm{can}}_X = \ &   -\int_{\H}\bigg[  \E (X^a\ac_a)  + \J^a \vor_{ab} X^b -\p_A \v[X^A]
- \left( {\T}^{ab} + {\P} q^{ab} \right) \sD_a X_b + \btheta X [\rho] \bigg]\volH \\
= \ & - \int_{\H} X^A \bigg[ (\v +\theta)[\p_A]+ \E \ac_A   + \J^B \vor_{BA}   
+ (\sD_B +\ac_B) \left( {\T}^{BA} + {\P} q^{BA} \right)  + \btheta \sD_A [\rho]  \bigg]\volH\\
&  + \int_{\pa \H} \bigg( \left(X^A\p_A\right) \volS + \left(\P X^A   + \T^{A}{}_B X^B \right)  \bvol_{ A}  \bigg)\\
= \ & -\int_\H X^A G_{A n} \volH + \int_{\pa \H} X^A \EM_A{}^a \bvol_a
\end{aligned}
\end{equation} 
where we referred the expression \eqref{Einstein} for $G_{An}$ and the decomposition of the stress tensor \eqref{sCT}. The horizontal diffeomorphism generates the Carrollian momentum conservation (i.e., the Damour equation), $G_{An} =0$. On-shell the charge reads
\be 
Q_{X}^{\mathrm{can}} \ \EOM  \int_{\pa \H} X^A \bigg( \p_A \volS +  \T_{A}{}^B    \bvol_{ B}  \bigg) = \int_{\pa \H} X^A  \EM_A{}^a   \bvol_a.
\ee
This expression is again in  perfect agreement with the  results of \cite{Freidel:2022vjq,Ciambelli:2023mir}.

Once combined the charge $Q^{\mathrm{can}}_T$ and $Q^{\mathrm{can}}_X$, we can write the canonical charge associated to tangential diffeomorphism $\xit = T \v + X^A e_A$ as 
\begin{align}
Q^{\mathrm{can}}_{\xit} = -\int_\H \xit^i G_{i n} \volH + \int_{\pa \H} \xit^j \EM_j{}^i \bvol_i \ \EOM \int_{\pa \H} \xit^j \EM_j{}^i \bvol_i. \lb{Q-tan}
\end{align}


\subsubsection{Shift symmetry}
The shift symmetry corresponds to a Lorentz transformation preserving the stretching $\rho$ and the normal $n_a$, while shifts the Ehresmann connection $k_a$ along a horizontal vector $Z = Z^A e_A$. This symmetry acts on the sCarrollian structure as
\be
\delta_Z\rho=0,\qquad \delta_{Z} k_a =Z_a, \qquad \delta_{Z} v^a = - 2\rho Z^a, \quad \text{and} \quad \delta_{Z} q_{ab}  = 2\rho( k_a Z_b +k_b Z_a).
\ee
It also transforms the boundary Lagrangian anomalously \eqref{Delta-News-1}, 
\be
\Delta_Z \bm{\ell}_\H = - \left(Z^A\sD_A \rho\right) \volH.
\ee
We find that the associated canonical charge \eqref{Q-can} is 
\be 
Q_{Z}^{\mathrm{can}}= \int_{\H} Z^a \left(-\J_a  + 2\rho \p_a    - \sD_a \rho  \right)\volH =0
\ee
The vanishing of the charge follows from the heat current, $\J_a$, defining relation \eqref{heat-gen}. It means that the shift symmetry is \emph{pure gauge} and therefore does not play any role in the construction of the charges.\footnote{The fact that shift is pure gauge is a central part of the analysis done in \cite{Freidel:coming}.}

\subsubsection{Rescaling symmetry}
Next, we consider the rescaling symmetry labelling by $W = k[R]$ (which is non-zero even if we fix $R=0$ on one surface). This symmetry acts on the sCarrollian fields as
\begin{equation}
\begin{aligned}
\delta_{\W}  k_a &=  \W k_a    ,\qq
\delta_{\W}  \v^a &= - \W v^a ,\qq
\delta_{\W}  \rho &= - 2 \W \rho ,\qq \text{and} \qq
\delta_{\W} q_{ab} & =0.
\end{aligned}
\end{equation} 
The anomaly of the boundary Lagrangian is also non-vanishing for this symmetry (see \eqref{Delta-News-1}),
\begin{flalign}
\Delta_\W \bm{\ell}_\H  =  \left(\W \theta \right) \volH - \dH \left(  \W\volS \right).
\end{flalign}
It is then straightforward to show that 
\begin{equation}
\begin{aligned}
I_{\W} \Theta^{\mathrm{can}}_{\H} 
& = -\int_{\H}\bigg[  \E \W    -\btheta (2\rho \W) \bigg]\volH = -\int_{\H}\left(  \theta \W  \right)\volH \\
\int_\H \Delta_W \bm{\ell}_\H &=  \int_{\H}\left( 
\W \theta   
 \right)\volH   - \int_{\pa \H} \left( \W  \volS \right),
\end{aligned}
\end{equation} 
where we recalled the expression for the energy $\E = \theta + 2\rho\btheta$. The anomaly of the boundary Lagrangian exactly cancels the bulk symplectic contribution.
Therefore, we obtain the canonical charge associated to the rescaling symmetry,
\begin{equation}
\begin{aligned}
Q_W^{\mathrm{can}} 
= \ &  - \int_{\pa \H} \left( \W  \volS  \right),
\end{aligned}
\end{equation} 
The charge is directly a corner charge without the need to impose a constraint. The rescaling symmetry, therefore, represents an edge mode symmetry, with the corresponding charge aspect being the area element (see \cite{Adami:2021nnf,Freidel:2021fxf,Geiller:2024amx} for the similar results in the null case).

\subsubsection{Transverse translation}
Eventually, we consider the transverse translation generated by a transverse vector field $\xi = R k$. First, if we consider the case where $(T,X^A,Z^A) =0$, we can write (by solving the constraints \eqref{radial-constraint}) the transformation parameter as $R = R_0 + r W$, where both $R_0$ and $W$ are functions on the surface $\H$. This means that the rescaling symmetry can be thought of as a part of the transverse translation, generated by $\xi = W rk$. Since we have already considered the rescaling symmetry, we will now consider pure transverse translation generated by $\xi = R k$ with the condition $W = k[R] =0$. The symmetry acts on the sCarrollian structure as,
\begin{equation}
\begin{aligned}
\delta_R  k_a &=  0    ,\cr
\delta_R  \v^a &=  - (\sD^a-\Upsilon^a)R ,\cr
\delta_R  \rho &=  - (v- \kappa )[R] ,\cr
\delta_R q_{ab} & = 2 R \bar\theta_{ab} + 2   k_{(a} (\sD-\Upsilon)_{b )}R.
\end{aligned}
\end{equation} 
Like the previous two symmetries, the boundary Lagrangian transforms anomalously under the transverse translation. The anomaly is given by  \eqref{Delta-News-1},
\begin{align}
\Delta_{R} \bm{\ell}_\H = - \bar{\theta} v[R]  \volH - \dH \left(  \sD^A R \bvol_A \right). 
\end{align}
In this case, the term $\iota_\xi \exd \bm{\ell}_\H$ is non-zero and is given by 
\begin{align}
\iota_\xi \exd \bm{\ell}_\H = R(k+\btheta)[\Wein] \volH.
\end{align}
We then need to evaluate the contraction with the pre-symplectic potential, 
\begin{equation}
\begin{aligned}
I_R \Theta^{\mathrm{can}}_{\H} 
&=  \int_{\H} \bigg[  \p^A  (\sD_A-\Upsilon_A)R   
+ R  \left({\T}^{AB} \bar\sigma_{AB}+\bar\theta(\P-\kappa) \right)  + \btheta v[R] \bigg]\volH. 
\end{aligned}
\end{equation} 
Finally, we obtain the canonical Noether charge associated to this symmetry,
\begin{equation}
\begin{aligned}
Q_R^{\mathrm{can}} = \ &  \int_{\H} R \bigg[ k [\Wein] +(\E+\P)\btheta   - (\sD_A +2\p_A+2\ac_A) \p^A +\T^{AB}\bs_{BA} \bigg]\volH\\
&  - \int_{\pa \H} (\sD^A - \p^A)R  \bvol_{ A}  \\
= \ &  -\int_{\H} R R_{k n}\volH  - \int_{\pa \H}  \left(\sD^A - \p^A\right)R  \bvol_{ A},
\end{aligned}
\end{equation} 
where we recalled the expression \eqref{R-kn} for the Ricci tensor $R_{k n}$. Note that we used the Stokes theorem \eqref{Stokes} and invoked some relations, such as $\Wein = \E+\kappa $ and $\Upsilon^A + \ac^A = 2(\p^A + \ac^A)$ in the derivation. Deeming that the canonical charge is the corner term imposes the vacuum Einstein equation $R_{k n} =0$, and the on-shell charge is
\begin{equation}
\begin{aligned}
Q_R^{\mathrm{can}} \ \EOM  - \int_{\pa \H}  \left(\sD^A - \p^A\right)R  \bvol_{ A}.
\end{aligned}
\end{equation} 
From this expression, we see that the charge vanishes if the cut is along $\sv=\mathrm{constant}$. This can be traced back to the condition $\bar\kappa=0$, since we expect $\bar\kappa$ to appear instead as the charge aspect. 

\section{Radial Evolution of Canonical Charge and Spin-2 Equation} \lb{sec:spin-2}
In the previous section, we have discussed the diffeomorphism symmetries preserving the background rigging structure and derived the corresponding conservation laws, i.e., the Einstein equations, and the associated canonical Noether charges. One notable equation is $R_{k n} =0$, associated to the transverse translation $\xi = R k$. It controls the radial evolution of the news scalar $\Wein$ (or equivalently, the time evolution of the extrinsic curvature $\btheta$). 

Some equations presented in \eqref{Einstein} are left unexplored from the symmetry perspective. Among these, the spin-2 equations, specifically the horizontal components $G_{\la AB \ra} = 0$, are particularly challenging. In principle, the equations are associated with some spin-2 symmetries generated by symmetric and trace-free horizontal tensor fields of degree two. However, since the symmetries in our setup are generated by the diffeomorphism vector field $\xi$ \eqref{diffeo}, it is to be expected that the spin-2 tensors we are seeking are constructed out of the components of $\xi$. More specifically, we will show that the spin-2 symmetries generating these components of the Einstein equations are labelled by $T \s^{AB} +\sD^{\la A}X^{B \ra}$. Interestingly, this feature is unveiled when considering the transverse (radial) evolution of the canonical charges. 

In this section, we will derive the radial evolution equations for both the time translation charge $Q^{\mathrm{can}}_T$ and the horizontal diffeomorphism charge $Q^{\mathrm{can}}_X$. We will start with the latter and present two methods of derivation: the covariant derivation and the explicit derivation. The covariant derivation is concise and easier to follow, as it derives from the 4-dimensional Bianchi identity of the spacetime $\M$. In contrast, the explicit derivation relies solely on the canonical information of the 3-dimensional stretched horizon $\H$, specifically the canonical potential \eqref{potential-can}, therefore aligning itself perfectly with the concept of holography. While the logic in the explicit derivation is straightforward, it involves lengthy technical computations. Both methods, however, lead to the same final results.

In the following sections, we will present the covariant derivation for both the horizontal diffeomorphism charge and the time translation charge. As a demonstration, the explicit derivation for the horizontal diffeomorphism sector is presented in Appendix \ref{app:spin2-derive}.

\subsection{Radial evolution of the horizontal diffeomorphism charge}

We begin with the canonical charge associated with horizon diffeomorphism, $Q^{\mathrm{can}}_X$ associated to the horizontal diffeomorphism labeled by $X^A$ such that $\Lie_k X^A = 0$ (in other words, $Z^A =0$). The expression for the charge is given by the equation \eqref{Q-X}. 

First, let us introduce the notations for the field-dependent transformation parameters derived from the components $X^A$,
\begin{flalign}
& X_\sD^{AB} := \sD^{A} X^{B}, \ \  \mr{X}^A := \Lie_\v X^A, \ \ X_\ac : = X^A \ac_A, \  \ X_\vor^A := \vor^A{}_B X^B, \  \ X_\rho := X^A \sD_A \rho &&  
\end{flalign}
As usual, the symmetric components of $X_\sD^{AB}$ decompose as $X_\sD^{(AB)} = X_\sD^{\langle AB \rangle} + \frac{1}{2} X_\sD q^{AB}$, where $X_\sD^{\langle AB \rangle}$ is the symmetric traceless components and $X_\sD = q_{AB} X_\sD^{AB}$ is its trace, and we denote with $X_\sD^{[AB]} = \sD^{[A}X^{B]}$ its antisymmetric components. The transformation of the sCarrollian structure under the horizontal diffeomorphism can be rewritten in terms of these new parameters as
\begin{align}
\delta_X k_\v = X_\ac,
\ \ \ \delta_X k_A =X_{\vor A}, 
\ \ \ \delta_X \v^A = - \mr{X}^A, 
\ \ \ \delta_X q_{AB} =  2X_{\sD (AB)},
\ \ \ \text{and} \ \ \ \delta_X \rho = X_\rho. \lb{X-spin2}
\end{align}

To evaluate the Lie derivative of $Q^{\mathrm{can}}_X$ along the radial direction, we need to start from the definition of the canonical charge \eqref{Q-can}. This involves explicitly computing
\begin{align}
\Lie_k Q^{\mathrm{can}}_X  = \Lie_k \left( I_{X} \Theta^{\mathrm{can}}_\H \right), 
\end{align}
which requires determining the radial evolution of both the conjugate canonical momenta and the variation of the sCarrollian structure. This is what we mean by an explicit derivation (see Appendix \ref{app:spin2-derive}).

On the other hand, the covariant derivation starts from the already derived form \eqref{Q-X} of the charge. This approach necessitates understanding the radial evolution of the components of the Einstein tensor, specifically the component $G_{An}$ in this case. This information can be inferred from the Bianchi identity $\nabla_b G_a{}^b = 0$. We find that the radial evolution of the (off-shell) horizontal diffeomorphism charge is given by 
\begin{equation}
\lb{Lie-k-QX}
 \begin{aligned}
 \Lie_k Q^{\mathrm{can}}_X = & \int_\H \left[ - X^{AB}_\sD G_{AB} - X_\ac G_{\v k}- \mr{X}^A G_{k A} - X_\vor^A G_{A n} + X_\rho G_{kk}  \right] \volH  \\
 & + \int_{\pa \H} \left[ X^A\big(G_{kA} + (\Lie_k +\btheta)\p_A\big)\volS + X^B\left( G_B{}^A + (\Lie_k +\btheta) \T_B{}^A + \p_B \Upsilon^A \right)\bvol_A \right].
 \end{aligned}
 \end{equation}
Observe that the bulk piece already contains $G_{AB}$, the spin-2 components of the Einstein equation. The remaining task is to write the corner term in terms of the symmetry parameters \eqref{X-spin2} using the expressions for the Einstein tensor $G_{kA}$ and $G_{AB}$, already presented in \eqref{Einstein}.

\paragraph{Null boundary:}
To see clearly how the radial evolution $\Lie_k Q^{\mathrm{can}}_X$ can be interpreted as the spin-2 charge, let us focus on the case of the null boundary $\N$ and assume the vanishing of the Carrollian vorticity (which can always be achieved by utilizing the shift symmetry). This assumption, $\vor_{AB} = 0$, corresponds to the choice of foliation (i.e., a global section of the bundle) of $\N = \bigcup_\sv \S_\sv$, where the leaves $\S_\sv$ are identified with the 2-sphere $\S$. We choose the boundary $\pa \N$ to coincide with the cut $\S_\sv$, and ensure that $\bvol_A \stackrel{\sss \pa \N}{=} 0$. This is called the \emph{Bondi slices} \cite{Freidel:2022bai}. By doing so, the above equation is greatly simplified while still retaining the essential information about the spin-2 equation and charge. We simply have  
 \begin{equation}
 \begin{aligned}
 \Lie_k Q^{\mathrm{can}}_X &\Neq  -\int_\N \left(  X^{AB}_\sD G_{AB} + X_\ac G_{\v k}+ \mr{X}^A G_{k A} \right) \volN  + \int_{\S_\sv}  X^A\big(G_{kA} + (\Lie_k +\btheta)\p_A\big)\volS \\
 & =  -\int_\N \left(  X^{AB}_\sD G_{AB} + X_\ac G_{\v k}+ \mr{X}^A G_{k A} \right) \volN  + \int_{\S_\sv}  X^A\sD_B (\btheta_A{}^B - \btheta \delta_A^B)\volS, 
 \end{aligned}
 \end{equation}
 where we substituted the expression for $G_{kA}$ on the $\N$ given in \eqref{Einstein-null}. By integration by parts and the fact that $\int_{\S_\sv}\sD_A V^A \volS = 0$ for any horizontal vector $V^A$ (see the formula \eqref{Stokes-S}), we finally obtain 
\begin{equation}
\begin{aligned}
 \Lie_k Q^{\mathrm{can}}_X = \ & -\int_\N \left(X^{\la AB \ra}_\sD G_{\la AB\ra} +  X_\sD(\tfrac12 q^{AB}G_{AB}) +X_\ac G_{\v k}+ \mr{X}^A G_{k A} \right) \volN \\
 & - \int_{\S_\sv} \left( X_\sD^{\la AB \ra} \bs_{AB} - \tfrac12 X_\sD \btheta \right)\volS.
\end{aligned}
\end{equation}
We now see that by demanding that $\Lie_k Q^{\mathrm{can}}_X$ is a corner term, we derive the spin-2 equation $G_{\la AB \ra} =0$, as well as the trace part $\tfrac12 q^{AB}G_{AB} =0$, the $G_{\v k} =0$ equation, and the $G_{k A} =0$ equation. On-shell, the charge is
\begin{equation}
\begin{aligned}
 \Lie_k Q^{\mathrm{can}}_X \ \EOM - \int_{\S_\sv} \left( X_\sD^{\la AB \ra} \bs_{AB} - \tfrac12 X_\sD \btheta \right)\volS.
\end{aligned}
\end{equation}
Notice that the symmetry parameters $X_\sD^{\la AB \ra}$ and $X_\sD$ are field-dependent and dynamical. 

For comparison, the radial evolution of the Noether charge associated with $\xi = X^A e_A$ in the Einstein-Hilbert theory is presented in Appendix \ref{app:radial-EH}.



\subsubsection{Covariant Derivation}


We now explain in details how to obtain the equation \eqref{Lie-k-QX}. First, let us recall the expression of the canonical charge associated with the horizontal diffeomorphism $\xi = X^A e_A$, where $\Lie_k X^A =0$. It is given by \eqref{Q-X}, 
\begin{align}
Q^{\mathrm{can}}_X =  -\int_\H X^A G_{A n} \volH + \int_{\pa \H} X^A \left( \p_A \volS +  \T_{A}{}^B    \bvol_{ B}  \right).
\end{align}
As we have already mentioned, $\Lie_k Q^{\mathrm{can}}_X$ involves the transverse evolution of the Einstein tensor, more specifically $(\Lie_k + \btheta)G_{An}$. Such term can be evaluated with the help of the spacetime Bianchi identity as follows:  

Starting from the Bianchi identity of the Einstein tensor, $\nabla_b G_{a}{}^b =0$, we focus on the horizontal components, that is $e_A{}^a \nabla_b G_{a}{}^b$. Expressing the metric in terms of the frame fields \eqref{4Dmetric}, we write the divergence of the Einstein tensor as
\begin{equation}
\begin{aligned}
e_A{}^a \nabla_b G_{a}{}^b &= e_A{}^a (n_b k^c + k_b \v^c + e_b{}^C e_C{}^c)\nabla_c G_{a}{}^b \\
&= e_A{}^a n_b \nabla_{k} G_{a}{}^b + e_A{}^a k_b \nabla_\v G_{a}{}^b + e_A{}^a  e_b{}^C \nabla_{e_C} G_{a}{}^b.
\end{aligned}
\end{equation}
We evaluate each term separately. We begin with the first term that, by using the Leibniz rule, can be written as 
\begin{equation}
\begin{aligned}
e_A{}^a n_b \nabla_{k} G_{a}{}^b & = k[G_{A n}] - G_{Aa}\nabla_{k} n^a - G_{na} \nabla_{k} e_A{}^a \\
& = \cL_k G_{n A}  -\p^B G_{AB} - \kappa G_{A k} - \btheta_A{}^B G_{B n} +\p_A G_{k n}.
\end{aligned}
\end{equation}
We used that $\nabla_k n^a =\kappa k^b + \p^b$ and $\nabla_k e_A{}^a =\btheta_A{}^b - \Omega_A{}^b - \p_A k^b$, see appendix \ref{App:derivative}.
Similarly, for the second term, we can evaluate it on the stretched horizon $\H$ as
 \begin{equation}
\begin{aligned}
e_A{}^a k^b \nabla_{\v} G_{ab} & = \v[G_{k A}] - G_{Aa}\nabla_{\v} k^a - G_{k a} \nabla_{\v} e_A{}^a \\
& = \big(\v + \kappa\big)[ G_{k A} ] + (\p^B + \ac^B)G_{AB} - (\ac_A + \p_A) G_{\v k}- \theta_A{}^B G_{k B} -\J_A G_{kk}. 
\end{aligned}
\end{equation} 
Where we used that $\nabla_v k^a =-(\p^b+\ac^b)$ and $\nabla_v e_A{}^a =(\p_A+\ac_A)v^b + \theta_A{}^b +\J_A k^b$.
The last term can be expressed as follows,
\begin{equation}
\begin{aligned}
e_A{}^a  e_B{}^b \nabla_{e_C} G_{ab} &= e_C[G_{AB}] - G_{Aa} \nabla_{e_C} e_B{}^a - G_{Ba} \nabla_{e_C} e_A{}^a \\
&= e_C [G_{AB}] -\Gamma^D_{CB} G_{AD} - \Gamma^D_{CA} G_{BD} + \btheta_{CB} G_{A\v}+ \btheta_{CA} G_{B\v} \\
& \ \ \ \ + (\theta_{CB}+2\rho \btheta_{CB}) G_{k A}  + (\theta_{CA}+2\rho \btheta_{CA}) G_{k B} \\
&= \sD_C G_{AB}  + \btheta_{CB} G_{A n} + \theta_{CB} G_{k A}  +\btheta_{CA} G_{B n} + \theta_{CA} G_{k B}.
\end{aligned}
\end{equation}
Where we used that 
$\nabla_C e_A{}^b = \Gamma^D_{CA} e_D{}^b
- \btheta_{CA}  \v^b  - \NN_{CA}  k^b$.
Putting these results together, the spacetime Bianchi identity $e_A{}^a \nabla^b G_{ab} = 0$ imposes the following constraint on the transverse derivative of $G_{A n}$,
\begin{equation}
\lb{Bianchi}
\boxed{
\begin{aligned}
- \left(\cL_k+ \btheta \right) G_{A n} = \left( \sD_B +\ac_B\right)G_{A}{}^B + \left(\cL_\v +\theta\right) G_{k A}  - \ac_A G_{\v k} + \vor_A{}^B G_{B \v}  +\sD_A \rho G_{kk}.
\end{aligned}
}
\end{equation}
From this result, one can easily show, with the help of Stokes theorem \eqref{Stokes}, that the radial evolution of the charge is \eqref{Lie-k-QX},
 \begin{align}
 \Lie_k Q^{\mathrm{can}}_X = \ & \int_\H \left[ - X^{AB}_\sD G_{AB} - X_\ac G_{\v k}- \mr{X}^A G_{k A} - X_\vor^A G_{A n} + X_\rho G_{kk}  \right] \volH  \\
 & + \int_{\pa \H} \left[ X^A\big(G_{kA} + (\Lie_k +\btheta)\p_A\big)\volS + X^B\left( G_B{}^A + (\Lie_k +\btheta) \T_B{}^A + \p_B \Upsilon^A \right)\bvol_A \right], \nonumber
 \end{align}
where we used the formulae for the Lie derivative of the volume forms: $\Lie_k \volS = \btheta \volS + \Upsilon^A \bm{\epsilon}_A$ and $\Lie_k (V^A \bm{\epsilon}_A) = \left(\cL_k + \btheta \right) V^A \bm{\epsilon}_A$ for a horizontal vector $V^A$ (see the proof in \eqref{Lie-k-vol}).


\subsection{Radial evolution of the time translation charge}

We have computed the radial evolution of the diffeomorphism charge, that is $\Lie_k Q^{\sss \mathrm{can}}_X$, and showed that it can be interpreted as the spin-2 charge. This feature can be traced back to the horizontal component $e_A{}^a \nabla_b G_a{}^b = 0$ of the spacetime Bianchi identity. There are, however, two remaining components of the Bianchi identity to be explored, namely the component $\v^a\nabla_b G_a{}^b = 0$ and $k^a\nabla_b G_a{}^b = 0$. The former is essential for the computation of the radial evolution of the time translation charge $Q^{\mathrm{can}}_T$.

Let us first consider the vertical components $\v^a\nabla_b G_a{}^b = 0$. It can be rewritten, using the Leibniz rule, as follows:
\begin{equation}
\begin{aligned}
&\v^a\nabla_b G_a{}^b  \\
& \ = \nabla_a G_\v{}^a - G_b{}^a \nabla_a \v^b \\
& \ =  \nabla_a \left( G_{\v k} \v^a + G_{\v n} k^a + G_\v{}^A e_A{}^a \right) - G_b{}^a \nabla_a \v^b \\
& \ =  (\Lie_\v + \theta) G_{\v k} + (\Lie_k + \btheta)G_{\v n} + (\sD_A+\ac_A)G_\v{}^A  - G_b{}^a \left( \theta_a{}^b + 2n_a  \p^b  +2 \rho k_a \ac^b  - v[\rho]k_a  k^b \right) \\
& \ =  (\Lie_\v + \theta) G_{\v k} + (\Lie_k + \btheta)G_{\v n} + (\sD_A+\ac_A)G_\v{}^A  -  \theta^{AB}G_{AB} - 2 \p^A G_{A n} -2 \rho \ac^A G_{kA} + \v[\rho] G_{kk}
\end{aligned}
\end{equation}
where to obtain the third equality, we used the divergence formulae $\nabla_a \v^a = \theta$, $\nabla_a k^a =\btheta$, and $\nabla_a (X^A e_A{}^a) = (\sD_A + \ac_A)X^A$. We also used the decomposition \eqref{nabla-l} of $\nabla_a \v^b$ and the relation $n^a = \v^a + 2\rho k^a$, together with the symmetric property of the Einstein tensor $G_{ab}$. The Bianchi identity $\v^a\nabla_b G_a{}^b = 0$ allows us to express the radial evolution of $G_{\v n}$ as
\begin{equation}
\lb{Bianchi-T}
\boxed{
\begin{aligned}
-(\Lie_k + \btheta)G_{\v n} = & (\Lie_\v + \theta) G_{\v k}  + (\sD_A+\ac_A)G_\v{}^A  -  \theta^{AB}G_{AB} - 2 \p^A G_{A n} -2 \rho \ac^A G_{kA} \\
& + \v[\rho] G_{kk}.
\end{aligned}
}
\end{equation}
The last two terms vanish on $\N$. This result and the Stokes theorem \eqref{Stokes} allow us to write the radial evolution of the constraint associated to the time translation, $\xi^a = T \v^a$, as
\begin{equation}
\lb{Bianchi-v}
\begin{aligned}
-\Lie_k ( TG_{\v n} \volH) =& \left(- v[T]G_{\v k} - \sD^AT G_{A\v} -  T\theta^{(AB)}G_{AB} - 2 T\p^A G_{A n} -2 \rho T\ac^A G_{kA}  + T\v[\rho] G_{kk} \right)\volH  \\
 & + \dH \left( TG_{\v k} \volS + TG_\v{}^A \bvol_A \right).
\end{aligned}
\end{equation}
Next, by recalling the expression \eqref{Q-T-off} for the off-shell time translation charge
\be 
Q_{T}^{\mathrm{can}} =
- \int_\H T G_{\v n} \volH - \int_{\pa \H} T\left(  \E  \volS +  \J^A  \bvol_{ A}  \right),
\ee
we can show that it evolves along the radial direction as
\begin{equation}
\begin{aligned}
\Lie_k Q_{T}^{\mathrm{can}} = &
- \int_\H \Lie_k(T G_{\v n} \volH) + \int_{\pa \H} T\left(  (\Lie_k+\btheta)\kappa  \volS -  (\Lie_k\J^A +\btheta\J^A - \kappa \Upsilon^A)  \bvol_{ A}  \right) \\
= & - \int_\H \left(v[T]G_{\v k} + \sD^AT G_{A\v} +  T\theta^{AB}G_{AB} + 2 T\p^A G_{A n} + 2 \rho T\ac^A G_{kA}  - T\v[\rho] G_{kk} \right)\volH \\
& -  \int_{\pa \H} T\left(  (\Lie_k \E+\btheta \E - G_{\v k})  \volS + (\Lie_k\J^A +\btheta\J^A + \E \Upsilon^A - G_\v{}^A)  \bvol_{ A}  \right) \\
\end{aligned}
\end{equation}
where we used again the formulae \eqref{Lie-k-vol} for the Lie derivative of the volume forms. 

\paragraph{Null boundary:} Similarly to the horizontal diffeomorphism charge, we evaluate $\Lie_k Q^{\mathrm{can}}_T$ on the null boundary $\N$ with the corner being the Bondi slices $\S_\sv$. 
By recalling the expression for $G_{\v k}$ given in \eqref{Einstein-null}, the radial evolution of the time translation charge can be written as
\begin{equation}
\begin{aligned}
\Lie_k Q_{T}^{\mathrm{can}} 
= & - \int_\N \left(v[T]G_{\v k} + (\sD^AT + 2\p^AT)  G_{A\v} +  T\s^{AB}G_{\la AB \ra} +T\theta (\tfrac{1}{2}q^{AB}G_{AB}) \right)\volN \\
& -  \int_{\S_\sv} \left( T \s^{AB} \bs_{AB} - \tfrac12 T \theta \btheta   -\p_A (\sD^A + 2 \p^A) T + T \mathscr{L}\right)\volS, \\
\end{aligned}
\end{equation}
where we also used the integration by parts \eqref{Stokes-S}. The last term,
\begin{align}
\mathscr{L} = \tfrac{1}{2}\RC  - \P\btheta  +\p_A\p^A - \T^{AB}\bs_{AB}, 
\end{align}
is precisely the bulk Lagrangian \eqref{L-bulk} evaluated on the null boundary. 
On-shell, we have that
\begin{align}
\Lie_k Q_{T}^{\mathrm{can}} \ \EOM  -  \int_{\S_\sv} \left( T \s^{AB} \bs_{AB} - \tfrac12 T \theta \btheta   -\p_A (\sD^A + 2 \p^A) T + T\mathscr{L} \right)\volS.
\end{align}

\subsection{Radial evolution of the tangential diffeomorphism charge}

Having separately derived the radial evolution of the horizontal diffeomorphism charge $Q^{\mathrm{can}}_X$ and the time translation charge $Q^{\mathrm{can}}_T$, we can combine them and write down the radial evolution of the canonical charge associated with tangential diffeomorphism, $Q^{\mathrm{can}}_{\xit}$. However, in this case, one needs to take into account the relation \eqref{radial-constraint} between $T$ and $X^A$, that is $\Lie_k X^A = -T \Upsilon^A$ for $Z^A =0$. In other words, $\Lie_k \xit =0$. 

First, we note that the combined Bianchi identity \eqref{Bianchi} and \eqref{Bianchi-v} can be expressed in terms of the variation of the sCarrollian structure under the tangential diffeomorphism $\xit$, 
\begin{equation}
\lb{Bianchi-tan}
\boxed{
\begin{aligned}
-\Lie_k \big( \xit^aG_{a n} \volH\big) =& \left(-G_{\v k} \delta_{\xit} k_\v - G_\v{}^A \delta_{\xit} k_A + G_{kA} \delta_{\xit} \v^A -\tfrac{1}{2}G^{AB}\delta_{\xit}q_{AB} +  G_{kk} \delta_{\xit}\rho \right)\volH & \\
 & + \dH \left( \xit^jG_j{}^i \bvol_i \right).
\end{aligned}
}
\end{equation}
The corner term can also be written as $\xit^bG_b{}^a \bvol_a = \xit^aG_{ak} \volS + \xit^aG_a{}^A \bvol_A$. With this result, the radial derivative of the tangential diffeomorphism charge \eqref{Q-tan} is given by 
\begin{equation}
\boxed{
\begin{aligned}
\Lie_k Q^{\mathrm{can}}_{\xit} = \ & \int_\H \bigg[-G_{\v k} \delta_{\xit} k_\v - G_\v{}^A \delta_{\xit} k_A + G_{kA} \delta_{\xit} \v^A -\tfrac{1}{2}G^{AB}\delta_{\xit}q_{AB} +  G_{kk} \delta_{\xit}\rho \bigg]\volH \\
& + \int_{\pa \H} \xit^j\bigg[ G_j{}^i + \left(\Lie_k + \btheta\right)\EM_j{}^i \bigg]\bvol_i. 
\end{aligned}
}
\end{equation}

We thus need to write the corner piece in terms of variations of the sCarrollian structure, in the same manner as the bulk piece. 

\paragraph{Null boundary:} As we have shown for both canonical charges $Q^{\mathrm{can}}_X$ and $Q^{\mathrm{can}}_T$, the situation simplifies when considering the null boundary $\N$. Again, by using the Einstein tensor $G_{\v k}$ and $G_{kA}$ given in \eqref{Einstein-null} and integration by parts \eqref{Stokes-S}, the radial evolution of the canonical charge associated to $\xit$ evaluated on the $\N$ can be written as
\begin{equation}
\begin{aligned}
 \Lie_k Q^{\mathrm{can}}_{\xit} = \ & -\int_\N \left(\mathbb{S}^{\la AB \ra} G_{\la AB\ra} +  \mathbb{S}(\tfrac12 q^{AB}G_{AB}) +\mathbb{R} G_{\v k}+  \mathbb{X}^A G_{\v A}+ \mathbb{Y}^A G_{k A} \right) \volN \\
 & - \int_{\S_\sv} \left( \mathbb{S}^{\la AB \ra} \bs_{AB} - \tfrac12 \mathbb{S}\btheta - \mathbb{X}^A \p_A + T\mathscr{L} \right)\volS,
\end{aligned}
\end{equation}
where we defined the field-dependent and dynamical transformation parameters as follows:
\begin{alignat}{3}
&\mathbb{S}_{(AB)} &&= T \theta_{(AB)} + \sD_{(A} X_{B)} &&=  \tfrac12 \delta_{\xit}q_{AB} \\
- &\mathbb{X}^A &&= (\sD^A-\ac^A)T &&= \delta_{\xit} k_A \\
- &\mathbb{Y}^A &&= -\Lie_\v X^A &&=  \delta_{\xit} \v^A \\
&\mathbb{R} &&= v[T] + X^A\ac_A &&= \delta_{\xit}k_\v
\end{alignat}
It turns out these new symmetry parameters are just the variation of the sCarrollian structure. These mean. This means we can write the radial evolution as the contraction, 
\begin{align}
\Lie_k Q^{\mathrm{can}}_{\xit} &= I_{\xit} \widehat{\Theta}_\N + \left( I_{\xit} \widehat{\Theta}_{\pa \N} - \int_{\pa\N} \iota_{\xit}\bm{\mathscr{L}} \right),
\end{align}
where $\bm{\mathscr{L}} := \mathscr{L} \volH$ and we defined
\begin{equation}
\begin{alignedat}{2}
&\widehat{\Theta}_\N &&:=  -\int_\N \left(  G_{\v k} \delta k_v -   G_\v{}^ A\delta k_A - G_{k A}\delta \v^A + \tfrac12 G^{AB}\delta q_{AB}\right) \volN \\
&\widehat{\Theta}_{\pa\N} &&:= - \int_{\S_\sv} \left( \tfrac12\btheta^{AB} \delta q_{AB} + \p^A \delta k_A \right)\volS.
\end{alignedat}
\end{equation}
The bulk term is zero on-shell, and we finally obtain
\begin{align}
\Lie_k Q^{\mathrm{can}}_{\xit} \ \EOM \  I_{\xit} \widehat{\Theta}_{\pa \N} - \int_{\pa\N} \iota_{\xit}\bm{\mathscr{L}}. 
\end{align}

\paragraph{The transverse component:}
For completeness, the last component of the Bianchi identity is the transverse component $k^a\nabla_b G_a{}^b = 0$. In the same manner, we can use the Leibniz rule and the decomposition \eqref{nabla-k} of $\nabla_a k^b$ to write $k^a\nabla_b G_a{}^b$ as
\begin{equation}
\begin{aligned}
&k^a\nabla_b G_a{}^b  \\
& \ = \nabla_a G_k{}^a - G_b{}^a \nabla_a k^b \\
& \ =  \nabla_a \left( G_{k k} \v^a + G_{k n} k^a + G_k{}^A e_A{}^a \right) - G_b{}^a \nabla_a k^b \\
& \ =  (\Lie_\v + \theta) G_{k k} + (\Lie_k + \btheta)G_{k n} + (\sD_A+\ac_A)G_k{}^A  - G_b{}^a \left( \btheta_a{}^b -\kappa k_ak^b - k_a (2\p^b + \ac^b) \right) \\
& \ = (\Lie_\v + \theta) G_{k k} + (\Lie_k + \btheta)G_{k n} + (\sD_A+\ac_A)G_k{}^A  - \btheta^{(AB)}G_{AB} + \kappa G_{kk} +  (2\p^A + \ac^A) G_{kA}. 
\end{aligned}
\end{equation}
After imposing the identity $k^a\nabla_b G_a{}^b = 0$, the radial evolution of the component $G_{kn}$ of the Einstein tensor is given by
\begin{equation}
\begin{aligned}
- (\Lie_k + \btheta)G_{k n} = (\Lie_\v + \theta) G_{k k}  + (\sD_A+\ac_A)G_k{}^A  - \btheta^{(AB)}G_{AB} + \kappa G_{kk} +  (2\p^A + \ac^A) G_{kA}. 
\end{aligned}
\end{equation}


\section{Conclusions}

We have completed the framework developed in \cite{Freidel:2022vjq}, providing a comprehensive toolbox for studying the geometry, dynamics, phase space, and symmetries of Carrollian stretched horizons. In our definition, a Carrollian stretched horizon is a causal manifold endowed with a sCarrollian structure, which generalizes the Carrollian structure conventionally defined only on null surfaces \cite{Duval:2014uoa, Ciambelli:2019lap}. The key extension from the Carrollian structure is the introduction of the \emph{stretching}, $\rho := -\tfrac12 h_{ij} \v^i \v^j$, which renders $h_{ij}$ non-degenerate in the vertical direction when it does not vanish. The (null) Carroll case is obtained when the stretching vanishes, making $\rho$ a crucial part of the phase space of sCarrollian stretched horizons. We have defined the notions of sCarrollian connection and sCarrollian stress tensor, demonstrating how these structures can be derived from a null rigging structure when treating the stretched horizon as a hypersurface embedded in a higher-dimensional spacetime. This approach unifies the intrinsic (sCarrollian) and embedding (stretched horizon) perspectives, providing a universal framework for general causal surfaces, whether timelike or null. In a sense, we have integrated sCarrollian aspects into the membrane paradigm viewpoint.

We expressed the Einstein equations in sCarrollian variables, including all components, and addressed the dual interpretation of some equations, either as time evolution or radial evolution. The canonical phase space of the sCarrollian surface was also discussed. Through Noether's theorem (using the modern covariant phase space formalism), we computed the Noether charge associated with diffeomorphism preserving the background rigging structure. For horizontal diffeomorphism $\xit = T\v + X^A e_A$ and transverse translations $\xi_{\perp} = R k$, their canonical charges include bulk terms that are components of the Einstein equation. Starting from the sCarrollian canonical pre-symplectic potential, which is derivable from but not directly connected to the Einstein-Hilbert theory in spacetime, this result demonstrates that the Einstein equation emerges holographically from the sCarrollian data on the surface. The symmetries of the sCarrollian structure preserving the metric $h_{ij}$ include rescaling $W$ and shift symmetry $Z_A$; the former is an edge mode symmetry, while the latter is gauge. Furthermore, we computed the evolution of the canonical charge along the transverse (radial) direction, which can be interpreted as the charge for the spin-2 sector of the Einstein equation, specifically the component $G_{\la AB \ra} = 0$ with the corresponding charge aspect being the connection's shear tensor $\bs_{AB}$.

The tools and concepts presented in this work not only enhance our understanding of the geometric and dynamic properties of stretched horizons but also offer a robust analytical framework for future research in this area. We expect this comprehensive toolbox to serve as an indispensable resource for studying a wide range of physical scenarios involving causal surfaces. 

Several directions remain to be explored:

$i)$ \emph{Hydrodynamics and Thermodynamics}:  In the context of the membrane paradigm \cite{Damour:1978cg, thorne1986black, Price:1986yy}, the correspondence between gravity and fluid dynamics is typically established through the horizon's stress tensor and its conservation laws, specifically the Raychaudhuri equation $G_{\v n} = 0$ and the Damour equation $G_{A n} = 0$. This correspondence also holds in our sCarrollian framework. However, on the gravity side, there are two additional quantities and their associated evolution equations whose connections with hydrodynamics have not been fully clarified. These are the sCarrollian connection's expansion $\btheta$ and shear tensor $\bs_{AB}$, governed by the equations $G_{kn} = 0$ and $G_{\la AB \ra} = 0$, respectively. The shear tensor, $\bs_{AB}$, corresponds to the gravitational shear at asymptotic null infinity \cite{Freidel:2024tpl}, encoding information about radiation at infinity when the ruling is chosen to be hypersurface orthogonal. Understanding the properties of both $\btheta$ and $\bs_{AB}$ from a fluid dynamics perspective may provide valuable insights into the behavior of gravitational systems that emit radiation. Additionally, viewing these phenomena through the lens of fluid dynamics may enhance our understanding of the thermodynamics of stretched horizons. A good starting point would be to examine isolated horizons \cite{Ashtekar:2000hw, Ashtekar:2001jb, Lewandowski:2006mx, Ashtekar:2021wld, Ashtekar:2024bpi, Ashtekar:2024mme}, which resemble thermodynamic equilibrium systems.

$ii)$ \emph{Infinity as a stretched horizon}: It would be interesting to develop more the correspondence between the description of asymptotic infinity as a stretched horizon and the hydrodynamical picture presented here. The fact that the same structure appears in both approach is very promising. One expects, for instance, that the stretched horizon framework would be useful to describe null and spacelike infinity in the same token.
On that front, it would be interesting to understand if our structure  connects with the works by Ashtekar and Khera \cite{Ashtekar:2023wfn, Ashtekar:2023zul}, which introduced a unified framework to describe null and spatial infinity.

$iii)$ \emph{Revisiting fluid/gravity correspondence}: The fluid/gravity correspondence, which originated in the context of AdS/CFT holography \cite{Bhattacharyya:2007vjd, Son:2007vk, Rangamani:2009xk, Hubeny:2010wp, Hubeny:2011hd}, relies on the timelike boundary of asymptotically AdS spacetimes, where the holographic fluids reside. Typically, reflective boundary conditions are imposed at this boundary, preventing the notion of gravitational radiation from leaking out. However, our sCarrollian construction works for a timelike surface and permits the notion of radiation. It would certainly be interesting to explore how our framework fits into the fluid/gravity correspondence. Attempts to incorporate radiation degrees of freedom into asymptotically AdS spacetimes, as well as their corresponding flat limit, have already been pursued in the context of fluid/gravity correspondence \cite{Ciambelli:2018wre,Campoleoni:2023fug}, and more notably in the context of $\Lambda$-BMS construction in a series of works by Compère, Fiorucci, and Ruzziconi \cite{Compere:2019bua,Compere:2020lrt,Fiorucci:2020xto}, and more recently in \cite{Ciambelli:2024kre}. Additionally, a hydrodynamical description of horizons in AdS has been attempted in \cite{Knysh:2024asf}.

$iv)$ \emph{Spin-2 equations and charges}: We have also demonstrated how the spin-2 equation and charge arise when considering the radial evolution of the tangential diffeomorphism charge, which consists of spin-0 and spin-1 charges. Additionally, we show that this mechanism can be traced back to the Bianchi identity $\nabla_b G_a{}^b = 0$, which represents the conservation laws for the Einstein tensor. This feature warrants detailed exploration. Interestingly, we speculate that, as the Bianchi identity relates the time derivative, the radial derivative, and the horizontal derivative of different components of the Einstein tensor (see \eqref{Bianchi} and \eqref{Bianchi-T}), this behavior should hold when considering the time evolution of the Noether charge defined on the spacelike surface. At infinity, the recurrence relations between charges of different spins and their implications for celestial holography have been studied extensively in \cite{Freidel:2021qpz, Freidel:2021dfs, Freidel:2021ytz}. Our results may help shed light on the definition of higher-spin charges at finite distances.

$v)$ \emph{Quantization}: Quantizing subsystems of gravity enclosed by null boundaries requires knowledge about their dynamics, symmetries, and charges. This has been previously explored in \cite{Reisenberger:2007pq, Reisenberger:2007ku, Reisenberger:2012zq, Fuchs:2017jyk, Reisenberger:2018xkn}, and more recently in \cite{Ciambelli:2023mir, Wieland:2024dop, Wieland:2024kzw}. We hope that our framework can provide new insights into this endeavor and serve as a valuable toolbox.

\section*{Acknowledgements}
We would like to give a special thanks to Luca Ciambelli, with whom we have shared many discussions about Carrollian geometry and shift symmetry. We would also like to thank Rob G. Leigh and Luca Ciambelli for their collaboration on a related subject, and Simone Speziale, Mait\'a Micol and Nicolas Cresto for interesting discussions. PJ would like to acknowledge the support from the Perimeter Institute visitor program, which helped in completing this work. Research at Perimeter Institute is supported by the Government of Canada through the Department of Innovation, Science and Economic Development and by the Province of Ontario through the Ministry of Colleges and Universities. This work was supported by the Simons Collaboration on Celestial Holography.

\appendix
\section{sCarrollian Geometry and Connection}

\subsection{sCarrollian Connection}\label{App:SCarrolcon}

In this Appendix, we show the detailed derivations of formulae relating to the sCarrollian connection presented in the main text. We start by establishing \eqref{Kva}.
On one hand, we have 
\bee
(D_i h_{jk}) \v^j \v^k &= D_i(-2\rho) - h_{jk} D_i(\v^j \v^k)\cr
&= -2( D_i\rho - 2\rho k_j D_i \v^j)\cr
&= -2 (D_i \rho - 2 \omega_i \rho),
\eee
where we used $h_{ij}\v^j = -2\rho k_i$ and the definition $\omega_i := k_j D_i \v^j$.
On the other hand, we have that 
\bee
(D_i h_{jk}) v^jv^k 
&= -2 \mK_{vi}.
\eee
Hence, the equation \eqref{Kva} follows.

Next, we establish the identity between the curvature and the generalized news tensor given by \eqref{KN}.
We first evaluate
\bea
D_i(2\rho k_j) &=& 2D_i\rho k_j + 2\rho D_i k_j \cr
&=& 2D_i\rho k_j + 2\rho (D_i k_k) v^k k_j 
+ 2\rho (D_i k_k)q^k{}_j\cr
&=& (2D_i\rho -2 \omega_i \rho) k_j 
+ 2\rho (D_i k_k)q^k{}_j, 
\label{Drk}
\eea
where we used the horizontal decomposition, $\delta_i^j = k_i \v^j + q_i{}^j$, and the Leibniz rule to evaluate $(D_i k_k) \v^k = - k_k D_i \v^k = -\omega_i$. Then we also compute
\bee
(D_iv^k) h_{kj} &=
D_i(-2\rho k_j) - v^k D_ih_{kj}\cr
&= - D_i(2\rho k_j) + \mK_{iv} k_j + \mK_{ij}\cr
&= - D_i(2\rho k_j) + (D_i \rho-2 \omega_i\rho) k_j + \mK_{ij}.\label{Dvh}
\eee
Combining \eqref{Drk} and \eqref{Dvh}, we obtain,
\begin{align}
 \Wein_i{}^kh_{kj} &=
 (D_iv^k) h_{kj} + 2\rho (D_i k_k)q^k{}_j \cr
 &=  D_i(2\rho k_j) -(2D_i\rho -2 \omega_i \rho) k_j - D_i(2\rho k_j) + (D_i \rho-2\omega_i\rho) k_j + \mK_{ij} \cr
 &= -D_i\rho k_j+ \mK_{ij},
\end{align}
which proves \eqref{KN}.

We now evaluate the different components of the connection. 
For instance, from \eqref{Liev}, one gets that 
\be 
D_v k_i= \cL_v k_i -  k_j D_i v^j
= -\varphi_i - \omega_i.
\ee
Furthermore, contracting \eqref{Dvh} with $v^i$, we get 
\bea
D_v v^j h_{ji}
&=& -D_v(2\rho k_i) - v^j D_v h_{ji} \cr
&= & - D_v(2\rho k_i) + (D_v \rho-2\omega_v\rho) k_i + \mK_{vi}\cr
&= & - (D_v \rho+2\kappa \rho) k_i 
-2\rho  D_vk_i +  D_i \rho -2\omega_i\rho \cr
&= & - (D_v \rho+2\kappa\rho) k_i 
+2\rho \varphi_i +  D_i \rho \cr
&=& - 2\kappa \rho k_i + q_i{}^j (D_i \rho + 2 \ac_i \rho),
\eea 
where we recalled $\omega_i = \kappa k_i + \p_i$, which yielded $\omega_\v = \kappa$. 
After projection, this gives 
\be 
\EuScript{A}_i = (D_v v^j) q_{ji} = q_i{}^j (D_j \rho + 2\varphi_j \rho). 
\ee 

\subsection{Jacobi identity} \lb{App:Jacobi}
The Jacobi identities of the commutators \eqref{com-4}, also corresponding to the nilpotent property of the exterior derivative, namely $\exd^2 \bm{k} =0$ and $\exd^2 \bm{e}^A =0$, impose the following conditions 
\begin{subequations}
\lb{Jacobi}
\begin{align}
\v[\vor_{AB}] - \left( \sD_A\ac_B - \sD_B\ac_A \right) &= 0, \lb{Jacobi-1}  \\
\v[\Omega_A{}^B] + \left(e_A + \ac_A \right)[\Upsilon^B] & =0, \lb{Jacobi-2} \\
k[\ac_A]  + \Omega_A{}^B \ac_B + \vor_{AB} \Upsilon^B & =0, \lb{Jacobi-3} \\
k[\vor_{AB}] + \Omega_A{}^C \vor_{CB} + \Omega_B{}^C \vor_{AC} &= 0, \lb{Jacobi-4} \\
e_A[\Omega_B{}^C] - e_B[\Omega_A{}^C]  + \vor_{AB} \Upsilon^C & =0, \lb{Jacobi-5} \\
\left(\sD_A - \ac_A\right)\vor_{BC}+ (\text{cyclic permutation on} \ A, B, C)& =0. \lb{Jacobi-6}
\end{align}
\end{subequations}
It is useful to note that the equations \eqref{Jacobi-1}, \eqref{Jacobi-2}, and \eqref{Jacobi-3} can be written, respectively, as the Lie derivative as $\Lie_\v \vor_{AB} = 2 \sD_{[A} \ac_{B]}$, $\Lie_k \ac_A = -\vor_{AB} \Upsilon^B$, and $\Lie_k \vor_{AB} = 0$. 

\subsection{Covariant derivatives} \lb{App:derivative}
Most of our computations rely on the decomposition of the spacetime covariant derivatives $( \nabla_a \v^b, \nabla_a k^b, \nabla_a e_A{}^b)$ in terms of Carrollian objects. We present their expressions here for the reader's convenience,
\begin{subequations}
\begin{align}
\nabla_a \v^b = \ & \theta_a{}^b + (\p_a + \kappa k_a) \v^b - k_a\left(\J^b-2 (\p^b+\ac^b)\rho \right)  + 
\left( \J_a   -k_a (\v-2\kappa)[\rho] \right)k^b 
\nonumber \\
& + n_a\left(  \p^b -\kappa  k^b \right), \lb{nabla-l}\\
\nabla_a k^b = \ & \btheta_a{}^b -(\p_a + \kappa k_a)k^b - k_a (\p^b + \ac^b),  \lb{nabla-k}\\ 
\nabla_a e_A{}^b = \ & \Gamma^b_{aA} +(
- \btheta_{aA}  + (\p_A + \ac_A) k_a )\v^b + \theta_A{}^b k_a - \left(\NN_{aA} - \J_A k_a \right) k^b \nonumber \\
& +n_a \left( \btheta_A{}^b - \Omega_A{}^b - \p_A k^b\right), \lb{nabla-eA}
\end{align} 
\end{subequations}
where the Christoffel-Carroll symbol is $\Gamma^A_{BC} = q^{AD} g(e_D, \nabla_{e_B} e_C)$ and the Carrollian heat current \eqref{heat-gen} is given by $\J_a = -(\sD_a - 2\p_a)\rho$. Furthermore, as the normal vector is given by the combination $n^a = \v^a + 2\rho k^a$, one can straightforwardly derive the decomposition of the covariant derivative of $n^a$. Using that 
$ \nabla_a n^b = \nabla_a \v^b + 2\rho \nabla_a k^b + 2\nabla_a \rho k^b$, we get
\begin{align}
\nabla_a n^b = \ & \NN_a{}^b+ 
(\p_a + \kappa k_a) \v^b 
- k_a \J^b  + (\v[\rho]k_a +\sD_a \rho) k^b+n_a\left( \kappa k^b + \p^b \right). \lb{nabla-n}
\end{align} 
We note again that $\NN_{ab} = \T_{ab} + \frac{1}{2} \E q_{ab}$. The spacetime divergences are given by  
\begin{align}
\nabla_a \v^a = \theta, \qquad \nabla_a k^a = \btheta,  \qquad \text{and} \qquad \nabla_a e_A{}^a  = \Gamma^B_{BA}  + \ac_A. 
\end{align}
Projecting the previous relation, we get that 
\be 
\nabla_k n^b =    \left( \kappa k^b + \p^b \right)\qquad  \text{and} \qquad
\Pi_a{}^c\nabla_a k^b \Pi_b{}^c = \mathsf{\bar{K}}_a{}^b = 
\btheta_a{}^b  - k_a (\p^b + \ac^b).
\ee

\subsection{Horizontal covariant derivative}
To discuss the horizontal covariant derivative $\sD_A$ defined in \eqref{sD-def}, let us look more closely at the horizontal basis. First, the covariant derivative of a horizontal basis vector along another horizontal basis vector can be computed straightforwardly from \eqref{nabla-eA} and is given by 
\begin{align}
\nabla_{e_A} e_B{}^a = \Gamma^a_{AB} - \btheta_{AB} \v^a - \NN_{AB} k^a. \lb{nabla-vec-e}
\end{align}
Similarly for the horizontal basis covector, we have 
\begin{align}
\nabla_{e_B} e_a{}^A = -\Gamma^A_{Ba} - \btheta_B{}^A n_a - \theta_B{}^A k_a. \lb{nabla-form-e}
\end{align}
For a generic horizontal vector fields $X^a := X^A e_A{}^a$, we show using \eqref{nabla-vec-e} that
\begin{equation}
\lb{nabla-e-X}
\begin{aligned}
\nabla_{e_A} X^a & = e_A[ X^B] e_B{}^a + X^B \nabla_{e_A} e_B{}^a \\
& =  e_A[X^B] e_B{}^a + \Gamma^a_{AB} X^B - X^B\btheta_{AB} \v^a - X^B\NN_{AB} k^a\\
& =  (\sD_A X^B)e_B{}^a - X^B\btheta_{AB} \v^a - X^B\NN_{AB} k^a.
\end{aligned}
\end{equation}
Once projected onto the horizontal subspace, we obtain
\begin{align}
(\nabla_{e_A} X^a) e_a{}^B & = e_A[ X^B] + \Gamma^B_{AC}X^C = \sD_A X^B. \lb{sD-X} 
\end{align}
Furthermore, the spacetime divergence of the horizontal vector is 
\begin{equation}
\begin{aligned}
\nabla_a \left( X^A e_A{}^a \right) &= e_A[ X^A] + X^A \nabla_a e_A{}^a = \left(\sD_A +\ac_A \right)X^A. \lb{divergence}
\end{aligned}
\end{equation} 

\subsubsection{Radial derivative of $\sD_B X^A$}
We are interested in expanding the horizontal covariant derivative $\sD_B X^A$ around a stretched horizon $\H$. This can be obtained by means of the derivative along the transverse direction $k = \pa_r$ as follows,   
\begin{equation}
\begin{aligned}
k \left[\sD_B X^A\right] &= k\left[e_B [X^A] + \Gamma_{BC}^A X^C \right] \\
&=  \sD_B \big( k[X^A] \big) - \Omega_B{}^C e_C[X^A] + k\left[ \Gamma_{BC}^A\right] X^C
\end{aligned}
\end{equation}
where we used the commutation relation \eqref{com-4}, $[e_A, k] = \Omega_A{}^B e_B$. The computation then lies in the tedious, but straightforward, evaluation of the radial derivative of the Christoffel-Carroll symbols, $k\left[ \Gamma_{BC}^A\right]$. Recalling the definition of the Christoffel-Carroll symbols in terms of the sphere metric, the expression for the extrinsic curvature \eqref{theta-bar}, and the commutation relation \eqref{com-4}, we show that 
\begin{equation}
\begin{aligned}
k\left[ \Gamma_{BC}^A\right] &= \tfrac{1}{2}k[q^{AD}]\left( e_B[q_{DC}]+e_C[q_{BD}] - e_D[q_{BC}] \right) + \tfrac{1}{2}q^{AD} k\left[ e_B[q_{DC}]+e_C[q_{BD}] - e_D[q_{BC}] \right] \\
& = - q^{AE}k[q_{ED}] \Gamma^D_{BC} +  \tfrac{1}{2}q^{AD} \left( e_B[k[q_{DC}]]+e_C[k[q_{BD}]] - e_D[k[q_{BC}]]\right) \\
& \ \ \ \  -  \tfrac{1}{2}q^{AD} \left( \Omega_B{}^E e_E[q_{DC}]+\Omega_C{}^Ee_E[q_{BD}] - \Omega_D{}^E e_E[q_{BC}] \right) \\
& =q^{AD} \left( \sD_B\btheta_{(CD)}+\sD_C\btheta_{(BD)}  - \sD_D\btheta_{(BC)}\right)  \\
& \ \ \ \ -\tfrac{1}{2}q^{AD} \left( 2\sD_{[B}\Omega_{D]C}+2\sD_{[C}\Omega_{D]B} -2 \sD_{[B}\Omega_{C]D} + 2 \sD_B \Omega_{CD}\right) \\
& \ \ \ \  -  \tfrac{1}{2}q^{AD} \left( \Omega_B{}^E e_E[q_{DC}]+\Omega_C{}^Ee_E[q_{BD}] - \Omega_D{}^E e_E[q_{BC}] \right). 
\end{aligned}
\end{equation}
The antisymmetric derivative terms can be evaluated with the help of the Jacobi identity \eqref{Jacobi-5}, and one can easily verify that 
\begin{align}
2 \sD_{[B} \Omega_{C]}{}^A = \vor_{CB} \Upsilon^A - 2\Omega_{[B}{}^D \Gamma_{C]D}^A. 
\end{align}
The last term in the above expression for $k\left[ \Gamma_{BC}^A\right]$ can be written in terms of $\Gamma_{BC}^A$ using the metric compatibility condition, $\sD_Aq_{BC} =0$, as
\begin{equation}
\begin{aligned}
&\Omega_B{}^E e_E[q_{DC}]+\Omega_C{}^Ee_E[q_{BD}] - \Omega_D{}^E e_E[q_{BC}]   \\
& \qquad = \Omega_B{}^E \left(\Gamma^F_{ED} q_{FC} + \Gamma^F_{EC} q_{FD} \right) - \Omega_D{}^E \left( \Gamma^F_{EB} q_{FC} \right) + (B \leftrightarrow C).
\end{aligned}
\end{equation}
Putting everything together, we obtain
\begin{equation}
\begin{aligned}
k\left[ \Gamma_{BC}^A\right] & =q^{AD} \left( \sD_B\btheta_{(CD)}+\sD_C\btheta_{(BD)}  - \sD_D\btheta_{(BC)}\right) - \sD_B \Omega_C{}^A \\
& \ \ \ \ - \tfrac{1}{2}\left( w^A{}_B \Upsilon_C + w^A{}_C \Upsilon_B + \vor_{BC} \Upsilon^A \right) - \Omega_B{}^D \Gamma^A_{CD}.
\end{aligned}
\end{equation}
With this result combined, together with the trace-traceless split \eqref{sigma-bar}, we finally arrive at the expression 
\begin{equation}
\begin{aligned}
k \left[\sD_B X^A\right] & = \sD_B \big( k[X^A] \big) - \Omega_B{}^C \sD_CX^A - X^C \sD_B \Omega_C{}^A - \tfrac{1}{2}\left( w^A{}_B \Upsilon_C + w^A{}_C \Upsilon_B + \vor_{BC} \Upsilon^A \right)X^C \\
& \ \ \ \ +X^C \left( \sD_B \btheta_{(C}{}^{A)} + \sD_C \btheta_{(B}{}^{A)} - \sD^A \btheta_{(BC)} \right), 
\end{aligned}
\end{equation}
or in terms of Lie derivative, 
\begin{equation}
\begin{aligned}
\Lie_k \left(\sD_B X^A\right) & = \sD_B \big( \Lie_k X^A \big) - \tfrac{1}{2}\left( w^A{}_B \Upsilon_C + w^A{}_C \Upsilon_B + \vor_{BC} \Upsilon^A \right)X^C \\
& \ \ \ \ + \left( \sD_B \btheta_{(C}{}^{A)} + \sD_C \btheta_{(B}{}^{A)} - \sD^A \btheta_{(BC)} \right)X^C.
\end{aligned}
\end{equation}
Its trace is given by 
\begin{equation}
\begin{aligned}
k \left[\sD_A X^A\right] & = \sD_A \left( k[X^A] - X^B \Omega_B{}^A \right) -  \vor_{AB} \Upsilon^A X^B +X^A\sD_A \btheta \\
\Lie_k \left(\sD_A X^A\right) &= \sD_A \left( \Lie_k X^A \right) -  \vor_{AB} \Upsilon^A X^B +X^A\sD_A \btheta.
\end{aligned}
\end{equation}
Lastly, let us evaluate the radial derivative of $2\sD_{(A} X_{B)} = q_{AC} \sD_B X^C +q_{BC}\sD_A X^C$, which will become useful later. Using the above result, one can verify that 
\begin{equation}
\lb{k-DX}
\begin{aligned}
k\left[ 2\sD_{(A} X_{B)} \right] & = k[q_{AC}] \sD_B X^C +q_{AC} k \left[\sD_B X^C\right] + (A \leftrightarrow B) \\
& = \sD_A \big(q_{BC} k[X^C] - X^C \Omega_{CB}\big) + \vor_{CA} \Upsilon_B X^C + X^C \sD_C \btheta_{(AB)} + 2\btheta_{(AC)} \sD_B X^C \\
& \ \ \ \ - \Omega_A{}^C\left(\sD_C X_B + \sD_{B} X_C\right) + (A \leftrightarrow B).
\end{aligned}
\end{equation}
Alternatively, it can be written using the Lie derivative as
\begin{equation}
\lb{Lie-k-DX}
\begin{aligned}
\Lie_k\left( 2\sD_{(A} X_{B)} \right)
& = \sD_A \big(q_{BC} \Lie_kX^C \big) + \vor_{CA} \Upsilon_B X^C + X^C \sD_C \btheta_{(AB)} + 2\btheta_{(AC)} \sD_B X^C + (A \leftrightarrow B).
\end{aligned}
\end{equation}

\subsubsection{Vertical derivative of $\sD_B X^A$}

The expression for the vertical derivative, $\v[\sD_B X^A]$, will also become handy later on. Using the commutation relations \eqref{com-4}, one can show that
\begin{equation}
\lb{ell-D-X}
\begin{aligned}
\v \left[\sD_B X^A\right] &= \v\left[e_B [X^A] + \Gamma_{BC}^A X^C \right] \\
&=  (\sD_B + \ac_B) \big( \v[X^A] \big) + \v\left[ \Gamma_{BC}^A\right] X^C.
\end{aligned}
\end{equation}
The vertical derivative of the Christoffel-Carroll symbols \eqref{Gamma} can be straightforwardly computed by recalling the tangential expansion, $\theta_{(AB)} = \frac{1}{2} \v[q_{AB}]$, and the commutators \eqref{com-4} as
\begin{equation}
\lb{ell-Gamma}
\begin{aligned}
\v\left[ \Gamma_{BC}^A\right] = \left(\sD_B + \ac_B\right) \theta^{(A}{}_{C)} + \left(\sD_C + \ac_C\right) \theta^{(A}{}_{B)} - \left(\sD^A + \ac^A\right)\theta_{(BC)}.
\end{aligned}
\end{equation}

\subsection{Lie derivative} \lb{App:Lie}

The Lie derivative along the vertical vector $\v$, the transverse null vector $k$, and a horizontal vector $X = X^Ae_A$ of the Carrollian metric $q_{ab}$ and its inverse $q^{ab}$ will be useful in our derivations. We list them here:
\begin{subequations}
\begin{align}
\Lie_\v q_{ab} &= 2\theta_{(ab)} + 2 \Upsilon_{(a} n_{b)} \\
\Lie_\v q^{ab} &= - 2\theta^{(ab)} + 2 \ac^{(a} \v^{b)} \\
\Lie_k q_{ab} &= 2\btheta_{(ab)} -2 \Upsilon_{(a} k_{b)} \\
\Lie_k q^{ab}  &= -2\btheta^{(ab)} \\
\Lie_X q_{ab} &= 2\sD_{(a} X_{b)} + 2\left(\Lie_\v X^c\right)q_{c(a} k_{b)} + 2\left(\Lie_k X^c\right)q_{c(a} n_{b)}   \\ 
\Lie_X q^{ab} &=  -2\sD^{(a} X^{b)} + 2X^c \vor_c{}^{(a} \v^{b)} 
\end{align}
\end{subequations}
where we recalled the notation \eqref{hor-Lie-def} and used $2\sD_{(a} X_{b)} = 2e_a{}^A e_b{}^B \sD_{(A} X_{B)} = e_a{}^A e_b{}^B \Lie_X q_{AB}$. These  equations are  obtained using that 
\begin{equation}
\begin{alignedat}{8}
&\cL_v v^a &&=0, \qquad &&\cL_v k^a &&=-\Upsilon^a, \qquad &&\cL_v n_a &&=0, \qquad &&\cL_v k_a &&=-\varphi_a, \\
&\cL_k v^a &&=\Upsilon_a, \qquad &&\cL_k k^a &&=0,\qquad
&&\cL_k n_a &&=0, \qquad &&\cL_k k_a &&=0.
\end{alignedat}
\end{equation}
From these results and the identities $2\theta_{(ab)}= q_a{}^cq_b{}^d \cL_vg_{ab} $,
 $2\bar\theta_{(ab)}= q_a{}^cq_b{}^d \cL_kg_{ab}$, we obtain for instance that 
 \be 
 (\cL_v q_{ab}) v^b=0, \qquad \text{and} \qquad (\cL_v q_{ab})k^b=- q_{ab} \cL_v k^b = \Upsilon_a.
 \ee 
 We proceed similarly for the other equalities.

Lastly, the Lie derivative along the transverse vector $k$ of the volume forms can be derived as follows:
\begin{subequations}
\lb{Lie-k-vol}
\begin{alignat}{4}
&\Lie_k \volH &&= - \Lie_k (\iota_k \volM) &&= -\iota_k \Lie_k \volM &&= \btheta \volH \\
&\Lie_k \volS && = \Lie_k (\iota_\v \volH) &&= \iota_{[k,\v]} \volH +\iota_\v \Lie_k \volH  &&= \btheta \volS + \Upsilon^A \bvol_A \\
&\Lie_k (X^A\bvol_A) && = \Lie_k (X^A\iota_{e_A} \volH) &&= k[X^A]\bvol_A+\iota_{[k,e_A]} \volH +\iota_X \Lie_k \volH  &&=  \left(\Lie_k +\btheta \right)X^A\bvol_A.
\end{alignat}
\end{subequations}
 
\subsection{Radial versus temporal derivative} \lb{App:RadT}
In this section, we derive the identities between the radial and temporal Lie derivatives \eqref{eq:RadT}. To begin with, the equations \eqref{eq:RadT-1} and \eqref{eq:RadT-3} can be seen as descended from \eqref{k-thetaAB}, which has already been proven in the main text:
\begin{equation}
\begin{aligned}
\Lie_k \theta_{(AB)}  =  \Lie_\v \btheta_{(AB)} + (\sD + \ac)_{(A}\Upsilon_{B)},
\end{aligned}
\end{equation}
Taking the trace and using that $\Lie_k q^{AB} = - 2 \btheta^{(AB)}$ and $\Lie_\v q^{AB} = - 2 \theta^{(AB)}$, we can derive \eqref{eq:RadT-1} as follows:
\begin{equation}
\begin{aligned}
\Lie_k \theta = \Lie_k (q^{AB} \theta_{AB}) & = q^{AB}\Lie_k \theta_{(AB)} + \theta_{(AB)} \Lie_k q^{AB} \\
& = q^{AB} \Lie_\v \btheta_{(AB)} + (\sD_A + \ac_A) \Upsilon^A - 2 \theta_{(AB)} \btheta^{(AB)} \\
& = \Lie_\v \btheta - \btheta_{(AB)} \Lie_\v q^{AB} + (\sD_A + \ac_A) \Upsilon^A - 2 \theta_{(AB)} \btheta^{(AB)} \\
& =  \Lie_\v \btheta + (\sD_A + \ac_A) \Upsilon^A.
\end{aligned}
\end{equation}
With these results, the equation \eqref{eq:RadT-3} can then be derived using $\s_{AB} = \theta_{(AB)} - \frac{1}{2}\theta q_{AB}$ and, again, $\Lie_k q_{AB} = 2 \btheta_{(AB)}$ and $\Lie_\v q_{AB} = 2 \theta_{(AB)}$:
\begin{equation}
\begin{aligned}
\Lie_k \s_{AB} &= \Lie_k \theta_{(AB)} - \tfrac{1}{2} q_{AB} \Lie_k \theta - \tfrac{1}{2} \theta \Lie_k q_{AB} \\
& =  \Lie_\v \btheta_{(AB)} - \tfrac{1}{2} q_{AB} \Lie_\v \btheta+ (\sD + \ac)_{\la A}\Upsilon_{B \ra} - \theta \btheta_{(AB)} \\
& = \Lie_\v \left( \btheta_{(AB)} - \tfrac{1}{2} q_{AB} \btheta \right) + \tfrac{1}{2}\btheta \Lie_\v q_{AB}+ (\sD + \ac)_{\la A}\Upsilon_{B \ra} - \theta \btheta_{(AB)} \\
& = \Lie_\v \bs_{AB} +\btheta\s_{AB} - \theta \bs_{AB}+ (\sD + \ac)_{\la A}\Upsilon_{B \ra}, 
\end{aligned}
\end{equation}
where to obtain the final equality, we recalled the definition, $\bs_{AB} = \btheta_{(AB)} - \frac{1}{2}\btheta q_{AB}$ and that $\btheta \theta_{AB} - \theta \btheta_{(AB)} = \btheta \s_{AB} - \theta\bs_{AB}$. 

To derive the remaining equations \eqref{eq:RadT-2}, \eqref{eq:RadT-4}, and \eqref{eq:RadT-5}, we can, as the simplest route, recall the relation $\NN_{AB} = \theta_{(AB)} + 2\rho \btheta_{(AB)} = \T_{AB} + \frac{1}{2} \E q_{AB}$ then utilize the previously obtained results \eqref{k-thetaAB}, \eqref{eq:RadT-1} and \eqref{eq:RadT-3}. Here, we derive them in an alternative way, starting from the derivation of $\Lie_k \NN_{AB}$ in a similar manner to \eqref{k-thetaAB}. On top of the equations presented in Appendix \ref{App:Lie}, we use that 
\bee
\cL_n k_b= -\varphi_b, \quad \cL_k v_b = -2 \kappa k_b, \qquad [k,n]= \Upsilon^A e_A + 2\kappa k.
\eee
One also uses that 
\be 
\NN_{ab} = (\cL_v + 2\rho \cL_k) q_{ab} - 2 \Upsilon_{(a}v_{b)}= \cL_n q_{ab} - 2 \Upsilon_{(a}v_{b)}.
\ee 
This implies the equation \eqref{eq:RadT-4}
\bee 
\Lie_k \NN_{AB} & = \tfrac{1}{2}e_A{}^a e_B{}^b
\Lie_k \left( 
\cL_n q_{ab} - 2 \Upsilon_{(a}v_{b)} \right)\cr
& = \tfrac{1}{2}e_A{}^a e_B{}^b \left( \Lie_{n} \Lie_k q_{ab} +  \Lie_{[k,n]}q_{ab}\right)  \\
  &= e_A{}^a e_B{}^b \left( \Lie_n \left(\btheta_{(ab)} - \Upsilon_{(a} k_{b)}\right) + \tfrac{1}{2}\Lie_\Upsilon q_{ab} + \kappa \Lie_k q_{ab}   \right)  \\
 & =  \Lie_n \btheta_{(AB)}  + (\sD + \varphi )_{(A}\Upsilon_{B)} - (\E + 2\P) \bar\theta_{AB},
\eee 
where we recalled the relation \eqref{pressure-gen}, $\frac{1}{2}\E + \P = -\kappa$. Taking the trace of $\Lie_k \NN_{AB}$, one can derive the radial derivative of the Carrollian energy \eqref{eq:RadT-3},  
\begin{equation}
\begin{aligned}
\Lie_k \E = \Lie_k (q^{AB} \NN_{AB}) & = q^{AB}\Lie_k \NN_{AB} + \NN_{AB} \Lie_k q^{AB} \\
& = q^{AB} \Lie_n \btheta_{(AB)} - (\E+2\P)\btheta + (\sD_A + \ac_A) \Upsilon^A - 2 \NN_{AB} \btheta^{(AB)} \\
& = \Lie_n \btheta - \btheta_{(AB)} \Lie_n q^{AB}  - (\E+2\P)\btheta + (\sD_A + \ac_A) \Upsilon^A - 2 \NN_{AB} \btheta^{(AB)} \\
& = \Lie_n \btheta - (\E+2\P)\btheta + (\sD_A + \ac_A) \Upsilon^A,
\end{aligned}
\end{equation}
where we used $\Lie_n q^{AB} = -2 \NN^{AB}$. Lastly, the radial derivative $\Lie_k \T_{AB}$ can be derived by recalling the definitions $\T_{AB} = \NN_{AB} - \frac{1}{2}\E q_{AB}$ and $\bs_{AB} = \btheta_{(AB)} - \frac{1}{2}\btheta q_{AB}$ as follows: 
\begin{equation}
\begin{aligned}
\Lie_k \T_{AB} &= \Lie_k \NN_{AB} - \tfrac{1}{2} q_{AB} \Lie_k \E - \tfrac{1}{2} \E \Lie_k q_{AB} \\
& =  \Lie_n \btheta_{(AB)} - \tfrac{1}{2} q_{AB} \Lie_n \btheta - (\E+2\P)\bs_{AB} + (\sD + \ac)_{\la A}\Upsilon_{B \ra} - \E \btheta_{(AB)} \\
& = \Lie_n \left( \btheta_{(AB)} - \tfrac{1}{2} q_{AB} \btheta \right) + \tfrac{1}{2}\btheta \Lie_n q_{AB}- (\E+2\P)\bs_{AB} + (\sD + \ac)_{\la A}\Upsilon_{B \ra} - \E \btheta_{(AB)} \\
& = \Lie_n \bs_{AB} - 2(\E+\P)\bs_{AB} +\btheta\T_{AB} + (\sD + \ac)_{\la A}\Upsilon_{B \ra}, 
\end{aligned}
\end{equation}
where the last equality followed from $\btheta \NN_{AB} - \E \btheta_{(AB)} = \btheta \T_{AB} - \E\bs_{AB}$. 

\subsection{Stokes theorem} \lb{App:Stokes}

In the main discussion, we need to perform integration on the stretched horizon $\H$. To derive the form of the Stokes theorem on the surface, let us consider a vector field $\zeta = F\v + X^Ae_A$ tangent to the surface. We compute the following surface exterior derivative $\dH \left( \iota_\zeta \volH \right)$ where $\volH$ is a volume form on $\H$. Using the identity $\dH \iota_k = -\iota_k \exd $, we can show that
\begin{equation}
\lb{Stokes0}
\begin{aligned}
\dH \left( \iota_\zeta \volH \right)  = \dH \left( \iota_k \iota_\zeta \volM \right)  &= - \iota_k \Lie_\zeta \volM \\
& = \nabla_a \left( F \v^a + X^A e_A{}^a \right) \volH \\
& = \left( (\Lie_\v+\theta)F + (\sD_A +\ac_A)X^A \right)\volH,
\end{aligned}
\end{equation}
where we also used the divergence formula \eqref{divergence} and that $\nabla_a \v^a = \theta$. To derive the Stokes theorem, we simply integrate the above formula on the stretched horizon $\H$. The formula is
\begin{equation}
\begin{aligned}
\int_\H\left( (\Lie_\v+\theta)F + (\sD_A +\ac_A)X^A \right)\volH = 
\int_{\pa \H} \left( F \volS  + X^A \bvol_A \right). \lb{Stokes}
\end{aligned}
\end{equation}

Let us comment on the specification of boundaries of $\H$. For example, if one picks a boundary sphere $\S$ to be identified with the cut $\sv = \text{constant}$ (as chosen in \cite{Ciambelli:2018ojf, Freidel:2022bai, Freidel:2022vjq}), the Stokes formula becomes 
\begin{equation}
\begin{aligned}
\int_{\H}\left( (\Lie_\v+\theta)F + (\sD_A +\ac_A)X^A \right)\volH = 
\int_{\S} \left( F  + \e^\a X^A\beta_A \right)\volS. 
\end{aligned}
\end{equation}
Alternatively, if one picks a \emph{Bondi slices} (see the discussion in \cite{Freidel:2022bai}) to be a boundary of $\H$ (which is only possible when $\vor_{AB} =0$), the Stokes theorem simplifies to 
\begin{equation}
\begin{aligned}
\int_{\H}\left( (\Lie_\v+\theta)F + (\sD_A +\ac_A)X^A \right)\volH = 
\int_{\S}  F \volS. 
\end{aligned}
\end{equation}
In this work, we will use the covariant formula \eqref{Stokes}.

It is also useful to have the version of Stokes theorem for the boundary of the surface $\H$. We show that the following relation holds for any horizontal vector $X = X^A e_A$, 
\begin{equation}
\begin{aligned}
\dH (\iota_X \volS) = \Lie_X \volS - \iota_X \dH \volS &= \Lie_X \iota_\v \volH - \theta\iota_X\volH \\
& = [\Lie_X, \iota_\v] \volH + \iota_\v \Lie_X \volH - \theta\iota_X\volH \\
& = -(\Lie_\v + \theta) X^A \bvol_A + \sD_A X^A \volS,
\end{aligned}
\end{equation}
where we recalled $\volS = \iota_\v \volH$ and $\bvol_A = \iota_{e_A} \volH$, and used that $\dH \volS = \theta \volH$. Since $\int_{\pa \H} \dH (\iota_X \volS) =0$, we finally obtain the relation 
\begin{equation}
\begin{aligned}
\int_{\pa \H} \left(\sD_A X^A \volS - (\Lie_\v + \theta) X^A \bvol_A \right) =0. \lb{Stokes-S}
\end{aligned}
\end{equation}



\section{Riemann-Carroll Curvature Tensor} \lb{app:Riemann}

The Riemann-Carroll curvature tensor will appear when expressing the boundary Einstein tensors $G_{k n}$ and $G_{AB}$ in terms of the Carrollian stretched horizon quantities. Its properties and the analog of the Gauss-Codazzi relations are also discussed in this section. 

The \emph{Riemann-Carroll tensor}, denoted by $\RC_{ABCD}$, of the surface $\H$ \cite{Ciambelli:2018xat} is defined through the commutator of the horizontal covariant  derivative of a horizontal vector
\begin{align}
[\sD_C, \sD_D] X^A = \RC^A{}_{BCD} X^B + \vor_{CD} \v[X^A]. \lb{Rie-Car}
\end{align}
The vorticity term $\vor_{AB}$ arises due to the anholonomy of the horizontal frame $e_A$, which can be seen from the commutator \eqref{com-4}, $[e_A, e_B] = \vor_{AB} \v$, also signifying the non-integrability of the horizontal subbundle of $T\H$ according to the Frobenius theorem. The commutator can be generalized to a horizontal tensor $\S_B{}^A$ of higher degree in the standard way,
\begin{align}
[\sD_C, \sD_D] \S_B{}^A = \RC^A{}_{ECD} \S_B{}^E -\RC^E{}_{BCD} \S_E{}^A  + \vor_{CD} \v[\S_B{}^A]. \lb{Rie-Car-1}
\end{align}
The Riemann-Carroll tensor is given in components by 
\begin{align}
\RC^A{}_{BCD} = \sD_C  \Gamma^A_{BD}-  \sD_D \Gamma^A_{BC}+ \Gamma^A_{CE}\Gamma^E_{BD} - \Gamma^A_{DE}\Gamma^E_{BC}.
\end{align}

Following from this definition, one can easily check the following algebraic properties of this curvature tensor:
\begin{align}
\RC^A{}_{BCD} = -\RC^A{}_{BDC}, \qquad \text{and} \qquad \RC^A{}_{[BCD]} = 0. \lb{Rie-property}
\end{align}
These properties are similar to those of the spacetime Riemann curvature tensor. Nonetheless, it is important to note that the Riemann-Carroll tensor is not fully antisymmetric in its first two indices and the symmetric components are given by 
\begin{align}
\RC_{(AB)CD} = \vor_{CD} \theta_{(AB)}. 
\end{align}
Due to this property, the \emph{Ricci-Carroll tensor}, defined as a contraction $\RC_{AB} := \RC^C{}_{ACB}$, is not fully symmetric, and its antisymmetric components are given by the Carrollian vorticity,
\begin{align}
\RC_{[AB]} = \tfrac{1}{2}\theta \vor_{AB}. \lb{Ric-anti}
\end{align}

\subsection{Differential Bianchi identity}
We next derive the differential Bianchi identity satisfied by the Riemann-Carroll tensor. This is achieved by considering the Jacobi identity of the horizontal covariant derivative,
\begin{align}
\big[\sD_B, [\sD_C, \sD_D] \big] X^A + (\text{cyclic permutation on\ } B,C,D) = 0. 
\end{align}
It is straightforward to show that 
\begin{equation}
\begin{aligned}
[\sD_B, [\sD_C, \sD_D]] X^A & = \sD_B, \left([\sD_C, \sD_D] X^A\right) - [\sD_C, \sD_D] (\sD_B X^A) \\
& = \sD_B \left( \RC^A{}_{ECD} X^E + \vor_{CD} \v[X^A]\right) -  \RC^A{}_{ECD} \sD_B X^E +  \RC^E{}_{BCD} \sD_E X^A \\
& \ \ \ \ - \vor_{CD} \v[\sD_B X^A] \\
& = \left( \sD_B  \RC^A{}_{ECD}  - \vor_{CD} \v[\Gamma^A_{BE}] \right) X^E + \v[X^A] (\sD_B - \ac_B )\vor_{CD} \\
& \ \ \ \ +  \RC^E{}_{BCD} \sD_E X^A, 
\end{aligned}
\end{equation}
where we made use of the vertical derivative of the horizontal covariant derivative \eqref{ell-D-X}. Recalling the Jacobi identity \eqref{Jacobi-6} and the property \eqref{Rie-property} of the Riemann-Carroll tensor, we finally obtain the differential Bianchi identity
\begin{align}
 \sD_B  \RC^A{}_{ECD}  - \vor_{CD}\v[\Gamma^A_{BE}] + (\text{cyclic permutation on\ } B,C,D) = 0.
\end{align}
We recall that $\v[\Gamma^A_{BE}]$ is given in terms of the derivative of the expansion tensor \eqref{ell-Gamma}. From this, one can derive the Bianchi identity of the Ricci-Carroll tensor,  
\begin{equation}
\begin{aligned}
\sD^B\RC_{BA}& = \tfrac{1}{2} \sD_A \RC - \vor^{BC}\left( \sD_C + \ac_C \right) \theta_{(AB)} - \vor_{AB}\left( \sD_C + \ac_C \right) (\theta^{(BC)} - \theta q^{BC} ) + \sD_B ( \theta^{(BC)}\vor_{CA} ), \lb{Bianchi-Ric}
\end{aligned}
\end{equation}
where $\RC$ is the \emph{Ricci-Carroll scalar} $\RC :=  q^{AB} \RC_{AB}$.

\subsection{Gauss-Codazzi equations for sCarrollian geometries} \lb{app:GC}
Next, let us discuss the analog of the Gauss-Codazzi equations for Carrollian geometries that will be useful for the derivation of the Einstein equation. Essentially, these equations express the relation between the fully horizontal components of the spacetime Riemann tensor $R_{ABCD} := e_A{}^a e_B{}^b e_C{}^c e_D{}^d R_{abcd}$ and the Riemann-Carroll tensor $\RC_{ABCD}$ \eqref{Rie-Car}. To derive the Gauss-Codazzi equations, we start from writing $R^A{}_{CBD}$ as the commutator
\begin{align}
R^A{}_{DBC}X^D = e_a{}^A e_B{}^b e_C{}^c[\nabla_b,\nabla_c]X^a = e_a{}^A e_C{}^c \nabla_{e_B}\nabla_c X^a - (B \leftrightarrow C). \lb{Riemann}
\end{align}
We first consider the following components of the covariant derivative,
\begin{align}
e_a{}^A e_C{}^c \nabla_{e_B} \nabla_c X^a & =e_B \left[e_a{}^A \nabla_{e_C} X^a\right] - (\nabla_{e_C} X^a)(\nabla_{e_B} e_a{}^A{}) - e_a{}^A{} (\nabla_b X^a)(\nabla_{e_B} e_C{}^b). 
\end{align}
The first term is simply $e_B \left[e_a{}^A \nabla_{e_C} X^a\right] = e_B\left[ \sD_C X^A\right]$ as following from \eqref{sD-X}. The second term can be written, with the help of \eqref{nabla-form-e} and \eqref{nabla-e-X}, as follows
\begin{equation}
\begin{aligned}
(\nabla_{e_C} X^a)(\nabla_{e_B} e_a^A) & = -(\nabla_{e_C} X^a) \left( \Gamma^A_{Ba} + \btheta_B{}^A n_a + \theta_B{}^A k_a \right)  \\
&= - \Gamma^A_{BD} \sD_C X^D  + \left(\btheta_B{}^A \NN_{CD}  +\theta_B{}^A \btheta_{CD} \right) X^D.
\end{aligned}
\end{equation}
The third term can also be written using \eqref{nabla-vec-e} as 
\begin{equation}
\begin{aligned}
e_a{}^A (\nabla_b X^a)(\nabla_{e_B} e_C{}^b) & = e_a{}^A (\nabla_b X^a) \left( \Gamma^b_{BC} - \btheta_{BC} \v^b - \NN_{BC} k^b \right) \\
&  = \Gamma^D_{BC}  \sD_D X^A- (e_a{}^A \nabla_\v X^a) \btheta_{BC} - (e_a{}^A \nabla_{k} X^a) \NN_{BC} \\
&  = \Gamma^D_{BC}  \sD_D X^A- \v[ X^A] \btheta_{BC} - (\Lie_ k X^A) \NN_{BC} - \left( \theta_D{}^A \btheta_{BC}  + \btheta_D{}^A\NN_{BC}\right)X^D. 
\end{aligned}
\end{equation}
Putting all the terms together, we obtain
\begin{equation}
\begin{aligned}
e^A{}_a  e_C{}^c \nabla_{e_B} \nabla_c X^a  = & \  \sD_B \sD_C X^A + \v[ X^A] \btheta_{BC}-\left(\btheta_B{}^A \NN_{CD}  +\theta_B{}^A \btheta_{CD} - \theta_D{}^A \btheta_{BC} \right) X^D  \\
& \ + \left( \Lie_k X^A + X^D \btheta_D{}^A \right)\NN_{BC}. \\
\end{aligned}
\end{equation}

Recalling \eqref{Riemann} and the definition of the Riemann-Carroll tensor \eqref{Rie-Car}, we can derive the Carrollian analog of the Gauss-Codazzi equations, 
\begin{equation}
\begin{aligned}
R_{CADB} =& \RC_{CADB} - \left(\btheta_{DC} \NN_{BA} +\theta_{DC} \btheta_{BA} - (D \leftrightarrow B)  \right)  - \vor_{DB} \theta_{AC},
\end{aligned}
\end{equation}
where we used the fact that the antisymmetric components of the expansion tensor and the extrinsic curvature tensor are related to the Carrollian vorticity as $\theta_{[AB]} = 2\rho \vor_{AB}$ and $\btheta_{[AB]} = -\tfrac{1}{2}\vor_{AB}$. Observe that if the Carrollian vorticity is zero, such as when one considers the case $\beta_A =0$, we recover the standard Gauss-Codazzi equations with $\RC_{ADBC}$ then coincides with the Riemann tensor on the corner sphere. The trace with $q^{AB}$ of the Riemann tensor, which will be useful in the subsequent derivations, is given by
\begin{equation}
\begin{aligned}
q^{CD}R_{CADB} &= \RC_{(AB)} - \btheta \theta_{(AB)} -\theta \btheta_{(AB)}   +\theta_B{}^C \btheta_{CA} + \theta_A{}^C \btheta_{CB} +2\rho (\btheta_{BC}\btheta_A{}^C-\btheta\btheta_{(AB)}) \\
&= \RC_{(AB)} - \btheta \NN_{AB} +2 \btheta_{C(A}\T_{B)}{}^C \lb{GC1} +2\rho \left( \btheta_{BC}\btheta_A{}^C -2 \btheta_{C(A}  \btheta_{B)}{}^C + \btheta \btheta_{(AB)} \right),\\
\end{aligned}
\end{equation}
and 
\begin{align}
q^{AB}q^{CD}R_{CADB} 
&= \RC - \btheta \E +2 \T^{AB}\bs_{AB}   +2\rho \left(\btheta_{AB}\btheta^{AB}-2\btheta_{AB} \btheta^{BA} +\btheta^2 \right) \lb{GC2}
\end{align}
where $\RC_{AB} := \RC_{CADB} q^{CD}$ is the Ricci-Caroll tensor and $\RC := \RC_{AB}q^{AB}$ is the Ricci-Carroll scalar. \\

\noindent \underline{Proof:} 
\begin{equation}
\begin{aligned}
q^{CD}R_{CADB} =& \RC_{AB} - \btheta (\theta_{BA} + 2\rho \btheta_{BA}) -\theta \btheta_{BA} + \btheta_B{}^C (\theta_{CA} + 2\rho \btheta_{CA}) +\theta_B{}^C \btheta_{CA} - \vor_{CB} \theta_A{}^C \\
=& \RC_{AB} - \btheta (\theta_{BA} + 2\rho \btheta_{BA}) -\theta \btheta_{BA} + \btheta_B{}^C (\theta_{CA}+2\rho \btheta_{CA}) +\theta_B{}^C \btheta_{CA} - 2\btheta_{[BC]} \theta_A{}^C \\
=& \RC_{(AB)} - \btheta \theta_{(AB)}  -\theta \btheta_{(AB)}   +\theta_B{}^C \btheta_{CA} + \theta_A{}^C \btheta_{CB}  +2\rho \left(\btheta_{BC}\btheta_A{}^C-\btheta\btheta_{(AB)}\right)
\end{aligned}
\end{equation}
where we used that $2\RC_{[AB]} = \theta \vor_{AB} =  2\theta \btheta_{[BA]}$. In terms of Carrollian variables, we have that
\begin{equation}
\begin{aligned}
q^{CD}R_{CADB} 
=&\ \RC_{(AB)} - \btheta \NN_{AB} -\E \btheta_{(AB)}   +(\NN_B{}^C -2\rho \btheta_B{}^C) \btheta_{CA}  + (\NN_A{}^C -2\rho \btheta_A{}^C) \btheta_{CB} \\
&\ +2\rho \left( \btheta_{BC}\btheta_A{}^C + \btheta \btheta_{(AB)} \right)\\
=&\ \RC_{(AB)} - \btheta \NN_{AB}+2 \btheta_{C(A}\T_{B)}{}^C   +2\rho \left( \btheta_{BC}\btheta_A{}^C -2 \btheta_{C(A}  \btheta_{B)}{}^C + \btheta \btheta_{(AB)} \right).
\end{aligned}
\end{equation}

\section{Derivation of Einstein equations}  \lb{app-Einstein}

In this appendix, we provide the reader the detailed derivation of the complete set of the vacuum Einstein equations, $G_{ab} =0$, evaluated on the Carrollian stretched horizon $\H$, which boils down to expressing different components of the Einstein tensor $G_{ab}$ in terms of the sCarrollian geometry and the conjugate momenta on $\H$. 

\subsection*{The $(\v n)$-component}

Let us consider first the $(\v n)$-component of the Einstein tensor, $G_{\v n}$. Due to the orthogonality of $\v$ and $n$, it is simply given by the corresponding Ricci tensor,
\begin{align}
G_{\v n} = R_{\v n}  = \left(\v^a \nabla_b \nabla_a n^b - \v^a \nabla_a \nabla_b n^b\right).
\end{align}
Using the decomposition \eqref{nabla-l} and \eqref{nabla-n}, we can write the Einstein tensor as
\begin{equation}
\begin{aligned}
G_{\v n}
& = \nabla_a \left( \nabla_\v n^a \right) - (\nabla_a \v^b)(\nabla_b n^a) - \v[\nabla_a n^a] \\
& = \nabla_a \left( \kappa v^a - \J^a + v[\rho]k^a \right) - \left(\theta^{ab} + 2\rho \btheta^{ab} \right)\theta_{ab} + \left( \J^a - 2\rho (\p^a + \ac^a) \right)\p_a - \p^a \sD_a \rho - \v[\E+2\kappa]\\
& = - \v[\E]+\kappa\theta - (\sD_a+2\ac_a)\J^a- \left(\theta^{ab} + 2\rho \btheta^{ab} \right)\theta_{ab} + \btheta \v[\rho] + [k,\v][\rho] \\
& \ \ \ \  +\left( \J^a - 2\rho\p_a \right)(\p^a+\ac^a) - \p^a \sD_a \rho \\
& = - \v[\E]+\kappa\theta - (\sD_a+2\ac_a)\J^a- \left(\theta^{ab} + 2\rho \btheta^{ab} \right)\theta_{ab} + \btheta \v[\rho]  -(\p^a+\ac^a)\sD_a \rho +(\Upsilon^a- \p^a) \sD_a \rho \\
& = - (\v+\theta)[\E]-\P\theta - (\sD_a+2\ac_a)\J^a- \T^{ab}\s_{ab} + \btheta \v[\rho].
\end{aligned}
\end{equation}
To elaborate, on the third line, we used that $\kappa= k[\rho]$ to write $k[\v[\rho]] - \v[\kappa] = [k,\v][\rho]$. The fourth equality followed from the commutator $[k,\v] = \Upsilon^A e_A$ and the expression for the heat current \eqref{heat-gen}, $\J_a = -(\sD_a - 2\p_a)\rho$. To obtain the last equality, we used the relation $\p^a +\ac^a = \Upsilon^a - \p^a$, the trace-traceless decomposition $\theta^{ab} + 2\rho \btheta^{ab} = \T^{ab} + \frac{1}{2}\E q^{ab}$, and the definition of the pressure \eqref{pressure-gen}, $\P = - \kappa - \frac{1}{2}\E$. The last term involving $\v[\rho]$ vanishes on the null surface $\N$.  

Here, we derived $G_{\v n}$ using directly the decomposition of the covariant derivatives. It is however worth mentioning that we can alternatively derive this expression using the key result of \cite{Freidel:2022vjq}. That is, the Einstein tensor is given in terms of the divergence of the Carrollian stress tensor,
\begin{align}
\Pi_a{}^c G_{cd} n^d = D_b \EM_a{}^b - \left(\bar{K}_a{}^b - \bar{K} \Pi_a{}^b\right)D_b \rho, \lb{G=DT}
\end{align}
where $\bar{K}_a{}^b := \Pi_a{}^c (\nabla_c k^d) \Pi_d{}^b$. Let us evaluate each term in $G_{\v n} = \v^a \Pi_a{}^c G_{cd} n^d $ separately. The first term can be written using \eqref{sCT} and \eqref{nabla-l} as follows, 
\begin{equation}
\begin{aligned}
\v^a D_b \EM_a{}^b &= D_a (\v^b \EM_b{}^a) - \EM_b{}^a D_a \v^b \\
&= - D_a (\E \v^a + \J^a) - \EM_b{}^a \left( \theta_a{}^b + (\p_a + \kappa k_b) \v^a - k_a (\J^b - 2\rho(\p^b +\ac^b) \right) \\
& = - \v[\E] - \E (\theta +\kappa) - (\sD_a +\p_a+\ac_a) \J^a-\T^{ab}\s_{ab} - \P\theta + \p_a \J^a + \E \kappa \\
& \ \ \ \ + \p_a\left(\J^a - 2\rho(\p^a +\ac^a)\right)\\
& = - (\v+\theta)[\E] - \P\theta - (\sD_a+ 2\ac_a) \J^a-\T^{ab}\s_{ab} + \left(\J^a - 2\rho\p_a\right)(\p^a +\ac^a)\\
& = - (\v+\theta)[\E] - \P\theta - (\sD_a+ 2\ac_a) \J^a-\T^{ab}\s_{ab} -(\p^a +\ac^a)\sD_a \rho,\\
\end{aligned}
\end{equation}
where we used $D_a \v^a = \theta + \kappa$ and $D_a \J^a = (\sD_a + \p_a +\ac_a) \J^a$ to get the third equality, and used \eqref{heat-gen} to arrive at the final equation. The last terms vanishes when considering the case where $\rho$ is constant, as did in \cite{Freidel:2022vjq} (equation (87) there\footnote{More precisely, in \cite{Freidel:2022vjq}, we used the relation $A^a\p_a -\ac_a\J^a= (\p^a+\ac^a)\sD_a\rho$,
where $A^a := e_{Aa} \nabla_\v \v^a = (\sD_A + 2\ac_A)\rho$ is the tangential acceleration, and assumed $D_a\rho =0$. This means $A^a\p_a = \ac_a \J^a$.}). 

Using \eqref{nabla-k}, one can show that 
\begin{align}
\v^a \left(\bar{K}_a{}^b - \bar{K} \Pi_a{}^b\right)D_b \rho = -(\p^a +\ac^a)\sD_a \rho - \btheta \v[\rho].
\end{align}
Therefore, the Einstein tensor computed using \eqref{G=DT} is
\begin{equation}
\begin{aligned}
G_{\v n} = - (\v+\theta)[\E]-\P\theta - (\sD_a+2\ac_a)\J^a- \T^{ab}\s_{ab} + \btheta \v[\rho],
\end{aligned}
\end{equation}
in agreement with the previous, straightforward, derivation. Note the cancellation of the term $(\p^a +\ac^a)\sD_a \rho$. 

\subsection*{The $(A n)$-components}
In a similar manner, the components $G_{A n}$ of the Einstein tensors are simply given by the corresponding components of the Ricci tensor,
\begin{align}
G_{A n} = R_{A n}  = \left(e_A{}^a \nabla_b \nabla_a n^b - e_A{}^a \nabla_a \nabla_b n^b\right).
\end{align}
Again, by using the decomposition of covariant derivative \eqref{nabla-l} and \eqref{nabla-eA}, one can evaluate the Einstein tensors on $\H$ as follows
\begin{equation}
\begin{aligned}
G_{An} & = \nabla_a (\nabla_{e_A}n^a) - \nabla_a n^b \nabla_b e_A{}^a - e_A[\nabla_a n^a] \\
& = \nabla_a \left( \theta_A{}^a + 2\rho\btheta_A{}^a +\p_A \v^a + \sD_A \rho k^a \right) -  \Gamma^B_{CA} \left(\theta_C{}^B + 2\rho\btheta_C{}^B \right)- \kappa \ac_A \\
& \ \ \ \ + \p^B\left(2\theta_{[BA]} + 2\rho \btheta_{BA} \right) - \J^B \btheta_{BA} - \left( \btheta_A{}^B - \Omega_A{}^B \right) \sD_B \rho  - \sD_A(\E + 2\kappa) \\
& = (\v + \theta)[\p_A] + \E \ac_A + \left( \sD_B  +\ac_B \right) (\T_A{}^B +\P \delta_A^B) + \btheta \sD_A \rho \\
& \ \ \ \ + [k,e_A][\rho]  + \p^B\left(2\theta_{[BA]} + 2\rho \btheta_{BA} \right) - \J^B \btheta_{BA} - \left( \btheta_A{}^B - \Omega_A{}^B \right) \sD_B \rho,
\end{aligned}
\end{equation}
where, to arrive at the last equality, we employed the split $\theta^{ab} + 2\rho \btheta^{ab} = \T^{ab} + \frac{1}{2}\E q^{ab}$, and the pressure $\P = - \kappa - \frac{1}{2}\E$. We also wrote $k[e_A[\rho]] - \sD_A[\kappa] = [k,e_A][\rho]$ as following from $\kappa = k[\rho]$. Next, by making use of the commutator \eqref{com-4}, $[k,e_A] = -\Omega_A{}^B e_B$, and recalling the antisymmetric components $\theta_{[AB]} = \rho \vor_{AB}$ and $2\btheta_{[AB]} = - \vor_{AB}$, we can show that
\begin{equation}
\begin{aligned}
&[k,e_A][\rho] + 2\p^B\left(\theta_{[BA]} + \rho \btheta_{BA} \right) - \J^B \btheta_{BA} - \left( \btheta_A{}^B - \Omega_A{}^B \right) \sD_B \rho  \\
& \qquad = - \J^B \btheta_{BA} - \btheta_A{}^B\left( \sD_B  - 2 \p_B\right)\rho \\
& \qquad =  - \vor_{AB} \J^B,
\end{aligned}
\end{equation}
where we also used $\J_A = -(\sD_A - 2\p_A)\rho$. Putting everything together, we finally obtain the following expression of the Einstein tensor $G_{An}$ in terms of Carrollian fluid variables,
\begin{equation}
\begin{aligned}
G_{A n} = (\v + \theta)[\p_A] + \E \ac_A - \vor_{AB} \J^B + \left( \sD_B  +\ac_B \right) (\T_A{}^B +\P \delta_A^B) + \btheta \sD_A \rho. 
\end{aligned}
\end{equation}

Similar to the previous components, we can derive $G_{An}$ from the conservation equations \eqref{G=DT} of the Carrollian stress tensor. First, using \eqref{sCT} and \eqref{nabla-eA}, we show that 
\begin{equation}
\begin{aligned}
e_A{}^a D_b \EM_a{}^b &= D_a (e_A{}^b \EM_b{}^a) - \EM_b{}^a D_a e_A{}^b \\
&= D_a (\p_A \v^a + \T_A{}^a + \P e_A{}^a) - \EM_b{}^a \left( \Gamma^b_{aA} +(
- \btheta_{aA}  + (\p_A + \ac_A) k_a )\v^b + \theta_A{}^b k_a \right) \\
& = (\v + \theta + \kappa)[\p_A] + \left( \sD_B +\p_B +\ac_B \right) (\T_A{}^B +\P \delta_A^B) - \J^B \btheta_{BA} + \E (\p_A + \ac_A) - \theta_A{}^B \p_B \\
& = (\v + \theta)[\p_A] + \E \ac_A - \vor_{AB} \J^B+ \left( \sD_B +\ac_B \right) (\T_A{}^B +\P \delta_A^B) \\
& \ \ \ \ - \J^B \btheta_{AB} +  \left(\T_A{}^B +\P \delta_A^B- \theta_A{}^B + (\E+\kappa) \delta_A^B\right) \p_B.
\end{aligned}
\end{equation}
Recalling that $\theta^{ab} + 2\rho \btheta^{ab} = \T^{ab} + \frac{1}{2}\E q^{ab}$ and $\P = - \kappa - \frac{1}{2}\E$, we show that
\begin{equation}
\begin{aligned}
- \J^B \btheta_{AB} +  \left(\T_A{}^B +\P \delta_A^B- \theta_A{}^B + (\E+\kappa) \delta_A^B\right) \p_B = - (\J^B - 2\rho\p^B) \btheta_{AB} = \btheta_A{}^B \sD_B \rho,
\end{aligned}
\end{equation}
after substituting $\J_A = -(\sD_A - 2\p_A)\rho$. This thus yields,
\begin{equation}
\begin{aligned}
e_A{}^a D_b \EM_a{}^b = (\v + \theta)[\p_A] + \E \ac_A - \vor_{AB} \J^B+ \left( \sD_B +\ac_B \right) (\T_A{}^B +\P \delta_A^B) + \btheta_A{}^B \sD_B \rho.
\end{aligned}
\end{equation}
Indeed, when $\rho = \mathrm{constant}$, the last term vanishes, hence reproducing the result of \cite{Freidel:2022vjq}. It follows from \eqref{nabla-k} that, 
\begin{align}
e_A{}^a \left(K_a{}^b - K \Pi_a{}^b\right)D_b \rho = (\btheta_A{}^B - \btheta \delta_A^B) \sD_B\rho.
\end{align}
Therefore, the horizontal components of \eqref{G=DT} reproduce the Einstein tensor 
\begin{equation}
\begin{aligned}
G_{A n} = (\v + \theta)[\p_A] + \E \ac_A + \J^B\vor_{BA} + \left( \sD_B  +\ac_B \right) (\T_A{}^B +\P \delta_A^B) + \btheta \sD_A \rho. 
\end{aligned}
\end{equation}

\subsection*{The $(k k)$-component}
For this component, it follows from the null-ness of the vector $k$ that $G_{k k} = R_{k k }$. We can use \eqref{nabla-k} to write 
\begin{equation}
\lb{G_kk}
\begin{aligned}
G_{k k}  &=  \left(k^a \nabla_b \nabla_a k^b - k^a \nabla_a \nabla_b k^b\right)\\
& = \left(\nabla_a (\nabla_k k^a) - \nabla_a k^b \nabla_b k^a - k[\nabla_a k^a] \right) \\
& = - \left( \Lie_k\btheta  + \btheta_{AB} \btheta^{BA} \right).
\end{aligned}
\end{equation}
The Einstein equation $G_{k k} =0$ imposes $\Lie_k\btheta  + \btheta_{AB}\btheta^{BA} = 0$, constraining the radial evolution of the expansion $\btheta$. 


\subsection*{The $(k n)$-component}
For the $(k n)$-component, its expression in terms of the Carrollian geometry of the stretched horizon $\H$ requires the Gauss-Codazzi equations derived in section \eqref{app:GC}. First, this component of the Einstein tensor is given by
\begin{align}
G_{k n} = R_{k n} - \tfrac{1}{2} R.
\end{align}
With the decomposition \eqref{4Dmetric} of the spacetime metric, the spacetime Ricci scalar on $\N$ can be expressed as
\begin{equation}
\begin{aligned}
R = g^{ab} R_{ab}& =(n^a k^b + k^a \v^b + q^{AB}e_A{}^a e_B{}^b) R_{ab} \\
&= R_{k n} + R_{k \v} + q^{AB}(n^a k^b + k^a \v^b + q^{ab}) R_{aAbB} \\
& = R_{k n} + R_{k \v} + q^{AB} (R_{n A k B} + R_{\v A k B}) + q^{AB} q^{CD}  R_{CADB}.
\end{aligned}
\end{equation}
This therefore allows us to write the Einstein tensor $G_{k n}$ as
\begin{equation}
\lb{G-kn-def} 
\begin{aligned}
G_{k n} &= \rho R_{k k} - \tfrac{1}{2}q^{AB}(R_{n A k B} + R_{\v A k B}) - \tfrac{1}{2}q^{AB} q^{CD} R_{CADB} \\
& = -q^{AB}R_{\v A k B} - \tfrac{1}{2}q^{AB} q^{CD} R_{CADB}
\end{aligned}
\end{equation}
where we used $n^a = \v^a + 2\rho k^a$ and that $q^{AB}R_{kAkB} = R_{kk}$ as following from the metric decomposition \eqref{4Dmetric}. Indeed, the second term is clearly in the form of the scalar Gauss-Codazzi relation \eqref{GC2}. To evaluate the first term, we can begin by writing it as a commutator
\begin{equation}
\lb{Rie_vAkB}
\begin{aligned}
R_{\v A k B} & =  e_{Aa} e_B{}^b \v^c [\nabla_b, \nabla_c] k^a  = e_{Aa} e_B{}^b \v^c (\nabla_b \nabla_c k^a - \nabla_c \nabla_b k^a).
\end{aligned}
\end{equation}
Using the decomposition of covariant derivatives \eqref{nabla-l} and \eqref{nabla-k}, we can write the following term as
\begin{equation}
\begin{aligned}
e_{Aa} \v^b \nabla_{e_B} \nabla_b k^a & = e_{Aa} \nabla_{e_B} (\nabla_\v k^a) - e_{Aa} (\nabla_b k^a) (\nabla_{e_B} \v^b) \\
& = -e_{Aa} \nabla_{e_B} \left( \p^a + \ac^a + \kappa k^a \right)- e_{Aa} (\nabla_b k^a) \left( \theta_B{}^b  + \p_B \v^b +\J_B k^b \right) \\
& = -(\sD_B -\p_B ) (\p_A + \ac_A) - \kappa \btheta_{BA} - \theta_B{}^C\btheta_{CA},
\end{aligned}
\end{equation}
where we also used the equation \eqref{nabla-e-X} for $\nabla_{e_B} \left( \p^a + \ac^a\right)$. Next, we apply the decomposition \eqref{nabla-k} and \eqref{nabla-eA} to write
\begin{equation}
\begin{aligned}
-e_{Aa} e_B{}^b \nabla_\v \nabla_b k^a &= - \v[e_{Aa} \nabla_{e_B} k^a] + e_{Aa} (\nabla_b k^a) (\nabla_\v e_B{}^b) + (\nabla_{e_B} k^a)(\nabla_\v e_{Aa}) \\
& = -\v[\btheta_{BA}] + \left( \btheta_{aA} - (\p_A +\ac_A) k_a \right) \nabla_\v e_B{}^a +\left( \btheta_{Ba} - \p_B k_a \right) \nabla_\v e_A{}^a \\
& = - \v[\btheta_{BA}] - (\p_A + \ac_A)(2\p_B+\ac_B)  + \theta_B{}^C\btheta_{CA} + \theta_A{}^C \btheta_{BC}. 
\end{aligned}
\end{equation}
Collecting these results together, the components \eqref{Rie_vAkB} of the Riemann tensor reads,
\begin{equation}
\lb{Rie-vAkB-1}
\begin{aligned}
R_{\v A k B} &= - (\v + \kappa) [\btheta_{BA}] -(\sD_B +\p_B + \ac_B ) (\p_A + \ac_A) + \theta_A{}^C \btheta_{BC}.
\end{aligned}
\end{equation}
Its trace is thus given by
\begin{equation}
\begin{aligned}
q^{AB} R_{\v A k B} &= - (\v + \kappa) [\btheta] -(\sD_A +\p_A + \ac_A ) (\p^A + \ac^A)  - \theta^{BA}\btheta_{AB},
\end{aligned}
\end{equation}
where we used that $\v[q^{AB}] = -2 \theta^{(AB)}$. Finally, the Einstein tensor $G_{kn}$ computed from \eqref{G-kn-def} can be expressed as,
\begin{equation}
\begin{aligned}
G_{k n} &= (\v + \kappa) [\btheta] +(\sD_A +\p_A + \ac_A ) (\p^A + \ac^A) + \theta^{BA}\btheta_{AB} - \tfrac{1}{2}q^{AB} q^{CD} R_{CADB} \\
& = (\v+\Wein) [\btheta] +(\sD_A +\p_A + \ac_A ) (\p^A + \ac^A) - \tfrac{1}{2} \RC  - \rho \left( \btheta_{AB} \btheta^{AB} + \btheta^2 \right)
\end{aligned}
\end{equation}
where we used \eqref{GC2} to derive the final equation and recalled $\Wein = \kappa + \E = \frac{1}{2}\E - \P$. Notice that the last term does not appear on the null surface $\N$. 

Using the previously derived $G_{ k k}$ component and the identities \eqref{eq:RadT-2}, one can also show that the $G_{k \v}$ component can be written as
\begin{equation}
\begin{aligned}
G_{k \v} = G_{k n} - 2\rho G_{k k} = \Lie_k \E + \btheta \E -\kappa\btheta - (\sD -\p)_A \p^A -\tfrac{1}{2}\RC + \rho \left(2\btheta_{AB}\btheta^{BA} -\btheta_{AB} \btheta^{AB} - \btheta^2\right).
\end{aligned}
\end{equation}

\subsection*{The $(AB)$-components}

Next, we consider the fully horizontal components of the Einstein tensor. It is given by 
\begin{equation}
\begin{aligned}
G_{A B} &= R_{A B} - \tfrac{1}{2} R q_{AB} = R_{A B} - \tfrac{1}{2}\left( R_{k n} + R_{k \v}  + q^{CD} R_{C D} \right)q_{AB}.
\end{aligned}
\end{equation}
These components can be decomposed into the symmetric traceless part and the trace part as\footnote{In $D$-dimensional spacetime, the relation is $q^{AB} G_{A B} = (D-4) G_{k n} - (D-3) R_{k n} - R_{k \v}$.}
\begin{align}
G_{\langle A B \rangle} = R_{\langle A B \rangle}, \qquad \text{and} \qquad q^{AB} G_{A B} = - \left( R_{k n} + R_{k \v} \right).
\end{align}

\paragraph{Symmetric trace-free components:} Let us first focus on the symmetric trace-free components of $G_{AB}$, which boils down to the evaluation of the horizontal components of the spacetime Ricci tensor, $R_{A B} = e_A{}^a e_B{}^b R_{ab}$. It can be expressed as
\begin{equation}
\begin{aligned}
R_{A B} &=    R_{n A k B} + R_{k A \v B} +q^{CD} R_{C A D B}  \\
& = R_{\v A k B} + R_{\v B k A} + 2\rho R_{k A k B}  +q^{CD} R_{C A D B}. 
\end{aligned}
\end{equation}
We have already evaluated the first two terms, which are given by \eqref{Rie-vAkB-1}. We then have
\begin{equation}
\begin{aligned}
R_{\v A k B} + R_{\v B k A}=  - 2(\Lie_\v + \kappa) \btheta_{(AB)} -2(\sD +\p + \ac )_{(A} (\p + \ac)_{B)} + 2\theta_{(A}{}^C \btheta_{B)C}.
\end{aligned}
\end{equation}
The term $R_{kAkB}$ can be evaluated by writing $R_{k A k B} = e_{Aa} e_B{}^b k^c [\nabla_b,\nabla_c] k^a$ as follows:
\begin{equation}
\begin{aligned}
R_{k A k B} 
& = e_{Aa} k^b \nabla_{e_B} \nabla_b k^a - e_{Aa} e_B{}^b \nabla_k \nabla_b k^a \\
& = - k[e_{Aa} \nabla_{e_B} k^a]+ (\nabla_{e_B} k^a) (\nabla_k e_{Aa}) - (e_{Aa}\nabla_{[e_B,k]} k^a) (\nabla_k e_B{}^b) \\
& = - k[\btheta_{BA}] - \Omega_B{}^C \btheta_{CA} - \Omega_A{}^C \btheta_{BC} + \btheta_A{}^C \btheta_{BC} \\
& = - \Lie_k \btheta_{(AB)} + \btheta_A{}^C \btheta_{BC},
\end{aligned}
\end{equation}
where we employed the covariant derivatives \eqref{nabla-k} and \eqref{nabla-eA}. We also used that $\Lie_k \vor_{AB} = 2\Lie_k \btheta_{[BA]} = 0$, according to the Jacobi identity \eqref{Jacobi-5}. We thus have
\begin{equation}
\lb{R-AB-0}
\begin{aligned}
-R_{A B} & = 2\left(\Lie_\v + \kappa\right) \btheta_{(AB)} +2(\sD +\p + \ac )_{(A} (\p + \ac)_{B)} -2\theta_{(A}{}^C \btheta_{B)C} + 2\rho( \Lie_k \btheta_{(AB)} - \btheta_A{}^C \btheta_{BC}) \\
& \ \ \ \ - q^{CD} R_{C A D B} \\
 & = 2\left(\Lie_\v +\rho \Lie_k + \kappa\right) \btheta_{(AB)} +2(\sD +\p + \ac )_{(A} (\p + \ac)_{B)} -4\T_{(A}{}^C \bs_{B)C} - \btheta \T_{AB} - \E \bs_{AB}  \\
& \ \ \ \ - \RC_{(AB)} + 2\rho(2 \btheta_{C(A} \btheta_{B)}{}^C - \btheta \btheta_{(AB)}) \\
\end{aligned}
\end{equation}
where we invoked the Gauss-Codazzi equation \eqref{GC1} to obtain the second equality.

It is important to appreciate that one can express $R_{AB}$ in many different ways, thank to the relation between $\theta_{AB}$ and $\btheta_{AB}$ \eqref{k-thetaAB},
\begin{align}
\Lie_k \theta_{(AB)} = \Lie_\v \btheta_{(AB)} + (\sD+\ac)_{(A} \Upsilon_{B)}.
\end{align}
For instance, we can manipulate the first term in \eqref{R-AB-0} as follows:
\begin{equation}
\begin{aligned}
2\left(\Lie_\v +\rho \Lie_k + \kappa\right) \btheta_{(AB)} &= 2\Lie_\v \btheta_{(AB)} +  \Lie_k\left(2\rho\btheta_{(AB)}\right)\\
& = \Lie_\v \btheta_{(AB)} + \Lie_k \left( \theta_{AB} + 2\rho \btheta_{AB}  \right) - (\sD+\ac)_{(A} \Upsilon_{B)} \\
& = \Lie_\v \btheta_{(AB)} + \Lie_k \NN_{AB} - \left(\sD+\ac\right)_{(A} \left(2\p+\ac\right)_{B)}.
\end{aligned}
\end{equation}
Thus, we can express the tensor $R_{AB}$ as
\begin{equation}
\lb{R-AB-1}
\begin{aligned}
-R_{A B} 
 & = \Lie_\v \btheta_{(AB)} + \Lie_k  \NN_{AB}  +(\sD + \ac )_{(A}\ac_{B)} + 2(\p+\ac)_{(A} \p_{B)}  \\
& \ \ \ \ -4\T_{(A}{}^C \bs_{B)C} - \btheta \T_{AB} - \E \bs_{AB} - \RC_{(AB)} + 2\rho\left(2 \btheta_{C(A} \btheta_{B)}{}^C - \btheta \btheta_{(AB)}\right). \\
\end{aligned}
\end{equation}
The trace is given by 
\begin{equation}
\begin{aligned}
-q^{AB} R_{AB} & = \Lie_\v \btheta + \Lie_k \E +2\btheta\E +(\sD + \ac )_A \ac^A+2(\p+\ac)_A \p^A - \RC  \\
& \ \ \ \ - 2\rho\left(\tfrac{1}{2}\vor_{AB} \vor^{BA}+ \btheta^2\right), 
\end{aligned}
\end{equation}
where we used $2\btheta_{[BA]} = \vor_{AB}$. With these results, the symmetric traceless components of the Einstein tensor $G_{\langle AB \rangle}$ become
\begin{equation}
\begin{aligned}
-G_{\langle A B \rangle} 
 & = \Lie_\v \bs_{AB} + \Lie_k \T_{AB} +(\sD + \ac )_{\langle A}\ac_{B \rangle} + 2(\p+\ac)_{\langle A} \p_{B\rangle} -4\T_{(A}{}^C \bs_{B)C} - \RC_{\langle AB \rangle}  \\
& \ \ \ \  + 2\rho\left(2 \btheta_{C \langle A} \btheta_{B \rangle}{}^C - 2\btheta \bs_{AB} + \bs_{CD} \bs^{CD} q_{AB} \right).
\end{aligned}
\end{equation}

\paragraph{Simplification in 4-dimension:} Some simplifications occur in four dimension as we have these nice identities,
\begin{align}
T_{C\la A} \bar{T}_{B \ra}{}^C = \tfrac{1}{2} \bar{T} T_{\la AB \ra}+ \tfrac{1}{2} T \bar{T}_{\la AB \ra}, \qq \text{and} \qq S_{C( A} \bar{S}_{B )}{}^C = \tfrac{1}{2} S^{CD} \bar{S}_{CD} q_{AB},
\end{align}
where $(T_{AB},\bar{T}_{AB})$ are any horizontal tensors of degree two, and $(T,\bar{T})$ and $(S_{AB},\bar{S}_{AB})$ are their respective trace and trace-free components. Thank to the identity $\T_{(A}{}^C \bs_{B)C} = \frac{1}{2}\T^{CD}\bs_{CD}q_{AB}$, we can write 
\begin{equation}
\begin{aligned}
\Lie_\v \bs_{\la AB \ra} + \Lie_k \T_{\la AB \ra} 
&= \Lie_\v \bs_{AB} + \Lie_k \T_{AB} - \tfrac{1}{2} q^{CD}\left( \Lie_\v \bs_{CD} + \Lie_k \T_{CD} \right)q_{AB} \\
&= \Lie_\v \bs_{AB} + \Lie_k \T_{AB}- 2\left(\T^{CD}\bs_{CD}-\rho \bs^{CD}\bs_{CD} \right)q_{AB} \\
&= \Lie_\v \bs_{AB} + \Lie_k \T_{AB}- 4\T_{(A}{}^C \bs_{B)C}+2\rho \bs^{CD}\bs_{CD}q_{AB},
\end{aligned}
\end{equation}
Finally, we obtain
\begin{equation}
\begin{aligned}
-G_{\langle A B \rangle} 
& = \Lie_\v \bs_{\la AB \ra} + \Lie_k \T_{\la AB\ra} +(\sD +2\p+\ac )_{\la A}\ac_{B \ra} + 2\p_{\la A} \p_{B \ra} - \RC_{\la AB \ra}.
\end{aligned}
\end{equation}

\paragraph{The trace part:} Computing the trace of the components $G_{AB}$ amounts to expressing the Ricci tensor $R_{\v k}$ and $R_{k n}$ in terms of the Carrollian geometry of $\H$. For $R_{\v k}$, it can be written using the commutator together with the decomposition \eqref{nabla-l} and \eqref{nabla-k} as follows
\begin{equation}
\begin{aligned}
- R_{\v k} 
&= \v^a [\nabla_a, \nabla_b] k^b \\
&= \v[\nabla_a k^a] - \nabla_a \left( \nabla_\v k^a \right) + (\nabla_b \v^a) (\nabla_a k^b)  \\
& = \v[\bar\theta] + k[\kappa] + \kappa \btheta + (\sD_A +\ac_A) (\p^A+\ac^A)-\p_A (2\p^A+\ac^A)+\theta^{AB}\btheta_{BA} \\
& = \v[\bar\theta] + k[\kappa] -\P \btheta + (\sD_A +\ac_A) (\p^A+\ac^A)-\p_A (2\p^A+\ac^A)+\T^{AB}\bs_{BA} - 2\rho \btheta^{AB}\btheta_{BA}
\end{aligned}
\end{equation} 
where we used $\theta_{AB} + 2\rho\btheta_{AB} = \T_{AB} +\frac{1}{2}\E q_{AB}$ and $-\P = \kappa + \frac{1}{2}\E$. 

We compute $R_{k n}$ using the similar trick, 
\begin{equation}
\begin{aligned}
- R_{k n} 
&= n^a [\nabla_a, \nabla_b] k^b \\
&= n[\nabla_a k^a] - \nabla_a \left( \nabla_n k^a \right) + (\nabla_b n^a) (\nabla_a k^b)  \\
&= n[\btheta] + k[\kappa] -\P \btheta + (\sD_A +\ac_A) (\p^A+\ac^A)-\p_A (2\p^A+\ac^A)+\T^{AB}\bs_{BA} 
\end{aligned}
\end{equation} 
One can easily check that $R_{kn}- R_{\v k} = -2\rho\left(k[\btheta] + \btheta^{AB}\btheta_{BA} \right) = 2\rho R_{kk}$ as it should be (see equation \eqref{G_kk}). Alternatively, we can use \eqref{k-thetaAB} to write $R_{k n}$ as
\begin{equation}
\begin{aligned}
- R_{k n} 
&=  k[\mathsf{N}] +(\P+\E) \btheta - (\sD_A +2\p_A+2\ac_A) \p^A+\T^{AB}\bs_{BA},
\end{aligned}
\end{equation} 
where we recalled $\mathsf{N} = \kappa + \E$. One can verify this equation by computing the commutator $R_{k n} = k^a [\nabla_b, \nabla_a] n^b$, then use the decomposition \eqref{nabla-k} and \eqref{nabla-n}.

The trace $q^{AB} G_{AB} = -(R_{\v k} + R_{k n})$ is thus
\begin{equation}
\begin{aligned}
\tfrac{1}{2}q^{AB} G_{AB} 
& = (\Lie_\v+\rho \Lie_k-\P)\btheta + \Lie_k\kappa  + (\sD_A +\ac_A) (\p^A+\ac^A)-\p_A (2\p^A+\ac^A)+\T^{AB}\bs_{BA} - \rho \btheta^{AB}\btheta_{BA} \\
& = \Lie_k \Wein +(\E+\P)\btheta   - (\sD_A +2\p_A+2\ac_A) \p^A +\T^{AB}\bs_{BA} + \rho G_{k k}.
\end{aligned}
\end{equation}

\paragraph{Double-null frame:} Let us mention some interesting observations. First, the tensor $R_{AB}$ \eqref{R-AB-0} contains the evolution of $\btheta_{(AB)}$ along the spacetime null vector $\ell:= \v + \rho k = n-\rho k$, that is $(\Lie_\v + \rho \Lie_k)\btheta_{(AB)} = \Lie_\ell \btheta_{(AB)}$. It is worth noting that, instead of the rigging-Carroll frame, the spacetime metric \eqref{4Dmetric} can be written in the double-null frame $(\ell, \bar{\ell})$ with $\bar{\ell} := k$ as $g_{ab} = 2\ell_{(a} \bar{\ell}_{b)} +q_{ab}$. Using \eqref{k-thetaAB}, one can show the following relation, 
\begin{align}
(\Lie_\ell + \kappa) \bar{\cL}_{(AB)} & =  \Lie_{\bar{\ell}} \cL_{(AB)} - (\sD+\ac)_{(A} \Upsilon_{B)},
\end{align}
where we define the extrinsic curvature tensors of the sphere in the double-null frame as
\begin{align}
\cL_{AB} : = g(e_B, \nabla_{e_A}\ell) = \theta_{AB} + \rho \btheta_{AB} \qquad \text{and} \qquad \bar{\cL}_{AB} : = g(e_B, \nabla_{e_A}\bar{\ell}) = \btheta_{AB}.
\end{align}
Neither of them are symmetric. With these new variables, we express $R_{AB}$ in two ways as
\begin{subequations}
\begin{align}
-R_{A B} & = 2\left(\Lie_\ell + \kappa + \tfrac{1}{2}\cL\right) \bar{\cL}_{(AB)} + \bar{\cL} \cL_{(AB)} +2(\sD +\p + \ac )_{(A} (\p + \ac)_{B)}   \\
& \ \ \ \ - 2\cL_{(A}{}^{C)}\bar{\cL}_{(BC)} - 2\cL_{(B}{}^{C)}\bar{\cL}_{(AC)} +\rho \vor_A{}^C \vor_{CB} \nonumber \\
& = 2\left(\Lie_{\bar{\ell}} + \tfrac{1}{2}\bar{\cL}\right) \cL_{(AB)} + \cL \bar{\cL}_{(AB)} -2(\sD -\p)_{(A} \p_{B)}   \\
& \ \ \ \ - 2\cL_{(A}{}^{C)}\bar{\cL}_{(BC)} - 2\cL_{(B}{}^{C)}\bar{\cL}_{(AC)} +\rho \vor_A{}^C \vor_{CB}. \nonumber
\end{align}
\end{subequations}

\subsection*{The $(k A)$-components}
For these components, we use the decomposition \eqref{nabla-k} and \eqref{nabla-eA}, and that $\btheta_{BA} - \btheta_{AB} = \vor_{AB}$, to write the Einstein tensor $G_{k A} = R_{k A}$ as
\begin{equation}
\begin{aligned}
{G}_{k A} &= \left(e_A{}^a \nabla_b \nabla_a k^b - e_A{}^a \nabla_a \nabla_b k^b \right) \\
&= \left( \nabla_a (\nabla_{e_A} k^a) - \nabla_a k^b \nabla_b e_A{}^a - e_A[\nabla_a k^a] \right) \\
& = \nabla_a (\btheta_A{}^a - \p_A k^a) - {\Gamma}_{BA}^C {\btheta}_C{}^B + \big({\btheta}_A{}^B - {\Omega}_A{}^B  \big){\p}_B 
- \big({\p}^B + {\ac}^B \big) {\btheta}_{BA} - \sD_A {\btheta} \\ 
& =  e_B[\btheta_A{}^B] + \btheta_A{}^B \nabla_a (e_B{}^a)   - {\Gamma}_{BA}^C {\btheta}_C{}^B
-  (k+\btheta)[\p_A]  + \big({\btheta}_A{}^B - {\Omega}_A{}^B  \big){\p}_B 
- \big({\p}^B + {\ac}^B \big) {\btheta}_{BA} - \sD_A {\btheta} \\ 
& =  (\sD_B+\ac_B)[\btheta_A{}^B]
-  (k+\btheta)[\p_A]  + \big({\btheta}_A{}^B - {\Omega}_A{}^B  \big){\p}_B 
- \big({\p}^B + {\ac}^B \big) {\btheta}_{BA} - \sD_A {\btheta} \\ 
& = -  (\Lie_k+\btheta)\p_A+  \sD_B (\btheta_A{}^B - \btheta \delta_A^B)
  - \vor_{AB}(\p^B+ \ac^B).
\end{aligned}
\end{equation}


\section{Derivation of the Pre-Symplectic Potential} \lb{App:potential}
We present here a derivation of the gravitational pre-symplectic potential \eqref{potential}. First, for the Einstein-Hilbert theory with the Lagrangian $\bm{L}^{\sss \mathrm{EH}} = \frac12 R \volM$, where $R$ denotes the spacetime Ricci scalar, the pre-symplectic potential $\Theta_{\sss \mathrm{EH}}$ is defined from the variation
\begin{align}
\delta \bm{L}^{\sss \mathrm{EH}} = \left(-\tfrac12 G^{ab}\delta g_{ab}\right)\volM + \exd\left( \Theta_{\sss \mathrm{EH}}^a \iota_a \volM \right) \qq \text{where} \qq \Theta_{\sss \mathrm{EH}}^a = \tfrac{1}{2} \left( g^{bc}\delta \Gamma^a_{bc} - g^{ab}\delta \Gamma^c_{bc} \right).
\end{align}
When the potential is pulled back to the stretched horizon $\H$, we have $\Theta^{\sss \mathrm{EH}}_\H = -\int_\H \Theta_{\sss \mathrm{EH}}^a n_a \volH$, since $\volH = -\iota_k \volM$. 

To derive \eqref{potential}, we start from the news scalar, $\Wein =  (\nabla_b n^a)\Pi_a{}^b = \nabla_a n^a - \kappa$, then take the field variation.  For the divergence term, we simply use $\nabla_a n^a = \frac{1}{2}\left(\nabla_a n^a + g^{ab}\nabla_a n_b \right)$ to show that 
\begin{equation}
\begin{aligned}
\delta (\nabla_a n^a) &=  \tfrac{1}{2}\left( \delta (\nabla_a n^a) -\nabla^a n^b \delta g_{ab} + g^{ab} \delta (\nabla_a n_b)  \right) \\
&=  \tfrac{1}{2}\left( \nabla_a \delta n^a -\nabla^a n^b \delta g_{ab} \right) + \tfrac{1}{2}\left( g^{ab}\delta \Gamma^c_{bc} - g^{bc}\delta \Gamma^a_{bc} \right)n_a \\
& = \tfrac{1}{2}\left( \nabla_a \delta \v^a + 2(k+\btheta)[\delta \rho] -\nabla^a n^b \delta g_{ab} \right) -\Theta_{\sss \mathrm{EH}}^an_a,
\end{aligned}
\end{equation}
where in the derivation, we used that $v^a=n^a + 2\rho k^a$ and that the rigging structure is preserved, i.e., $\delta n_a =\delta k^a =0$. To obtain the third equality, we also used $\nabla_a (\delta \rho k^a) = (k+\btheta)[\delta \rho]$. It also follows from $\delta \kappa = k[\delta \rho]$ that the variation of the news scalar is 
\begin{equation}
\begin{aligned}
\delta \Wein & = \tfrac12\nabla_a \delta \v^a + \btheta \delta \rho -\tfrac12\nabla^a n^b \delta g_{ab} -\Theta_{\sss \mathrm{EH}}^an_a.
\end{aligned}
\end{equation}
By using the variation of the volume form, $\delta \volH = -\iota_k \delta \volM = \left(\frac12 g^{ab}\delta g_{ab}\right) \volH$, together with the fact that $\delta \v^a$ is purely tangent to $\H$ (simply follows from $n_a \delta \v^a = - \v^a \delta n_a =0$), that is $\delta \v^a = (\delta v^b)\Pi_b{}^a$, we can write the pre-symplectic potential as
\begin{equation}
\begin{aligned}
\Theta^{\sss \mathrm{EH}}_\H = \int_\H  \bigg[\tfrac12\left(\nabla^a n^b - \Wein g^{ab}\right) \delta g_{ab} -\btheta \delta \rho \bigg] \volH+ \delta \int_\H \Wein \volH - \tfrac12\int_{\pa \H}\delta \v^a \bvol_a,
\end{aligned}
\end{equation}
where we employed the Stokes theorem \eqref{Stokes}. The pre-symplectic potential is already in the form of \eqref{potential}, where the canonical pre-symplectic potential is given by
\begin{equation}
\begin{aligned}
\Theta^{ \mathrm{can}}_\H &= \int_\H  \bigg[\tfrac12\left(\nabla^a n^b - \Wein g^{ab}\right) \delta g_{ab} -\btheta \delta \rho \bigg] \volH \\
&=  \int_{\H} \bigg[ 
-\left( \E \v^a + \J^a \right)  \delta k_a - \p_a \delta \v^a  + \tfrac{1}{2}(\T^{ab} + \P q^{ab}) \delta q_{ab}  -\btheta \delta \rho \bigg] \volH,
\end{aligned}
\end{equation}
where the final result was obtained by using the decomposition of the metric $\delta g_{ab}$ in terms of the variation of the sCarrollian structure $(\delta \v^a, \delta k_a, \delta q_{ab}, \delta \rho)$ and the decomposition \eqref{nabla-n} $\nabla^a n^b$, together with the relations $\NN^{ab} = \T^{ab} +\frac12 \E q^{ab}$ and $\Wein = \kappa + \E = \frac{1}{2}\E - \P$. 

\section{Symmetry Transformations} \lb{App:symmetries}

In this section, we derive the symmetry transformations presented in section \ref{sec:symmetry}. First, let us emphasize that the rigging structure is a background structure, inferring that our variation is restricted to satisfy 
\be
\delta n_a=0, \qq \text{and} \qq \delta k^a =0, \qq \text{so that} \qq \delta \Pi_a{}^b=0. 
\ee
These assumptions mean that 
\be
\delta k_a k^a =0,\qq \text{and} \qq \delta q_{ab} k^b=0,
\ee
which in turn imply that the variation is restricted to be such that
\begin{subequations}
\begin{align}
\delta  k_a = k_a \delta k_\v + e_a{}^A \delta k_A, \ \ \ \delta  \v^a = -(\delta k_\v)  v^a + \delta v^A  e_A{}^a, \ \ \ \delta q_{ab}  = -2   k_{(a} e_{b) A} \delta v^A + e_a{}^A e_b{}^B \delta q_{AB},
\end{align}
\end{subequations} 
where we defined $e_{bA} = e_b{}^B q_{B A}$ and used the notations: $\delta k_\v := \v^a \delta k_a$, $\delta k_A := e_A{}^a \delta k_a$, $\delta v^A := \delta v^a e_a{}^A$, and $\delta q_{AB} := e_A{}^a e_B{}^b\delta q_{ab}$.
The general variation of a spacetime field can be written as the Lie derivative action plus the anomaly action,
\be 
\delta_\xi =\Lie_\xi +\Delta_\xi.
\ee 
The fact that $n_a$ and $k^a$ are preserved by the variation means that their anomaly under infinitesimal diffeomorphism $\xi = T \v + R k + X^A e_A$ are 
\be
\Delta_\xi n_a=-\Lie_\xi n_a=-\pa_a R, \qq \text{and} \qq \Delta_\xi k^a =-\Lie_\xi k^a = \pa_r \xi^a= [k,\xi]^a. 
\ee
To understand the anomaly for the rigging vector $k^a$, we use that 
\bea
[k,\xi]
&=& k[T] \v  + k[R] k  + Z^A e_A,
\eea
where we introduced the combination 
\begin{align}
Z^A := \Lie_k X^A+  T \Upsilon^A \qq \text{or} \qq Z^a= \Lie_k X^a +T \Upsilon^a = \pa_r X^a +T \Upsilon^a, \lb{Z-def}
\end{align} 
where $Z^a=Z^A e_A{}^a$. The anomaly of the projector is then
\be
\Delta_\xi \Pi_a{}^b =  (D_a R)k^b - n_a (k[T]\v^b+Z^b). \lb{Delta-Pi}
\ee
One can therefore evaluate the anomaly of the Carrollian vector and Ehresmann connection  $(v^a,k_a)$ using that
$\Delta_\xi n^a = g^{ab} \Delta_\xi n_b$, 
$\Delta_\xi k_a = g_{ab} \Delta_\xi k^b$ and $ \Delta_\xi v^a= \Delta_\xi (n_c g^{cb}  \Pi_b{}^a) $.
This implies that 
\begin{align}
\Delta_\xi n^a &= -  \nabla^a R= n^a k[R] + k^a v[R] + \sD^a R     \\
\Delta_\xi k_a &= k[T] v_a +   k[R]  k_a + Z_a  \\
\Delta_\xi v^a &= - \pa_c R g^{cb}  \Pi_b{}^a  - n^b  \Delta_\xi n_b  k^a- 2\rho \Delta_\xi k^a \nonumber \\
&=  k^a n[R]- v^a k[R]  - \sD^a R -  2\rho (k[T] \v^a  + k[R] k^a  + Z^a) \nonumber \\
&= k^a \v[R]- v^a (k[R] +2\rho k[T]) - (\sD^a R+ 2\rho Z^a). 
\end{align}
Using $q_a{}^b = \Pi_a{}^b - k_a \v^b$, we can also compute the anomaly of the horizontal projector,
\begin{equation}
\lb{Delta-q-0}
\begin{aligned}
\Delta_\xi q_a{}^b &= \Delta_\xi\Pi_a{}^b - (\Delta_\xi k_a) \v^b - k_a (\Delta_\xi \v^b) \\
& = k^b \sD_a R +k_a\sD^b R -\v_a Z^b - Z_a \v^b - 2k[T]\v_a \v^b. 
\end{aligned}
\end{equation}

\subsection{Gauge-preserving conditions}

The variation $\delta_\xi$ and therefore the anomaly $\Delta_\xi$ needs to preserve the null condition $ g_{ab} k^ak^b=0$ and the gauge \eqref{gauge1}, implying that the acceleration  of the rigging vector $\Lie_k \bm{k}=0$ vanishes.
For the null condition, this infers 
\bea
\Delta_\xi (g_{ab} k^a k^b) = 2 k^a \Delta_\xi k_a = 2 k[T]=0.
\eea
For the acceleration, this implies that 
\bea
\Delta_\xi (\iota_k \rd \bm{k}) &=&  k[R] (\iota_k \rd \bm{k}) + Z^A (\iota_{e_A} \rd \bm{k}) + \iota_k \rd ( k[R]  \bm{k} + \bm{Z})\cr
&=& Z^A ( \varphi_A \bm{k} - \vor_{AB} \bm{e}^B) + k^2[R] \bm{k} + \Lie_k  \bm{Z}\cr
&=& \Lie_k \bm{Z} + \left( k^2[R]+Z^A \ac_A \right) \bm{k} - \vor_{AB} \bm{e}^B \cr
&=& 0.
\eea
where $k^2[R]=\pa_r^2R$. The Lie derivative $\Lie_k \bm{Z}$ is not completely horizontal and we can show, using \eqref{gauge2}, that
\begin{align}
\Lie_k  \bm{Z} =  \Lie_k (Z_A \bm{e}^A) &= k[Z_A] \bm{e}^A + Z_A (\iota_k  \rd \bm{e}^A)  = k[Z_A] \bm{e}^A + Z_A (- \Upsilon^A \bm{k} + \Omega_B{}^A \bm{e}^B)
\end{align}
Then, the gauge-preserving conditions are
\be
 k[T] = 0, \qquad k^2[R] -2 Z^A \p_A =0, \qq \text{and} \qq \Lie_k Z_A=  Z^B\vor_{BA}, \lb{gauge-preserving}
\ee
where we used $2\p_A = \Upsilon_A- \varphi_A$. Note that the last condition can be rewritten as $\Lie_k Z^A = -2Z^B \btheta_B{}^A$, while the second condition can be written as 
\be 
\pa_r R =\W, \qquad \text{and} \qquad \pa_r W =2Z^A\p_A,
\ee
where we introduced the Weyl transformation parameter $\W=\pa_r R$.

\subsection{Variations of the sCarrollian structure}
We now derive the variations of the sCarrollian structure $(\v^a, k_a, q_{ab},\rho)$ under the gauge-preserving diffeomorphism $\xi$. We first evaluate the Lie derivative of the Carrollian vector and the Ehresmann connection along $\xi$. They are given by
\begin{align}
\Lie_\xi k_a &=  \left(\v[T] + X^A\ac_A \right) k_a + \left( (\sD_A-  \ac_A)T +\vor_{AB} X^B  \right) {e}_a{}^A,\\
\Lie_\xi v^a&= -\v [R] k^a -  \left(\v [T] + X^A \varphi_A \right)  \v^a - \left(\v[X^A] - R\Upsilon^A\right)  e_A{}^a.
\end{align}
And therefore their variations under $\xi$ can be computed from $\delta_\xi = \Delta_\xi + \Lie_\xi$
\begin{align}
\delta_\xi k_a &= \left(\v[T] + X^A\ac_A+ k[R]\right) k_a + \left( (\sD_a-  \ac_a)[T] +\vor_{ab} X^b  + Z_a \right), \\
\delta_\xi \v^a &= - \left(\v [T] + X^A \varphi_A + k[R] \right)  \v^a -  \left(\v[X^A] + (\sD^A- \Upsilon^A) R + 2\rho Z^A \right)  e_A{}^a.
\end{align}

One can also evaluate the anomaly and the variation of the scalar norm $\rho$ as follows,
\bea
\Delta_\xi \rho &=& \tfrac12 g^{ab} \Delta_\xi (n_a n_b) = n^b \Delta_\xi n_b =- v[R] -2\rho k[R],\cr
\Lie_\xi \rho &=& R k[\rho] + T v[\rho]  + X^A \sD_A\rho\cr
\delta_\xi \rho &=& - (v +2\rho k- \kappa)[R] + (T v  + X^A e_A)[\rho]\eea
where we used the expression for the surface gravity $\kappa=k[\rho]$.

For the Carrollian metric $q_{ab}$, we use that the spacetime metric is anomaly-free, $\Delta_\xi g_{ab} =0$, and the form of $\Delta_\xi q_a{}^b$ given in \eqref{Delta-q-0}, and show that 
\bea
\lb{Delta-q}
\Delta_\xi q_{ab}  
&=& k_a (\sD_b R + 2\rho Z_b) - n_a Z_b + (a \leftrightarrow b) \cr
\Delta_\xi q^{ab}  
&=&  - v^a Z^b + k^a \sD^b R + (a \leftrightarrow b).
\eea
The Lie derivative of $q_{ab}$ follows the formulae given in Appendix \ref{App:Lie},
\bea
\Lie_\xi q_{ab} &=& 2T \left(\theta_{(ab)} + \Upsilon_{(a} n_{b)}  \right) + 2R \left(\btheta_{(ab)} -  \Upsilon_{(a} k_{b)}  \right) + 2 \sD_{(a} X_{b)} + 2 (\Lie_\v X^c) q_{c(a} k_{b)} + 2 (\Lie_k X^c) q_{c(a} n_{b)} \cr
&=& 2 \left( T \theta_{(ab)} + R \btheta_{(ab)} + \sD_{(a} X_{b)} \right) + 2 \left( \Lie_\v X^c - R\Upsilon^c\right) q_{c(a} k_{b)} + 2 Z_{(a} n_{b)} \cr
&=& 2 \left( T \theta_{(ab)} + R \btheta_{(ab)} + \sD_{(a} X_{b)} \right) - 2 \left( \Lie_\xi v^c\right) q_{c(a} k_{b)} + 2 Z_{(a} n_{b)}. 
\eea
Finally, its variation is 
\bea
\delta_\xi q_{ab}  &=& 2 \left( T \theta_{(ab)} + R \btheta_{(ab)} + \sD_{(a} X_{b)} \right) + 2 \left( \Lie_\v X^c +(\sD^c - \Upsilon^c) R + 2\rho Z^c \right) q_{c(a} k_{b)}  \cr
&=& 2 \left( T \theta_{(ab)} + R \btheta_{(ab)} + \sD_{(a} X_{b)} \right) - 2 \left( \delta_\xi v^c\right) q_{c(a} k_{b)}. 
\eea

Lastly, we consider the transformations of the frame fields $(e_A, \bm{e}^A)$. Firstly, we note that the transformations of the sCarrollian fields we derived previously are not enough to completely fix the transformations of both the frame $e_A$ and the coframe $\bm{e}^A$. This can be seen by considering $(\delta_\xi q_a{}^b) e_b{}^A$ and $e_A{}^b(\delta_\xi q_b{}^a)$, which can be written as
\begin{align}
(\delta_\xi q_a{}^b) e_b{}^A = \delta_\xi e_a{}^A - q_a{}^b (\delta_\xi e_b{}^A) \qq \text{and} \qq e_A{}^b(\delta_\xi q_b{}^a) = \delta_\xi e_A{}^a - (\delta_\xi e_A{}^b)q_b{}^a,
\end{align}
inferring that only their vertical and transverse components are determined from $\delta_\xi q_a{}^b$ (also true for the anomaly and the Lie derivative). To derive their transformations, we need to impose the gauge condition \eqref{gauge2}, that is $\dH( \delta_\xi \bm{e}^A) =0$. The solution to this gauge-preserving equation is that
\begin{align}
\delta_\xi \bm{e}^A = \dH f^A_\xi \qq \text{where} \qq \v[f^A_\xi] = - \delta_\xi \v^A =\left(\v[X^A] + (\sD^A- \Upsilon^A) R + 2\rho Z^A \right). \lb{delta-coframe}
\end{align}
Let us observe that the horizontal components of $\delta_\xi \bm{e}^A$, which are $e_B{}^a\delta_\xi e_a{}^A = e_B[f^A_\xi]$, are non-local as one first needs to integrate the above equation for $f^A_\xi$. From this result, one can derive the symmetry transformation of the frame $e_A{}^a$, given by
\begin{equation}
\lb{delta-frame}
\begin{aligned}
\delta_\xi e_A{}^a &= - (e_A{}^b \delta_\xi k_b) \v^a -(e_A{}^b \delta_\xi n_b) k^a - (e_A{}^b \delta_\xi e_b{}^B) e_B{}^a \\
& = - \left( (\sD_A-  \ac_A)T +\vor_{AB} X^B  + Z_A \right) \v^a - e_A[f^B_\xi] e_B{}^a.
\end{aligned}
\end{equation}

\subsection{Anomaly and variation of the stress tensor} \lb{app:var-em}
In this section, we use that $k[T]=0$.
Finding the anomaly and variation of the Carrollian stress tensor $\EM_a{}^b$ boils down to finding the anomaly and variation of the news tensor $\Wein_a{}^b$. By exploiting the fact that the spacetime covariant derivative commutes with the anomaly operator, $\Delta_\xi \nabla_a = \nabla_a \Delta_\xi$, we derive the anomaly of the news tensor as follows, 
\bea
\Delta_\xi \Wein_a{}^b & = & \left(\Delta_\xi \Pi_a{}^c\right) \nabla_c n^d \Pi_d{}^b + \Pi_a{}^c \nabla_c n^d  \left( \Delta_\xi \Pi_d{}^b \right) + \Pi_a{}^c \nabla_c \left(\Delta_\xi n^d\right) \Pi_d{}^b  \cr
& =& D_a R (\nabla_k n^c) \Pi_c{}^b - n_a Z^c \Wein_c{}^b  + \left(\Wein_a{}^c D_c R\right) k^b - D_a \rho Z^b - k[R] \Wein_a{}^b \cr
&& - D_a\left( k[R] \right) \v^b - \v[R]\Pi_a{}^c (\nabla_c k^d) \Pi_d{}^b - D_a \sD^b R \cr
& =& - k[R] \Wein_a{}^b - n_a Z^c \Wein_c{}^b  + \left(\Wein_a{}^c D_c R\right) k^b + D_a R \p^b - D_a \rho Z^b \cr
&&  - D_a\left( k[R] \right) \v^b - \v[R] (\btheta_a{}^b  - k_a (\p^b + \ac^b))  - D_a \sD^b R \cr
& =& - \left(\v^b D_a +\Wein_a{}^b\right)k[R] - \v[R] \left(\btheta_a{}^b- k_a\Upsilon^b \right)  + \left(\Wein_a{}^c D_c R\right) k^b + \sD_a R \p^b  \cr
&&  - D_a \sD^b R- n_a Z^c \Wein_c{}^b- D_a \rho Z^b,
\eea
 where we used the anomaly of the projector \eqref{Delta-Pi}.
 In the first equality we used that 
 \bee  
 \Pi_a{}^c (\nabla_c \nabla^d R)  \Pi_d{}^b &= \Pi_a{}^c \nabla_c( n^d k[R] + k^d v[R] + \sD^d R)  \Pi_d{}^b \cr
 &= k[R] \Wein_a{}^b+ D_a\left( k[R] \right) \v^b + \v[R]\Pi_a{}^c (\nabla_c k^d) \Pi_d{}^b + D_a \sD^b R 
 \eee
  The third equality was obtained using the decompositions \eqref{nabla-k} and \eqref{nabla-n} of the spacetime covariant derivatives $\nabla_a k^b$ and $\nabla_a n^b$, respectively. In the last equality, we used the relation \eqref{momentum-gen}, $2\p^a + \ac^a = \Upsilon^a$. We now evaluate the term $D_a \sD^b R$. Using the decomposition $\Pi_a{}^b = q_a{}^b + k_a\v^b$, we have that
 \begin{equation}
 \begin{aligned}
 D_a \sD^b R &= \Pi_a{}^c \nabla_c \left(\sD^d R\right) \Pi_d{}^b\\
 & = q_a{}^c \nabla_c (\sD^d R) q_d{}^b + k_d\left(q_a{}^c\nabla_c (\sD^d R) + k_a \nabla_\v (\sD^d R) \right)\v^b + k_a \nabla_\v (\sD^d R) q_d{}^b \\
  & = q_a{}^c \nabla_c (\sD^d R) q_d{}^b - (\sD_d R)\left(q_a{}^c\nabla_c k^d+ k_a \nabla_\v k^d \right)\v^b + k_a \nabla_\v (\sD^dR) q_d{}^b \\
  & = \sD_a \sD^b R  -  \left( \btheta_a{}^c - k_a (\p^c + \ac^c) \right)(\sD_cR)  \v^b + k_a \nabla_\v (\sD^dR) q_d{}^b, 
 \end{aligned}
 \end{equation}
where on the third equality, we used the Leibniz rule and that $k_a \sD^a R =0$, and we used \eqref{nabla-k} and \eqref{sD-X} to obtain the last equality. Note the notation, $\sD_a \sD^b R  = (\sD_A \sD^B R) e^A{}_a e_B{}^b$. The last term can be expressed as 
\begin{equation}
\begin{aligned}
\nabla_\v (\sD^a R) &= \nabla_\v (q^{AB} e_B [R] e_A{}^a)  \\
& = \left(\v[q^{AB}] e_B [R]  + q^{AB} \v \big[ e_B[R]\big] \right)e_A^a + q^{AB}e_B [R] \nabla_\v e_A{}^a \\
& = \left(-2\theta^{(AB)} \sD_B R + (\sD^A+ \ac^A)\v[R]\right)e_A{}^a   + (\sD^A R) \left( (\p_A+\ac_A)\v^a + \theta_A{}^a +\J_A k^a \right) \\
& = -\theta^{ab} \sD_b R + \left( \sD^a + \ac^a \right) \v[R] + \left( \p^b +\ac^b \right) \sD_b R \v^a + \J^b \sD_b R k^a.
\end{aligned}
\end{equation}
With this, we obtain, 
 \begin{equation}
 \begin{aligned}
 D_a \sD^b R 
 & = \sD_a \sD^b R  -  \left( \btheta_a{}^c - k_a (\p^c + \ac^c) \right)(\sD_c R)  \v^b - k_a \left(\theta^{bc}\sD_c R - ( \sD^b + \ac^b ) \v[R] \right).
 \end{aligned}
 \end{equation}
We finally arrive at the anomaly of the news tensor,
\begin{equation}
\lb{Delta-Wein}
\begin{split}
\Delta_\xi \Wein_a{}^b 
 =& - \left(\v^b D_a +\Wein_a{}^b\right)k[R] -  \left(\btheta_a{}^b+ k_a(\sD^b - 2\p^b) \right)\v[R]  + \left(\Wein_a{}^c D_c R\right) k^b   \cr
& +\sD_a R \p^b   - \sD_a \sD^b R + \left( \btheta_a{}^c - k_a (\p^c + \ac^c) \right)(\sD_c R)  \v^b - k_a \left(\theta^{bc}\sD_c R  \right) \cr
&  - n_a Z^c \Wein_c{}^b- D_a \rho Z^b. 
\end{split}
\end{equation}
The variation of the news tensor is then 
\begin{equation}
\lb{delta-Wein}
\begin{split}
\delta_\xi \Wein_a{}^b 
 = \ & \Lie_\xi \Wein_a{}^b - \left(\v^b D_a +\Wein_a{}^b\right)k[R] -  \left(\btheta_a{}^b+ k_a(\sD^b - 2\p^b) \right)\v[R]  +\sD_a R \p^b   \cr
&   - \sD_a \sD^b R + \left( \btheta_a{}^c - k_a (\p^c + \ac^c) \right)(\sD_c R)  \v^b - k_a \left(\theta^{bc}\sD_c R  \right) - D_a \rho Z^b \cr
& + \left(\Wein_a{}^c D_c R\right) k^b - n_a Z^c \Wein_c{}^b \\
= \ & \LH_\xi \Wein_a{}^b - \left(\v^b D_a +\Wein_a{}^b\right)k[R] -  \left(\btheta_a{}^b+ k_a(\sD^b - 2\p^b) \right)\v[R]  +\sD_a R \p^b   \cr
&   - \sD_a \sD^b R + \left( \btheta_a{}^c - k_a (\p^c + \ac^c) \right)(\sD_c R)  \v^b - k_a \left(\theta^{bc}\sD_c R  \right) - D_a \rho Z^b,\cr
= \ & \LH_\xi \Wein_a{}^b - \left(\v^b D_a +\Wein_a{}^b\right)\W - D_a \rho Z^b 
-\left( \sD_a \sD^b R + \btheta_a{}^b \v[R]  -\sD_a R \p^b\right)
\cr
&    
+ \left(\btheta_a{}^c\sD_c R  - k_a (\p^c + \ac^c)(\sD_c R)\right)  \v^b  -k_a\left[(\sD^b - 2\p^b)\v[R] +  \left(\theta^{bc}\sD_c R  \right) \right] 
\end{split}
\end{equation}
where we have  defined the Lie derivative projected onto the stretched horizon $\H$ as
\begin{align}
\LH_\xi \Wein_a{}^b := \Pi_a{}^c (\Lie_\xi \Wein_c{}^d) \Pi_d{}^b  = \Lie_\xi \Wein_a{}^b + (\Wein_a{}^c D_c R)k^b - n_a Z^c \Wein_c{}^b. 
\end{align}
This definition can be generalized to a tensor of arbitrary degree.

For a vector $\hat{\xi}$ tangent to $\H$, and radially constant, $[k,\hat\xi]=0$, we have that $\Lie_{\hat\xi}\Pi_a{}^b=0 $ and therefore 
\be
\LH_{\hat\xi} = \Lie_{\hat\xi}.
\ee 
On the other hand, when $\xi = Rk$ we have that 
\bee
\LH_{R k }\Wein_a{}^b &=  \Pi_a{}^c( R\nabla_k\Wein_c{}^d + \nabla_c(Rk^e) \Wein_e{}^d -\Wein_c{}^e \nabla_e(Rk^d) ) \Pi_d{}^b
\cr
&=
R\LH_{ k }\Wein_a{}^b = R \Lie_k \Wein_a{}^b.
\eee
In the last equality, we use that $\Lie_k n_a=0=\Lie_kk^a$.

The anomaly and the variation of the news scalar, $\Wein = \Wein_a{}^a$, is given by
\begin{align}
\Delta_\xi \Wein  & = - \left(\v +\Wein \right)\big[k[R]\big] - \btheta\v[R]    - (\sD_a + \ac_a) \sD^a R  - Z^a \sD_a \rho \lb{Delta-News} \\
\delta_\xi \Wein  & = \Lie_\xi \Wein- \left(\v +\Wein \right)\big[k[R]\big] - \btheta\v[R]    - (\sD_a + \ac_a) \sD^a R  - Z^a \sD_a \rho. \lb{delta-News} 
\end{align}

From these results, we can derive the variation of Carrollian fluid quantities by recalling that the news tensor decomposes as
\begin{align}
\Wein_a{}^b = \NN_a{}^b + (\p_a + \kappa k_a) \v^b - k_a \J^b,
\end{align}
where $\NN_a{}^b = \theta_a{}^b + 2\rho \btheta_a{}^b = \T_a{}^b  + \frac{1}{2}\E q_a{}^b$ and $\kappa = - \P - \frac{1}{2}\E$. 

\paragraph{Fluid energy and viscous stress tensor:} We first need to compute the anomaly $\Delta_\xi \NN_{ab}$. Using $\NN_{ab}= q_a{}^c \Wein_c{}^d q_{db}$, the anomaly of the news tensor \eqref{Delta-Wein} and the anomaly of the Carrollian metric \eqref{Delta-q}, we can show that 
\begin{equation}
\begin{aligned}
\Delta_\xi \NN_{ab} = \ & q_a{}^c (\Delta_\xi \Wein_c{}^d) q_{db} + q_a{}^c\Wein_c{}^d (\Delta_\xi q_{db}) + (\Delta_\xi q_a{}^c) \Wein_c{}^d q_{db} \\
= \ & - k[R] \NN_{ab} - \v[R]\btheta_{(ab)} - (\sD-2\p)_{(a} \sD_{b)} R 
+2\J_{(a} Z_{b)} + 2k_{(a}\NN_{b)}{}^c \sD_c R  -  2\v_{(a}\NN_{b)c} Z^c.
\end{aligned}
\end{equation}
Note that, to arrive at the result, we used $\sD_a\sD_b R = \sD_{(a}\sD_{b)} R + \sD_{[a}\sD_{b]} R = \sD_{(a}\sD_{b)} R + \frac{1}{2}\vor_{ab} \v[R]$, following from the commutator \eqref{com-4}, and that $\btheta_{ab} = \btheta_{(ab)} - \frac{1}{2}\vor_{ab}$. 

The anomaly and the variation of the fluid energy are 
\begin{align}
\Delta_\xi \E  & =  - k[R] \E- \v[R]\btheta - (\sD-2\p)_a \sD^a R 
+2\J_a Z^a \\
\delta_\xi \E  & = \Lie_\xi \E -  k[R] \E- \v[R]\btheta - (\sD-2\p)_a \sD^a R 
+2\J_a Z^a.
\end{align}
For the anomaly and the variation of $\T_{ab} = \NN_{ab} - \frac{1}{2}\E q_{ab}$, we can show that
\begin{align}
\Delta_\xi \T_{ab}   = \ &  - k[R] \T_{ab}- \v[R]\bs_{ab} - (\sD-2\p)_{\la a} \sD_{b \ra} R +2\J_{\la a} Z_{b \ra} + 2 k_{(a} \T_{b)}{}^c \sD_c R  - 2\v_{(a} \T_{b)c} Z^c\\
\delta_\xi \T_{ab}  = \ & \left(\LH_\xi - k[R] \right) \T_{ab} - \v[R]\bs_{ab} - (\sD-2\p)_{\la a} \sD_{b \ra} R +2\J_{\la a} Z_{b \ra} + 2 k_{(a} \T_{b)}{}^c (\sD_c R + 2\rho Z_c) \\
\delta_\xi \T_a{}^b  = \ & \left(\LH_\xi - k[R] \right) \T_a{}^b - \v[R]\bs_a{}^b - (\sD-2\p)_{\la a} \sD^{b \ra} R +2\J_{\la a} Z^{b \ra} +  k_{a} \T^{bc} (\sD_c R + 2\rho Z_c) - v^b \T_{ac} Z^c .
\end{align}

\paragraph{Fluid momentum:} It follows from $\p_a = q_a{}^b \Wein_b{}^c k_c$ that the anomaly of $\p_a$ is given by
\bea
\Delta_\xi \p_a 
&= & q_a{}^c (\Delta_\xi \Wein_c{}^b) k_b + (\Delta_\xi q_a{}^c )\Wein_c{}^b k_b + q_a{}^c \Wein_c{}^b (\Delta_\xi k_b) \cr
&=& -\sD_a k[R] + \left( \T_{ab} + (\E+\P)q_{ab} \right)Z^b +\btheta_a{}^b \sD_b R + k_a \p^b  \sD_b R  -\v_a(Z^b\p_b).  
\eea
Using $\LH_\xi \p_a := \Pi_a{}^b \Lie_\xi \p_b = \Lie_\xi \p_a - (Z^b \p_b) n_a$, the variation of the fluid momentum reads
\bea
\delta_\xi \p_a 
&=& \LH_\xi \p_a -\sD_a k[R] + \left( \T_{ab} + (\E+\P)q_{ab} \right)Z^b +\btheta_a{}^b \sD_b R + k_a \p^b  (\sD_b R +2\rho Z_b)  .  
\eea

\paragraph{Heat current:} Using the fact that $\J^a = - \v^c \Wein_c{}^b q_b{}^a$ one can show that the anomaly is given by
\begin{flalign}
\Delta_\xi \J^a 
&=   - \v^c (\Delta_\xi \EM_c{}^b) q_b{}^a - (\Delta_\xi \v^c)\EM_c{}^b q_b{}^a - \v^c \EM_c{}^b (\Delta_\xi q_b{}^a) & \nonumber \\
&=  - \left( \sD^a - 2\p^a \right) \v[R] + ( \T^{ab}+(\E+\P)q^{ab}) (\sD_b R+2\rho Z_b) - \theta^{ab}\sD_b R \nonumber \\ 
& \ \ \ \ - v[\rho] Z^a - \J^bZ_b \v^a  + (\J^b \sD_b R) k^a, 
\end{flalign}
and the corresponding variation is
\begin{flalign}
&\delta_\xi \J^a 
= \LH_\xi \J^a  - \left( \sD^a - 2\p^a \right) \v[R] + ( \T^{ab}+(\E+\P)q^{ab}) (\sD_b R+2\rho Z_b) - \theta^{ab}\sD_b R - v[\rho] Z^a - \J^bZ_b \v^a,  &&
\end{flalign}
where we used that $\LH_\xi \J^a := \Lie_\xi \J^b \Pi_b{}^a = \Lie_\xi \J^a + (\J^b \sD_b R) k^a$. 

\paragraph{Pressure:} Lastly, for the pressure term, we can use the relation $\Wein = \frac{1}{2}\E -\P$ to show that 
\bea
\Delta_\xi \P
&=& - k[R] \P + \left( k + \tfrac{1}{2}\btheta \right)\left[ v[R]\right] + \tfrac{1}{2}(\sD_a - 2\p_a) \sD^a R +2\rho Z^a \p_a \\ 
\delta_\xi \P
&=& \Lie_\xi \P  - k[R] \P + \left( k + \tfrac{1}{2}\btheta \right)\left[ v[R]\right] + \tfrac{1}{2}(\sD_a - 2\p_a) \sD^a R +2\rho Z^a \p_a.
\eea
In summary, we have that
\begin{alignat}{3}
& \delta_\xi \E  
&& = \Lie_\xi \E -  \W \E- \v[R]\btheta - (\sD-2\p)_a \sD^a R 
+2\J_a Z^a && \nonumber \\
& \delta_\xi \P 
&&=  \Lie_\xi \P  - \W \P + \left( k + \tfrac{1}{2}\btheta \right)\left[ v[R]\right] + \tfrac{1}{2}(\sD_a - 2\p_a) \sD^a R +2\rho Z^a \p_a && \nonumber\\
&\delta_\xi \p_a 
&&= \LH_\xi \p_a -\sD_a \W + \left( \T_{ab} + (\E+\P)q_{ab} \right)Z^b +\btheta_a{}^b \sD_b R + k_a \p^b  (\sD_b R +2\rho Z_b)&& \nonumber\\
&\delta_\xi \J^a 
&&= \LH_\xi \J^a  - \left( \sD^a - 2\p^a \right) \v[R] + ( \T^{ab}+(\E+\P)q^{ab}) (\sD_b R+2\rho Z_b) - \theta^{ab}\sD_b R - v[\rho] Z^a - \J^bZ_b \v^a &&\nonumber \\
&\delta_\xi \T_a{}^b  
&&=  \left(\LH_\xi - \W \right) \T_a{}^b - \v[R]\bs_a{}^b - (\sD-2\p)_{\la a} \sD^{b \ra} R +2\J_{\la a} Z^{b \ra} +  k_{a} \T^{bc} (\sD_c R + 2\rho Z_c) - v^b \T_{ac} Z^c. &&
\end{alignat}
As we have already elaborated, when considering the diffeomorphism purely tangent to the surface, such that $R =0$ and $Z^A =0$, the sCarrollian momenta transform as a tensor on the stretched horizon $\H$.

\section{Canonical and Einstein-Hilbert Charges} \lb{App:charges}

Here, we make a connection between the canonical charge discussed in the main text and the Noether charge computed with the Einstein-Hilbert theory, in turn justifying the formula \eqref{Q-can} for the canonical charge. 

For the Einstein-Hilbert theory, $\bm{L}^{\sss \mathrm{EH}} = \tfrac12 R \volM$, the Noether charge can be computed from the contraction \eqref{Noether},
\begin{align}
Q^{\sss \mathrm{EH}}_\xi := I_\xi \Theta^{\sss \mathrm{EH}}_\H - \int_\H \iota_\xi \bm{L}^{\sss \mathrm{EH}} = C^{\sss \mathrm{EH}}_\xi + \int_{\pa \H} \bm{q}^{\sss \mathrm{EH}}_\xi.  
\end{align}
The constraint term is the Einstein tensor, $C^{\sss \mathrm{EH}}_\xi = -\int_\H \xi^b G_b{}^a \iota_a \volM$, and the charge aspect is given by the Komar superpotential 
$ \bm{q}^{\sss \mathrm{EH}}_\xi = \nabla^a \xi^b \bm{\epsilon}_{ab}$ \cite{Komar1959}. Using the splitting \eqref{potential} of the Einstein-Hilbert pre-symplectic potential into the canonical term, the total variation term, and the corner term, we can show that
\begin{equation}
\begin{aligned}
 I_\xi \Theta^{\sss \mathrm{EH}}_\H &= Q^{\sss \mathrm{EH}}_\xi + \int_\H \iota_\xi \bm{L}^{\sss \mathrm{EH}} \cr &= I_\xi \Theta^{\mathrm{can}}_\H + \int_\H \delta_\xi \bm{\ell}_\H + \int_{\pa\H} I_\xi \bm{\vartheta}_{\pa \H}   \\
& = I_\xi \Theta^{\mathrm{can}}_\H + \int_\H \left(\Delta_\xi \bm{\ell}_\H + \iota_\xi \exd \bm{\ell}_\H\right) + \int_{\pa\H} \left(\iota_\xi \bm{\ell}_\H + I_\xi \bm{\vartheta}_{\pa \H} \right)\\
& = C^{\mathrm{can}}_\xi + \int_{\pa \H} \left( \bm{q}^{\mathrm{can}}_\xi +\iota_\xi \bm{\ell}_\H + I_\xi \bm{\vartheta}_{\pa \H} \right),
\end{aligned}
\end{equation}
where we used $\delta_\xi = \Delta_\xi + \Lie_\xi$ and the Cartan formula $\Lie_\xi = \exd \iota_\xi + \iota_\xi \exd$ to obtain the second equation, then used \eqref{symmetries-def} to obtain the third equality. With this, we can infer the relation between the canonical charge and the Einstein-Hilbert charge,
\begin{align}
C^{\mathrm{can}}_\xi &= C^{\sss \mathrm{EH}}_\xi + \int_\H \iota_\xi \bm{L}^{\sss \mathrm{EH}} \\
\bm{q}^{\mathrm{can}}_\xi &= \bm{q}^{\sss \mathrm{EH}}_\xi - \iota_\xi \bm{\ell}_\H - I_\xi \bm{\vartheta}_{\pa \H}. \lb{Q-EH}
\end{align}
The constraint term computed from the canonical pre-symplectic potential differs from the Einstein-Hilbert constraint by the term $\int_\H \iota_\xi \bm{L}^{\sss \mathrm{EH}}$, which vanishes on-shell. Note also that only the transverse diffeomorphism $\xi_{\perp}$ contributes to the difference. As expected, the canonical charge aspect differs from the Einstein-Hilbert (Komar) charge aspect by the contributions from the boundary Lagrangian and the corner symplectic.

\section{Explicit Derivation of the Spin-2 Charges} \lb{app:spin2-derive}

In this Appendix, we demonstrate how to obtain the radial evolution of the horizontal diffeomorphism charge, $\Lie_k Q^{\mathrm{can}}_X$, given in \eqref{Lie-k-QX} by explicitly evaluating the radial evolution of the canonical charge \eqref{Q-can} from the radial evolution of the pre-symplectic potential \eqref{potential}, 
\begin{align}
\Lie_k Q^{\mathrm{can}}_X  = \Lie_k \left( I_{X} \Theta^{\mathrm{can}}_\H \right).
\end{align}
To evaluate the radial derivative of the canonical pre-symplectic potential \eqref{potential-can}, we need to calculate the radial derivative of the variation of the sCarrollian structure, $\delta_X( v^a, k_a,  q_{ab},\rho)$. Their radial expansions can be obtained by taking the Lie derivative $\Lie_k$ on the variations \eqref{sCarrollian-transform} with the condition $\Lie_k X^A =0$. The results are, 
\begin{subequations}
\lb{symmetries-sub}
\begin{align}
\Lie_k\big(\delta_X k_\v \big)  & = X_{\vor A}\Upsilon^A  \\
\Lie_k\big(\delta_X k_A \big)  & = 0 \\
\Lie_k \big(\delta_X \v^A\big) & =  \Lie_X\Upsilon^A  + {\Upsilon}^AX_\ac  \\
\Lie_k\big(\delta_X q_{AB}\big) & = 2
\Lie_X \btheta_{(AB)}- 2 X_{\vor (A} \Upsilon_{B)} \\
\Lie_k (\delta_X \rho) &= X^A \sD_A \kappa.
\end{align}
\end{subequations}
Let us provide some explanations. These radial evolution equations followed\footnote{For the variation $\delta_X \v^A = -\v[X^A]$, one can show using $[k,\v] = \Upsilon^A e_A$ that 
\begin{align*}
\Lie_k (\delta_X \v^A) = -k\left[ \v[X^A] \right] + \v[X^B]\Omega_B{}^A = - \Upsilon [X^A]- \v\left[k[X^A]\right] + \v[X^B]\Omega_B{}^A.
\end{align*}
The condition $\Lie_k X^A = k[X^A] - X^B\Omega_B{}^A = 0$ together with the Jacobi identity $\v[\Omega_A{}^B] + \left(e_A + \ac_A \right)[\Upsilon^B] =0$ implies $\Lie_k (\delta_X \v^A) = (X[\Upsilon^A] - \Upsilon[X^A]) + X^B \ac_B \Upsilon^A$.} from the commutators \eqref{com-4} and the Jacobi identity \eqref{Jacobi}. For $\Lie_k (\delta_X q_{AB})$, where we needed to evaluate $\Lie_k (\sD_{(A} X_{B)})$, the formula is given by the equation \eqref{Lie-k-DX}. Furthermore, one can also check that $\Lie_k (\delta_X \ln \sqrt{q}) = \Lie_X \btheta - X_\vor^A \Upsilon_A$. 

We consider the radial derivative of each term in \eqref{potential-can} individually as follows: 
\paragraph{Energy term:} Using the transverse evolution of $\delta k_\v$ \eqref{symmetries-sub} and the Lie derivative $\Lie_k\volH = \btheta \volH$, we can easily show
\begin{equation}
\begin{aligned}
\Lie_k\big[\E (\delta_X k_\v) \volH \big]  = \left[ X_\ac (\Lie_k + \btheta)\E + X_\vor^A\E (2\p_A + \ac_A)  \right]\volH.
\end{aligned}
\end{equation}
\paragraph{Heat term:} In a similar manner, we have for the heat current term,
\begin{equation}
\begin{aligned}
 \Lie_k\big[\J^A (\delta_X k_A)\volH \big] 
& = X_{\vor A}\left[ q^{AB} \Lie_k \J_B - 2\bs^{AB}\J_B  \right]\volH \\
& = X_\vor^{A}\left[(2\p_A - \sD_A) \kappa + 2\rho \Lie_k \p_A - 2\bs_{AB}\J^B  \right]\volH,
\end{aligned}
\end{equation}
where we used the expression \eqref{heat-gen} for the heat current, $\J_A = (2\p_A - \sD_A) \rho$, and that $k[\rho] = \kappa$.

\paragraph{Momentum term:} For the momentum sector, we can show that 
\begin{equation}
\begin{aligned}
&\Lie_k \left[\p_A  (\delta_X \v^A)\volH \right] \\
&\ = \bigg[ -\mr{X}^A\left( \Lie_k\p_A +\btheta\p_A  \right) + X_\ac \left(\p \cdot \Upsilon \right) +\p_A \Lie_X\Upsilon^A  \bigg]\volH\\
& \  = - \bigg[X_\sD^{\langle AB \rangle}\left( (\sD + {\ac})_{\langle A} {\ac}_{B\rangle} +2 {\p}_{\langle A} ({\p} + {\ac})_{B \rangle} - {\RC}_{\langle AB \rangle}\right)+ X_\sD \left( -\tfrac{1}{2}(\sD+{\ac}) \cdot {\ac} + {\p}\cdot {\Upsilon}\right) \\
& \qq \qq + X_\ac \left((\sD - {\p}) \cdot {\p}  + \tfrac{1}{2}{\RC}\right) + \mr{X}^A \left( \Lie_k\p_A +{\btheta} {\p}_A + {\vor}_{AB}({\p}^B + {\ac}^B) +\tfrac{1}{2} \sD_B {\vor}_A{}^B\right)  \\
& \qq \qq -X_\vor^A\left( \Lie_\v\p_A- 2 {\s}_A{}^B {\p}_B + \sD_B (\s_A{}^B- \tfrac{1}{2}\theta \delta_A^B) \right) \bigg]\volH \\
& \ \ \ \ \ \ - \dH \bigg[  \bigg(X^B(\sD_B {\p}^A+{\vor}_{BC} \theta^{(AC)}+{\RC}^A{}_B) -X_\sD^{BA}({\p}_B + {\ac}_B) +X_\sD ({\p}^A + {\ac}^A)  \\
&   \qquad \qquad \qq -X^A \big(\sD\cdot{\p} + {\p} \cdot ({\p} + {\ac}) + \tfrac{1}{2} \RC\big)  + \tfrac{1}{2} \mr{X}^B {\vor}^A{}_B \bigg) \bm{\epsilon}_A -\tfrac{1}{2}X_\sD^{[AB]}{\vor}_{AB} \volS \bigg].
\end{aligned}
\end{equation}
To obtain the final equality, we needed to manipulate the Lie derivative term, $\p_A \Lie_X \Upsilon^A$, which was highly non-trivial. We provide the derivation separately in Appendix \ref{App:momentum}.

\paragraph{Shear term:} This sector also requires tedious manipulation of Lie derivative. The final result (see the derivation in Appendix \ref{App:shear}) is
\begin{equation}
\begin{aligned}
& -\tfrac{1}{2} \Lie_k\left[\T^{AB} (\delta_X q_{AB})\volH \right] \\
& \ = \bigg[-X_\sD^{(AB)} \left(\Lie_k \T_{AB} - \btheta \T_{AB} -4 \T_{(A}{}^C \bs_{B)C} \right)-\T^{AB}\Lie_X \btheta_{AB}  - X_{\vor}^A \left(\T_A{}^B\Upsilon_B \right)  \bigg] \volH\\
& \ =\bigg[-X_\sD^{\langle AB\rangle} \left(  \Lie_\v \bs_{\la AB \ra} + \Lie_k \T_{\la AB \ra} +\rho \left( 2\btheta_{C\la A} \btheta_{B \ra}{}^C +  \btheta_{C \la A} \btheta^C{}_{B\ra}  +  \btheta_{\la A}{}^C \btheta_{B \ra C} -4\btheta \bs_{AB}\right)\right) \\
& \qquad \qq + X_\sD \left( \T^{AB} \bs_{AB}- \rho \bs_{AB} \bs^{AB}   \right) + X_\ac \left(\rho \bs_{AB} \bs^{AB}  \right) + \mr{X}^A\sD_B \bs_{A}{}^{B}  + X_\rho \left( \bs_{AB} \bs^{AB} \right) \\
& \qquad \qq + X_{\vor}^A \left(\T_A{}^B\Upsilon_B\right)   \bigg] \volH    + \dH \left[ X_\sD^{\langle AB \rangle }\bs_{AB} \volS - \left( \mr{X}^B \bs_{B}{}^{A}+(\T_{BC}-\rho \bs_{BC}) \bs^{BC} X^A \right) \bm{\epsilon}_A \right].
\end{aligned}
\end{equation}

\paragraph{The remaining terms:} For the pressure and the extrinsic curvature terms, we have
\begin{equation}
\begin{aligned}
& -\Lie_k \left[\P (\delta_X \ln \sqrt{q}) \volH- \btheta \delta_X \rho \volH) \right]  \\
& \ = \left[- X_\sD (\Lie_k \P + \btheta \P) - \P (\Lie_X \btheta - X_\vor^A \Upsilon_A ) + X_\rho (\Lie_k\btheta + \btheta^2) - \btheta\Lie_X (\P +\tfrac{1}{2}\E) \right] \volH\\
& \ = \left[X_\sD (\tfrac{1}{2}\Lie_\v \btheta-\Lie_k \P + \tfrac{1}{2}\E\btheta - \tfrac{1}{2}\rho\btheta^2 ) - X_\ac (\kappa \btheta + \tfrac{1}{2}\rho \btheta^2) + X_\vor^A (\P\Upsilon_A ) + X_\rho (\Lie_k\btheta + \tfrac{1}{2}\btheta^2) - \tfrac{1}{2}\mr{X}^A\sD_A\btheta \right] \volH \\
& \qquad -\dH \left[ \tfrac{1}{2}X_\sD \btheta \volS - \left( \tfrac{1}{2}\btheta \mr{X}^A- (\P \btheta+\tfrac{1}{2}\rho \btheta^2) X^A\right) \bm{\epsilon}_A \right].
\end{aligned}
\end{equation}
Putting all these results together and referring to the expressions for different components of the Einstein tensor \eqref{Einstein}, we can derive \eqref{Lie-k-QX}. 

\subsection{On the momentum term} \lb{App:momentum}
We would like to show how to manipulate the term $\p_A \Lie_X \Upsilon^A$. First, recalling the expression for the fluid momentum, $\p_A =  \frac{1}{2} (\Upsilon_A - \ac_A)$, we write 
\begin{align}
 2\p_A \Lie_X \Upsilon^A \volH = \left( \Upsilon_A \Lie_X \Upsilon^A-\ac_A \Lie_X \Upsilon^A \right)\volH. \lb{pi-Upsilon-term}
\end{align}
The first term can be written as
\begin{equation}
\begin{aligned}
\Upsilon_A \Lie_X \Upsilon^A\volH &= \left(\Upsilon_A X^B \sD_B \Upsilon^A - \Upsilon_A \Upsilon_B \sD^B X^A\right)\volH \\
& =  \tfrac{1}{2}X^A \sD_A \left( \Upsilon \cdot \Upsilon \right)\volH - \Upsilon_A \Upsilon_B \sD^{(A} X^{B)} \volH \\
& = \left(-X_\sD^{\langle AB \rangle} \Upsilon_{\langle A} \Upsilon_{B \rangle} - X_\sD \left( \Upsilon \cdot \Upsilon\right) - \tfrac{1}{2}X_\ac\left( \Upsilon \cdot \Upsilon \right) \right)\volH + \dH \left(\tfrac{1}{2}  (\Upsilon \cdot \Upsilon) X^A \bm{\epsilon}_A \right) 
\end{aligned}
\end{equation}
where we used Stokes theorem \eqref{Stokes} to arrive at the last equality. 
The second term in \eqref{pi-Upsilon-term} can be expressed as
\begin{equation}
\begin{aligned}
\ac_A \Lie_X \Upsilon^A \volH&= \left(\ac_A X^B \sD_B \Upsilon^A - \ac_A \Upsilon_B \sD^B X^A\right)\volH \\
&= \left( \ac_A X^B \sD_B \Upsilon^A - \ac_B (\Upsilon_A +\ac_A) \sD^A X^B + X_\sD^{\langle A \rangle} \ac_{\langle A} \ac_{B \rangle} + \tfrac{1}{2} X_\sD (\ac \cdot \ac) \right)\volH, \\
\end{aligned}
\end{equation}
where we added and subtracted $\ac_B \ac_A \sD^A X^B$, then performed the trace-traceless splitting. We now need to manipulate the first two terms. Using the Stokes theorem \eqref{Stokes}, we can convert $\ac_A$ to the covariant derivative $\sD_A$ as
\begin{equation}
\lb{pi-Upsilon-1}
\begin{aligned}
&\left( \ac_A X^B \sD_B \Upsilon^A - \ac_B (\Upsilon_A +\ac_A) \sD^A X^B\right)\volH \\
&  = \left[- X^B \sD_A \sD_B \Upsilon^A + \sD^A X^B \sD_B \ac_A + (\Upsilon^A + \ac^A)\sD_B \sD_A X^B \right]\volH \\
&\ \ \ \  + \dH \left[ (X^B \sD_B \Upsilon^A - (\Upsilon^B + \ac^B) \sD_B X^A)\bm{\epsilon}_A\right] \\
&  = \left[2 \sD^{A} X^{B} \left(\sD_{(A} \ac_{B)}-  \sD_{[A} \ac_{B]}\right) - \sD^A X^B \sD_B \ac_A  - X^B \sD_A \sD_B \Upsilon^A + (\Upsilon^A + \ac^A)\sD_B \sD_A X^B \right]\volH \\
&\ \ \ \  + \dH \left[ (X^B \sD_B \Upsilon^A - (\Upsilon^B + \ac^B) \sD_B X^A)\bm{\epsilon}_A\right] \\
& = \bigg[\sD^{A} X^{B}\left(2 \sD_{(A} \ac_{B)}\right) -\sD^{A} X^{B} \cL_\v \vor_{AB} + \tfrac{1}{2}X^A \sD_A ( \ac \cdot \ac)  +X^B \sD_A \sD_B (\ac^A- \Upsilon^A)  \\
&\ \ \ \ + (\Upsilon^A + \ac^A)\sD_B \sD_A X^B \bigg]\volH  + \dH \left[ (X^B \sD_B (\Upsilon^A-\ac^A) - (\Upsilon^B + \ac^B) \sD_B X^A)\bm{\epsilon}_A\right] \\
& = 2\bigg[X_\sD^{\langle AB\rangle } \sD_{\langle A} \ac_{B\rangle} -X_\sD^{[AB]}\tfrac{1}{2} \cL_\v \vor_{AB}+ X_\sD\left( \tfrac{1}{2}\sD \cdot \ac-\tfrac{1}{4} \ac \cdot \ac\right)- \tfrac{1}{4}X_\ac( \ac \cdot \ac) -X^B \sD_A \sD_B \p^A \\
& \qquad  + (\p^A + \ac^A)\sD_B \sD_A X^B \bigg]\volH + \dH \left[ 2\left(X^B \sD_B \p^A - (\p^B + \ac^B) \sD_B X^A + \tfrac{1}{4}(\ac \cdot \ac)X^A\right)\bm{\epsilon}_A\right]
\end{aligned}
\end{equation}
where to get the second equality, we used the decomposition $\sD_B \ac_A = \sD_{(A} \ac_{B)} -\sD_{[A} \ac_{B]}$. To obtain the third equality, we recalled the Jacobi identity \eqref{Jacobi-1} that $2\sD_{[A} \ac_{B]} = \v[\vor_{AB}]$, used the Leibniz rule on $\sD^A X^B \sD_B \ac_A = \sD_A( X^B \sD_B \ac^A) -  X^B \sD_A\sD_B \ac^A$, then wrote $X^B \ac_A \sD_B \ac^A = \frac{1}{2} X^A \sD_A (\ac \cdot \ac)$. We used the Stokes theorem, the trace-traceless splitting $\sD_{(A} \ac_{B)} =\sD_{\langle A} \ac_{B\rangle} + \frac{1}{2} (\sD \cdot \ac) q_{AB}$, and the relation $2\p_A = \Upsilon_A - \ac_A$ to arrive at the final equation. 

We then evaluate the terms involving two horizontal derivatives. Recalling the definition of the Riemann-Carroll tensor \eqref{Rie-Car}, we show that
\begin{equation}
\begin{aligned}
&\left( -X^B \sD_A \sD_B {\p}^A + ({\p}^A + {\ac}^A)\sD_B \sD_A X^B\right)\volH \\
& = \left( -X^B [\sD_A, \sD_B] {\p}^A - X^A \sD_A (\sD \cdot {\p}) + ({\p}^A + {\ac}^A)[\sD_B, \sD_A] X^B + ({\p}^A + {\ac}^A) \sD_A (\sD \cdot X) \right)\volH  \\
& =  \bigg( -X_\vor^A\left( \cL_\v\p^A + \theta (\p_A+\ac_A)\right)+ \RC_{AB}{\ac}^AX^B +\mr{X}^A\vor_{AB}(\p^B + \ac^B) -X_\sD\left(\sD \cdot {\ac} + {\ac}\cdot ({\ac}+{\p})\right) \\
&\qquad  + X_\ac (\sD \cdot {\p})\bigg)\volH + \dH \left[  \left(X_\sD(\p^A + \ac^A)-X^A (\sD\cdot{\p}) \right) \bm{\epsilon}_A  \right] \\
& =  \bigg(  - X_\vor^A \left(\Lie_\v\p_A - 2\s_A{}^B\p_B + \theta \ac_A\right)+  \RC_{AB}{\ac}^AX^B +\mr{X}^A\vor_{AB}(\p^B + \ac^B) -X_\sD\left(\sD \cdot {\ac} + {\ac}\cdot ({\ac}+{\p})\right) \\
&\qquad + X_\ac (\sD \cdot {\p})\bigg)\volH+ \dH \left[  \left(X_\sD(\p^A + \ac^A)-X^A (\sD\cdot{\p}) \right) \bm{\epsilon}_A  \right]
\end{aligned}
\end{equation}
where we also used the Stokes theorem \eqref{Stokes}. The term $\RC_{AB}\ac^AX^B$ can be further manipulated using the Stokes theorem \eqref{Stokes} and the differential Bianchi identity \eqref{Bianchi-Ric} as follows
\begin{equation}
\begin{aligned}
\RC_{AB}{\ac}^AX^B \volH & = \bigg[-X^A\sD^B {\RC}_{BA} - {\RC}_{AB} \sD^A X^B \bigg] \volH + \dH \left( {\RC}^A{}_B X^B\bm{\epsilon}_A \right) \\
& = \bigg[- \tfrac{1}{2} X^A\sD_A {\RC} - X^A \sD_B \left( {\theta}^{(BC)} {\vor}_{CA}\right) -X_\vor^A (\sD_B + {\ac}_B) (\theta_{(A}{}^{B)} - \theta \delta_A{}^B) \\
& \ \ \ \ + X^A {\vor}^{BC} (\sD_C + {\ac}_C) {\theta}_{(AB)} - X_\sD^{\langle AB \rangle}{\RC}_{\langle AB \rangle} - \tfrac{1}{2} X_\sD {\RC} - \tfrac{1}{2} X_\sD^{[ AB ]} {\theta}{\vor}_{AB} \bigg]\volH \\
& \ \ \ \  + \dH \left( \RC^A{}_B X^B\bm{\epsilon}_A \right) \\
& = \bigg[- X_\sD^{\langle AB \rangle}\RC_{\langle AB \rangle} - \tfrac{1}{2} X_\sD^{[ AB ]} \theta\vor_{AB} + \tfrac{1}{2} X_\ac \RC-X_\vor^A \left( \sD_B (\theta_{(A}{}^{B)}- \theta \delta_A^B) - \theta \ac_A\right)   \\
& \ \ \ \ - X^A (\sD_B+\ac_B) \left( {\theta}^{(BC)} {\vor}_{CA}\right) + X^A {\vor}_B{}^C (\sD_C + {\ac}_C) {\theta}_{(A}{}^{B)} \bigg] \volH  \\
& \ \ \ \  + \dH \left(  ({\RC}^A{}_B X^B - \tfrac{1}{2} {\RC} X^A)\bm{\epsilon}_A \right),
\end{aligned}
\end{equation}
where we used the antisymmetric components of the Ricci-Carroll tensor \eqref{Ric-anti} on the second equality, and we used integration by parts and the Stokes theorem to arrive at the result. Combining with the previous result, we thus obtain 
\begin{equation}
\begin{aligned}
&\left[ -X^B \sD_A \sD_B {\p}^A + ({\p}^A + {\ac}^A)\sD_B \sD_A X^B\right]\volH \\
& =  \bigg[- X_\sD^{\langle AB \rangle}\RC_{\langle AB \rangle} - \tfrac{1}{2} X_\sD^{[ AB ]} \theta\vor_{AB} -X_\sD\left(\sD \cdot {\ac} + {\ac}\cdot ({\ac}+{\p})\right)+ X_\ac (\sD \cdot {\p} + \tfrac{1}{2}{\RC}) \\
& \qquad +\mr{X}^A\vor_{AB}(\p^B + \ac^B) -X_\vor^A\left( \Lie_\v \p_A- 2 \s_A{}^B \p_B + \sD_B (\btheta_{(A}{}^{B)]} - \tfrac{1}{2}\theta \delta_A^B) \right)  \\
& \qquad - X^A (\sD_B+{\ac}_B) ({\theta}^{(BC)} {\vor}_{CA})+X^A {\vor}_B{}^C (\sD_C + {\ac}_C) {\theta}_{(A}{}^{B)} \bigg]\volH \\
& \ \ \ \ + \dH \left(  \left({\RC}^A{}_B X^B +X_\sD({\p}^A + {\ac}^A)-X^A (\sD\cdot{\p} + \tfrac{1}{2} {\RC}) \right) \bm{\epsilon}_A  \right).  
\end{aligned}
\end{equation}
Before getting back to the equation \eqref{pi-Upsilon-1}, let us consider the following combinations
\begin{equation}
\begin{aligned}
&\bigg[- \tfrac{1}{2} X_\sD^{[ AB ]}(\cL_\v+ \theta)\vor_{AB} - X^A (\sD_B+{\ac}_B ) ({\theta}^{(BC)} {\vor}_{CA})+X^A {\vor}_B{}^C (\sD_C + {\ac}_C) {\theta}_{(A}{}^{B)} \bigg]\volH \\
& = \bigg[\tfrac{1}{2}\left(\cL_\v X_\sD^{[AB]}\right) \vor_{AB} - X^A (\sD_B+{\ac}_B ) ({\theta}^{(BC)} {\vor}_{CA})+X^A {\vor}_B{}^C (\sD_C + {\ac}_C) {\theta}_{(A}{}^{B)} \bigg]\volH \\
& \ \ \ \  +\dH \left( -\tfrac{1}{2} X_\sD^{[ AB ]}\vor_{AB}  \volS  \right) \\
& = \bigg[\tfrac{1}{2}\left(\cL_\v\sD_A X^B\right)\vor^A{}_B - {\vor}_{CA} {\theta}^{(BC)} \sD_B X^A - X^A (\sD_B+{\ac}_B ) ({\theta}^{(BC)} {\vor}_{CA})+X^A {\vor}_B{}^C (\sD_C + {\ac}_C) {\theta}_{(A}{}^{B)} \bigg]\volH \\
& \ \ \ \  +\dH \left( -\tfrac{1}{2} X_\sD^{[ AB ]}\vor_{AB}  \volS  \right) \\
& = \bigg[\tfrac{1}{2}{\vor}^A{}_B (\sD_A + {\ac}_A) \cL_\v X^B \bigg]\volH  +\dH \left( -\tfrac{1}{2} X_\sD^{[ AB ]}\vor_{AB} \volS+  X^C {\vor}_C{}^B {\theta}^{(A}{}_{B)}   \bm{\epsilon}_A  \right) \\
& = \bigg[\tfrac{1}{2} \mr{X}^A \sD_B {\vor}_A{}^B \bigg]\volH  +\dH \left(-\tfrac{1}{2} X_\sD^{[ AB ]}\vor_{AB}\volS +  \left(X^C {\vor}_C{}^B {\theta}_{(B}{}^{A)}  + \tfrac{1}{2} \mr{X}^B {\vor}^A{}_B \right) \bm{\epsilon}_A  \right), 
\end{aligned}
\end{equation}
where we used the Stokes theorem \eqref{Stokes} to obtain the first equality, then used that ${\v}[q^{AB}] = 2 {\theta}^{(AB)}$ to get the second equality. On the third equality, we employed the formula \eqref{ell-D-X} and the Stokes theorem. We applied again the Stokes theorem to arrive at the last equality. 

Assembling these results together, the equation \eqref{pi-Upsilon-1} becomes 
\begin{equation}
\begin{aligned}
&\tfrac{1}{2}\left( {\ac}_A X^B \sD_B {\Upsilon}^A - {\ac}_B ({\Upsilon}_A +{\ac}_A) \sD^A X^B\right)\volH \\
& = \bigg[X_\sD^{\langle AB \rangle}\left( \sD_{\langle A} {\ac}_{B\rangle} - {\RC}_{\langle AB \rangle}\right)+ X_\sD \left( -\tfrac{1}{2}\sD \cdot {\ac}-{\ac}\cdot ({\ac}+{\p})-\tfrac{1}{4} {\ac} \cdot {\ac}\right) \\
& \qquad +X_\ac \left(\sD \cdot {\p} - \tfrac{1}{4} {\ac} \cdot {\ac} + \tfrac{1}{2}{\RC} \right) + \mr{X}^A \left( {\vor}_{AB}({\p}^B + {\ac}^B) +\tfrac{1}{2} \sD_B {\vor}_A{}^B\right)  \\
&\qquad  -X_\vor^A\left( \cL_\v\p_A- 2 {\s}_A{}^B {\p}_B + \sD_B (\s_A{}^B- \tfrac{1}{2}\theta \delta_A^B) \right)   \bigg]\volH \\
& \ \ \ \  + \dH \bigg[  \bigg(X^B(\sD_B {\p}^A+{\vor}_B{}^C {\theta}_{(C}{}^{A)}+{\RC}^A{}_B) -({\p}^B + {\ac}^B)\sD_B X^A +X_\sD (\p^A + \ac^A)  \\
& \qquad \qquad \qquad-X^A (\sD\cdot{\p} + \tfrac{1}{2} {\RC} - \tfrac{1}{4} {\ac} \cdot {\ac})  + \tfrac{1}{2} \mr{X}^B {\vor}^A{}_B \bigg) \bm{\epsilon}_A -\tfrac{1}{2} X_\sD^{[ AB ]}\vor_{AB} \volS \bigg]
\end{aligned}
\end{equation}
Finally, by recalling the relation $2{\p}_A = {\Upsilon}_A - {\ac}_A$, we can write  \eqref{pi-Upsilon-term} in the form
\begin{equation}
\begin{aligned}
&{\p}_A \cL_X {\Upsilon}^A \volH \\
& = -\bigg[X_\sD^{\langle AB \rangle}\left( \sD_{\langle A} {\ac}_{B\rangle} + \tfrac{1}{2}{\Upsilon}_{\langle A} {\Upsilon}_{B \rangle} + \tfrac{1}{2} {\ac}_{\langle A} {\ac}_{B \rangle} - {\RC}_{\langle AB \rangle}\right)+ X_\sD \left( -\tfrac{1}{2}(\sD+{\ac}) \cdot {\ac} + {\p}\cdot {\Upsilon}\right) \\
& \qquad + X_\ac \left(\sD \cdot {\p} + \tfrac{1}{2}{\RC} + {\p} \cdot ({\p} + {\ac})\right) + \mr{X}^A \left( {\vor}_{AB}({\p}^B + {\ac}^B) +\tfrac{1}{2} \sD_B {\vor}_A{}^B\right)  \\
&\qquad -X_\vor^A\left( \cL_\v\p_A- 2 {\s}_A{}^B {\p}_B + \sD_B (\s_A{}^B- \tfrac{1}{2}\theta \delta_A^B) \right)\bigg]\volH \\
& \ \ \ \  - \dH \bigg[  \bigg(X^B(\sD_B {\p}^A+{\vor}_B{}^C \theta_{(C}{}^{A)}+{\RC}^A{}_B) -({\p}^B + {\ac}^B)\sD_B X^A +X_\sD ({\p}^A + {\ac}^A)  \\
& \qquad \qquad \qquad-X^A (\sD\cdot{\p} + \tfrac{1}{2} {\RC} + {\p} \cdot ({\p} + {\ac}))  + \tfrac{1}{2} \mr{X}^B {\vor}^A{}_B \bigg) \bm{\epsilon}_A -\tfrac{1}{2} X_\sD^{[ AB ]}\vor_{AB} \volS \bigg].
\end{aligned}
\end{equation}

\subsection{On the shear term} \lb{App:shear} 

To manipulate this term, let us start by considering,  
\begin{equation}
\begin{aligned}
&\btheta_{ab} \Lie_X \E^{ab} \\
& = \tfrac{1}{2}\btheta_{ab} \Lie_X \left( -\Lie_n q^{ab} +   2\v^{(a} \ac^{b)} - 4k^{(a} \sD^{b)}\rho  \right) \\
& = \tfrac{1}{2}\btheta_{ab}\left( -\Lie_n \Lie_X + \Lie_{[n,X]} \right)  q^{ab}  + \btheta_{(ab)} [X, \v]^a \ac^b -2 \btheta_{(ab)}[X, k]^a \sD^b \rho  \\
& = \btheta_{(AB)}\left(\cL_n \left( \sD^{A}X^{B}\right) + 2\rho X_\vor^A \Upsilon^B - \sD^A \cL_\v X^B - (X\cdot \ac)\theta^{AB} + 2 X[\rho] \btheta^{AB} - \ac^A \cL_\v X^B \right)  \\
& = -X_\sD^{(AB)} \left(  \cL_\v \btheta_{AB} +\theta \btheta_{AB} +\rho \left( 2\btheta_{CA} \btheta_{B}{}^C +  \btheta_{CA} \btheta^C{}_{B}  +  \btheta_{AC} \btheta_B{}^C\right)\right) - X_\sD \left( \rho \btheta_{(AB)} \btheta^{AB}  \right) \\
&\ \ \ \  - X_\ac \left( \btheta_{(AB)} (\theta^{AB} + \rho \btheta^{AB}) \right) + \mr{X}^A\sD_B \btheta_{(A}{}^{B)} + X_\rho \left( \btheta_{(AB)} \btheta^{AB} \right)  \\
& \ \ \ \ +\left( \cL_\v + \theta\right) ( X_\sD^{(AB)}\btheta_{AB} )  +  (\sD_A + \ac_A)  \left(- \mr{X}^B \btheta_{(B}{}^{A)}+\rho \btheta_{(BC)} \btheta^{BC} X^A \right)
\end{aligned} 
\end{equation}
where we imposed the condition $\cL_k X^A = 0$ on the third equality, and employed the radial expansion \eqref{Lie-k-DX} to obtain the fourth equation.

In a similar manner, we can write $q_{ab} \Lie_X \E^{ab}$ as
\begin{equation}
\begin{aligned}
&q_{ab} \Lie_X \E^{ab} \\
& = q_{AB}\left(\cL_n \left( \sD^{A}X^{B}\right) +2\rho X_\vor^A \Upsilon_A - (\sD^A+\ac^A) \cL_\v X^B - (X\cdot \ac)\theta^{AB} + 2 X[\rho] \btheta^{AB}  \right)  \\
& =   -2 X_\sD^{(AB)}\NN_{AB} - X_\sD \E- X_\ac \E + \left(\cL_\v+\theta\right) X_\sD+ (\sD_A+\ac_A)(-\mr{X}^A +2\rho\btheta X^A).
\end{aligned} 
\end{equation}
Using that $\cL_X \E = q_{ab}\cL_X \E^{ab} + \E^{ab} \cL_Xq_{ab}$, we show that
\begin{equation}
\begin{aligned}
\Lie_X \E =  - X_\sD \E- X_\ac \E + \left(\cL_\v+\theta\right) X_\sD+ (\sD_A+\ac_A)(-\mr{X}^A +2\rho\btheta X^A).
\end{aligned} 
\end{equation}
With these results, we have
\begin{equation}
\begin{aligned}
&\btheta_{ab} \Lie_X \T^{ab} \\
& = -X_\sD^{\langle AB\rangle} \left(  \cL_\v \bs_{\la AB \ra} + \btheta \T_{AB} +\rho \left( 2\btheta_{C\la A} \btheta_{B \ra}{}^C +  \btheta_{C \la A} \btheta^C{}_{B \ra}  +  \btheta_{\la AC} \btheta_{B \ra}{}^C-4\btheta \bs_{AB}\right)\right) \\
& \ \ \ \ - X_\sD \left( \T^{AB} \bs_{AB} +\rho \bs_{AB} \bs^{AB}   \right) - X_\ac \left( \T^{AB} \bs_{AB} -\rho \bs_{AB} \bs^{AB}  \right) \\
&\ \ \ \  + \mr{X}^A\sD_B \bs_{A}{}^{B}  + X_\rho \left( \bs_{AB} \bs^{AB} \right)  \\
& \ \ \ \ +\left( \cL_\v + \theta\right) ( X_\sD^{\langle AB \rangle }\bs_{AB} )  +  (\sD_A + \ac_A)  \left(- \mr{X}^B \bs_{B}{}^{A}+\rho \bs_{BC} \bs^{BC} X^A \right)
\end{aligned} 
\end{equation}
Using the Leibniz rule and the Stokes theorem \eqref{Stokes}, we finally arrive at the following result
\begin{equation}
\begin{aligned}
&-\left(\T^{AB} \Lie_X \btheta_{AB}\right) \volH \\
& = \left( \btheta_{AB} \Lie_X\T^{AB} + (X_\sD + X_\ac)(\T^{AB} \bs_{AB})-(\sD_A+\ac_A)(\T^{BC} \bs_{BC} X^A) \right)\volH \\
& =\bigg[-X_\sD^{\langle AB\rangle} \left(  \cL_\v \bs_{\la AB \ra} + \btheta \T_{AB} +\rho \left( 2\btheta_{C\la A} \btheta_{B \ra}{}^C +  \btheta_{C \la A} \btheta^C{}_{B \ra}  +  \btheta_{\la AC} \btheta_{B\ra}{}^C-4\btheta \bs_{AB}\right)\right) \\
& \qquad - X_\sD \left( \rho \bs_{AB} \bs^{AB}   \right) + X_\ac \left(\rho \bs_{AB} \bs^{AB}  \right) + \mr{X}^A\sD_B \bs_{A}{}^{B}  + X_\rho \left( \bs_{AB} \bs^{AB} \right)  \bigg] \volH  \\
&\qquad   + \dH \left[ X_\sD^{\langle AB \rangle }\bs_{AB} \volS - \left( \mr{X}^B \bs_{B}{}^{A}+(\T_{BC}-\rho \bs_{BC}) \bs^{BC} X^A \right) \bm{\epsilon}_A \right].
\end{aligned}
\end{equation}

\subsection{Radial evolution of the Einstein-Hilbert charge} \lb{app:radial-EH}

For comparison, let us also compute the radial evolution of the charge associated with the horizontal diffeomorphism in the Einstein-Hilbert theory. It follows from the relation \eqref{Q-EH} that the radial development of the Einstein-Hilbert charge $Q^{\sss \mathrm{EH}}_X$ differs from those of the canonical charge by the boundary Lagrangian term and the corner symplectic, 
\begin{equation}
\begin{aligned}
 \Lie_k Q^{\sss \mathrm{EH}}_X &=  \Lie_k Q^{\mathrm{can}}_X + \int_{\pa \H} \Lie_k (\Wein X^a \bvol_a) - \tfrac{1}{2}\int_{\pa \H} \Lie_k\left(\delta_X \v^a \bvol_a \right),
 \end{aligned}
\end{equation}
where the contraction of the bulk Einstein-Hilbert Lagrangian, $\iota_X \bm{L}^{\sss \mathrm{EH}}$, vanishes when pulled back on the surface $\H$. When considering the null case with vanishing vorticity and with $\pa \H = \S_\sv$, the boundary Lagrangian term is also zero, since $X^a \bvol_a \stackrel{\sss \S_\sv}{=} 0$.   

It then follows from $\delta_X \v^a \bvol_a = (k_a \delta_X \v^a) \volS = - \delta_X k_\v \volS$ that the corner contribution reads
\begin{equation}
\begin{aligned}
- \tfrac{1}{2}\int_{\S_\sv} \Lie_k\left(\delta_X \v^a \bvol_a \right) =  \tfrac{1}{2}\int_{\S_\sv} \Lie_k\left(\delta_X k_\v \volS \right) &= \tfrac{1}{2}\int_{\S_\sv} \left(\Lie_k (\delta_X k_\v) + \btheta \delta_X k_\v \right) \volS  = \tfrac{1}{2}\int_{\S_\sv} X_\ac \btheta \volS,
 \end{aligned}
\end{equation}
where we recalled the variation \eqref{sCarrollian-transform} under horizontal diffeomorphism, $\delta_X k_\v = X^A \ac_A$, and used the conditions that $\Lie_k X^A = 0$ and $\Lie_k \ac_A =0$ when the vorticity vanishes \eqref{Jacobi-3}. We finally obtain the radial equation for the Einstein-Hilbert charge
\begin{equation}
\begin{aligned}
 \Lie_k Q^{\sss \mathrm{EH}}_X = \ & -\int_\N \left(X^{\la AB \ra}_\sD G_{\la AB\ra} +  X_\sD(\tfrac12 q^{AB}G_{AB}) +X_\ac G_{\v k}+ \mr{X}^A G_{k A} \right) \volN \\
 & - \int_{\S_\sv} \left( X_\sD^{\la AB \ra} \bs_{AB} - \tfrac12 X_\sD \btheta - \tfrac12 X_\ac \btheta\right)\volS.
\end{aligned}
\end{equation}
The evolution equations $\tfrac{1}{2} q^{AB}G_{AB} = 0$ and $G_{\v k} = 0$ are associated with the charge aspect given by the transverse expansion $\btheta$.

\bibliography{Biblio.bib}
\bibliographystyle{Biblio}

\end{document}